\documentclass[onecolumn]{emulateapj}
\usepackage{graphics,graphicx,epsf,natbib}
\slugcomment{ApJ, accepted 28 Feb 2011}

\shorttitle{PAHs in Cluster Cool Core BCGs}
\shortauthors{Donahue et al.}

\begin{document}


\title{PAHs, Ionized Gas, and Molecular Hydrogen in \\
Brightest Cluster Galaxies of Cool Core Clusters of Galaxies}


\author{Megan Donahue}
\affil{Physics and Astronomy Dept., Michigan State University, East Lansing, MI, 48824 
\email{donahue@pa.msu.edu} }

\author{Genevi\`{e}ve E. de Messi\`{e}res, Robert W. O'Connell}
\affil{Astronomy Dept., University of Virginia, Charlottesville, VA}

\author{G. Mark Voit, Aaron Hoffer}
\affil{Physics and Astronomy Dept., Michigan State University, East Lansing, MI, 48824, \email{voit\@pa.msu.edu,hofferaa\@msu.edu} }

\author{Brian R. McNamara\altaffilmark{1,}\altaffilmark{2}}
\affil{Department of Physics and Astronomy, University of Waterloo, 200 University Avenue West, Waterloo, Ontario, N2L 3G1 Canada}
\altaffiltext{1}{Perimeter Institute for Theoretical Physics, 31 Caroline St. N., Waterloo, Ontario, N2L 2Y5, Canada}
\altaffiltext{2}{Harvard-Smithsonian Center for Astrophysics, 60 Garden Street, Cambridge, MA 02138, USA }

\author{Paul E. J. Nulsen}
\affil{Harvard-Smithsonian Center for Astrophysics, 60 Garden Street, Cambridge, MA 02138, USA}

\begin{abstract}

We present measurements of 5-25 $\mu$m emission features  
of brightest cluster galaxies (BCGs) with strong optical emission lines in a sample of 9 cool-core
clusters of galaxies observed with the Infrared Spectrograph on board the Spitzer Space
Telescope. These systems provide a view of dusty molecular gas and star formation, surrounded
by dense, X-ray emitting intracluster gas. Past work has shown that 
BCGs in cool-core clusters may host powerful radio sources, luminous optical emission line systems,
and excess UV, while BCGs in other clusters never show this activity. In this sample,   
we detect polycyclic aromatic hydrocarbons (PAHs), extremely luminous, rotationally-excited
molecular hydrogen line emission, forbidden line emission from ionized gas ([Ne~II] and [Ne~III]), and infrared 
continuum emission from warm dust and cool stars.  We show here that these BCGs exhibit
more luminous forbidden neon and H$_2$ rotational line emission than star-forming galaxies with
similar total infrared luminosities, as well as somewhat higher ratios of 70 $\mu$m / 24 $\mu$m luminosities. 
Our analysis suggests that while star formation
processes dominate the heating of the dust and PAHs, a heating process consistent with
suprathermal electron heating from the hot gas, distinct from 
star formation, is heating the molecular gas and contributing to the heating of the ionized
gas in the galaxies. The survival of PAHs and dust suggests that dusty gas is somehow shielded
from significant interaction with the X-ray gas.

\end{abstract}

\keywords{galaxies: clusters: intracluster medium, galaxies: starburst, dust, infrared: galaxies, }

\section{Introduction}

Infrared spectroscopy provides critical clues about the power sources of luminous
galaxies whose energy sources are shielded from visual inspection by layers of dust and gas
\citep{1998ApJ...498..579G,1998ARA&A..36..189K,2000A&A...359..887L}, including 
star formation activity  and AGN \citep[e.g.,][]{2001A&A...372..427R}. 
The Infrared Spectrograph (IRS) 
on board the Spitzer Space Telescope  \citep{2004ApJS..154...18H}, 
exploiting the sensitivity and spatial resolution of {\em Spitzer}, delivered stunning infrared spectra from galaxies
of many types. A key project,
Spitzer Infrared Nearby Galaxy Survey \citep{2003PASP..115..928K}, as a survey of nearby galaxies ($D<30$ Mpc), 
was limited to mainly spirals and a few ellipticals. 
Studies targeting the brightest infrared galaxies in the sky \citep[e.g., ][]{2009PASP..121..559A} included 
mainly the most luminous IR galaxies (LIRGs) and nearby 
IR-bright, star-forming galaxies. {\em Spitzer} programs such as these have produced a treasure of
infrared spectra of galaxies as well as improved and standardized
techniques for measuring infrared features \citep[e.g.,][]{2009ApJ...693.1821D}. These recent theoretical and
observational efforts have identified  useful infrared diagnostics, which now allow a physical interpretation
of the spectra based on models \citep{2006ApJ...646..161D,2007ApJ...667..149F,2007ApJ...656..770S}.

We explore here the infrared spectral signatures of Brightest Cluster Galaxies (BCGs).
BCGs are not common even in large samples of galaxies, 
because massive clusters themselves are rare (of order a few in a box 100 Mpc on a side), so in general, to
study an interestingly large and bright sample of BCGs, they must be specially targeted. 
Although most BCGs have red colors and are dust-free, suggesting
little star formation, some 15-25\% show evidence of significant star
formation (rates up to $\sim 100\, \rm{M}_\odot~\rm{yr}^{-1}$) in their UV and optical
continua \citep[e.g., ][]{1987MNRAS.224...75J,1989AJ.....98.2018M,1994ARA&A..32..277F,1999MNRAS.306..857C,
2005ApJ...635L...9H,2008ApJ...687..899R,2008MNRAS.389.1637B,2010ApJ...719.1844H,2010ApJ...715..881D}.

Star-forming BCGs seem to be exclusively found in 
the centers of clusters whose hot intracluster medium (ICM) cores exhibit gas cooling 
times shorter than about 1 billion years or low hot gas entropies ($K$) where $K = kT n_e^{-2/3} < 30$
keV cm$^{-2}$, \citep[e.g.][]{2008ApJ...683L.107C,2008ApJ...687..899R,2010A&A...513A..37H} 
These clusters, once known as cooling flows, are called ``cool-core" clusters. About half of the nearby
X-ray luminous clusters fall into this category. 
This trend for cooling flow clusters to host BCGs with powerful emission line nebulae 
was first found by \citet{1985ApJS...59..447H}, who noted a $\sim 14$ billion year 
threshold, limited by the far cruder X-ray data available at the time. 
Then, astronomers suspected that gas cooling from the hot phase was somehow
related to these nebular emission line systems although the emission lines themselves were too
bright to be generated directly by cooling gas. Almost two decades later, 
X-ray Multi-Mirror (XMM) spectroscopy failed to show [Fe~XVII] and [O~VII] emission lines, 
conclusively demonstrating that very 
little X-ray emitting gas existed at temperatures 1/2-1/3 the temperature of most of the 
ICM \citep[e.g.,][]{2003ApJ...590..207P}. 
However, the disproof of the simplest massive cooling flow model did not explain
why BCGs in these systems frequently exhibited properties indicative of activity: extended emission line systems
\citep{1989ApJ...338...48H}, including vibrationally-excited molecular hydrogen at $1000-2000$ K 
\citep{1994ASSL..190..169E, 1997MNRAS.284L...1J, 2000ApJ...545..670D}, CO masses indicating
cold H$_2$ at $\sim100$ K \citep{2001MNRAS.328..762E}, UV excesses 
\citep[most recently, ][]{2010ApJ...719.1619O, 2010ApJ...719.1844H, 2010ApJ...715..881D}, 
radio sources \citep[e.g., ][]{1990AJ.....99...14B, 2008ApJ...683L.107C}, 
IR emission from warm dust \citep{2006ApJ...647..922E,2007ApJ...670..231D,2008ApJ...681.1035O}.
It is important to note that the XMM spectral results ruled out the enormous X-ray cooling rates
inferred from the simple cooling flow model ($\sim100$s M$_\odot$ yr$^{-1}$), but do not provide limits near the 
cooling rates similar to the typical star formation rates estimated for these BCGs ($\sim1-10$s M$_\odot$ yr$^{-1}$), 
with quantities an order of magnitude higher for the most extreme systems.
Early Herschel results of a few classic examples of these active brightest 
cluster galaxies in cool-core clusters reveal far-IR spectra of similar sources 
that are consistent with the short wavelength {\em Spitzer} observations: strong peaks in 
broadband photometry from dust warmed by recently formed stars 
and powerful interstellar coolant lines of OI (63 $\mu$m) and CII (153 $\mu$m) 
\citep{2010A&A...518L..46E,2010A&A...518L..47E}, and {\em Spitzer} IRS measurements of individual BCGs reveal
that some, like Zwicky 3146, have not only have powerful IR emission from warm dust but unusually luminous molecular hydrogen 
\citep{2006ApJ...652L..21E}, while others, like NGC4696, have luminous molecular hydrogen but only
faint dust emission \citep{2008ApJ...684..270K}.

These brightest cluster galaxies also pose a challenge to galaxy formation
models. The so-called ``over-cooling" problem in galaxy formation simulations creates massive galaxies
that are bluer, even more luminous, and with higher star formation rates than observed 
\citep[see][]{2001MNRAS.326.1228B,2006MNRAS.365...11C,2006MNRAS.370..645B}. 
To remedy this situation, models must include AGN feedback in the form of non-radiative energy 
in addition to stellar feedback to quench the formation of stars \citep[e.g., ][]{2005ApJ...620L..79S}
and prevent the rapid cooling of hot intergalactic gas \citep[e.g., ][]{2001ApJ...554..261C,2007ARA&A..45..117M}. 
Furthermore, the accretion of hot gas has been proposed as the dominant mode for forming
the most massive ($>10^{11.4}~\rm{M}_\odot$) galaxies \citep[e.g., ][]{2005MNRAS.363....2K}.

Conveniently, BCGs provide a laboratory for this type of galaxy formation. 
Chandra observations clearly show AGN interactions with the hot ICM
in the form of cavities in the hot atmospheres of clusters and galaxies.  The mechanical energy
associated with these cavities is is sufficient to offset cooling \citep{2004ApJ...607..800B,2006MNRAS.373..959D}, 
and to somehow regulate or quench star formation in most systems \citep{2006ApJ...652..216R}.
However, energetic feedback from AGN is apparently unable to offset cooling entirely in all BCGs, and
these systems that are struggling to offset
rapid cooling are rich in cold gas and star formation \citep{2008ApJ...681.1035O}.  
The correlation between  star formation and short central cooling times in 
the hot gas shows that the gas fueling star formation may well have cooled from the hot ICM  
\citep{2008ApJ...683L.107C, 2008ApJ...687..899R}.
Assessing and decoding the state of gas, dust, and stars in these galaxies, using {\em Spitzer} spectra, 
will yield clues about which physical processes are most relevant 
in determining a system's appearance and star formation rate during accretion of hot gas.

We present here Spitzer Infrared Spectrograph (IRS) \citep{2004ApJS..154...18H} 
measurements of a sample of nine BCGs residing in cool-core clusters. 
We describe the measurement procedures, including the scaling we applied to the data to match aperture 
photometry, in \S2. We present our results in \S3. In \S4, we compare the full spectra to a set of simulated
time-averaged starburst spectral energy distributions (SEDs). 
In \S5 we compare the emission line and polycyclic aromatic hydrocarbons (PAH) ratios and correlations
seen in our sample to those seen in other types of galaxies. In \S6 we discuss the correlation and lack of
correlation in the various spectral components that suggest that at least two sources of heat must be
considered in order to interpret the observations of these systems. 
We assume $H_0=70$ km s$^{-1}$ Mpc$^{-1}$, and a flat, $\Omega_M=0.3$ cosmology throughout.

\section{Observations and Data Reductions}

%%%%%%%%%%%%%%%%

% Update this table using spitzer_obslog_table.pro.
\begin{deluxetable}{llcccrrcc}
\tabletypesize{\scriptsize}
\tablecaption{\label{obstable}Observation Log.}
\tablehead{
\colhead{Cluster} & \colhead{Redshift\tablenotemark{a}} & \colhead{{\it Spitzer}} & \colhead{IRS Mode} & \colhead{Obs Date} & \multicolumn{2}{c}{Duration (s)} & \multicolumn{2}{c}{\# of Slit Positions} \\
\colhead{BCG} & \colhead{} & \colhead{ID} & \colhead{} & \colhead{} & \colhead{SL} & \colhead{LL} & \colhead{SL} & \colhead{LL}
}
\startdata
    2A0335+096 & $  0.0347$ & $20345$ & $\rm{Staring}$ & $                 2006-09-16$ & $10156$ & $1698$ & $12$ & $12$ \\
     Abell 478 & $  0.0860$ & $20345$ & $\rm{Staring}$ & $                 2006-03-17$ & $10156$ & $1698$ & $12$ & $12$ \\
    Abell 1068 & $  0.1386$ & $ 3384$ & $\rm{Mapping}$ & $                 2005-04-21$ & $ 2925$ & $ 234$ & $ 6$ & $ 2$ \\
    Abell 1795 & $  0.0633$ & $ 3384$ & $\rm{Mapping}$ & $                 2005-02-07$ & $11701$ & $1879$ & $24$ & $16$ \\
    Abell 1835 & $  0.2520$ & $ 3384$ & $\rm{Mapping}$ & $                 2005-02-13$ & $ 2925$ & $ 234$ & $ 6$ & $ 2$ \\
    Abell 2597 & $  0.0821$ & $ 3384$ & $\rm{Mapping}$ & $2005-06-30\tablenotemark{b}$ & $ 6826$ & $ 704$ & $14$ & $ 6$ \\
       Hydra A & $  0.0549$ & $ 3384$ & $\rm{Staring}$ & $                 2005-12-14$ & $ 7313$ & $1509$ & $12$ & $12$ \\
MS 0735.6+7421 & $   0.216$ & $20345$ & $\rm{Staring}$ & $                 2006-04-25$ & $21280$ & $2768$ & $ 4$ & $ 4$ \\
   PKS 0745-19 & $  0.1028$ & $20345$ & $\rm{Staring}$ & $                 2006-05-16$ & $10640$ & $1384$ & $ 4$ & $ 4$
\enddata
\tablenotetext{a}{BCG redshift sources, from emission lines: 2A0335+096  \citep{2007AJ....134...14D}, Hydra A \citep{2004AJ....128.1558S}, Abell 1795 (CGCG 162-010) \citep{1993AJ....106..831H}, Abell 2597 (PKS 2322-12) \citep{1997ApJ...486..242V,2003astro.ph..6581C}, see also \url{http://www.mso.anu.edu.au/2dFGRS/}, Abell 478 (NVSS J041325+102754) \citep{1990ApJS...74....1Z}, PKS 0745-19 \citep{1978MNRAS.185..149H}, Abell 1068 (FIRST J104044.4+395712) \citep{1992MNRAS.259...67A}, MS0735 (ZwCl 1370 or BCG 4C $+74.12$) \citep{1991ApJS...76..813S}, Abell 1835 (SDSS J140102.07+025242.5) (SDSS DR2 ; see also \citet{1992MNRAS.259...67A}.}
\tablenotetext{b}{Also 2005-07-05.}
\end{deluxetable}

\begin{deluxetable}{lrcrccl}
\tabletypesize{\scriptsize}
\tablecaption{\label{obstable2}Spitzer MIPS/IRAC Observation Log.}
\tablehead{
\colhead{Cluster} & \colhead{IRAC AOR} & \colhead{Obs Date} & \colhead{Duration} & \colhead{MIPS AOR\tablenotemark{a}} & \colhead{Obs Date} & \colhead{Duration\tablenotemark{b}} \\
\colhead{BCG}     & \colhead{}         & \colhead{}         & \colhead{(s)}      & \colhead{}         & \colhead{}         & \colhead{(s)} 
}
\startdata
2A0335+096 & 18646528 & $2006-09-26$ & 108 & 18636544 & $2007-02-27$ & 400, 300 \\
Abell 478  & 11579904 & $2005-09-16$ & 1200 & 14944256 (1), 14944512 (2 \& 3) & $2006-02-22, 2006-03-02$ & 550, 1500, 1200 \\
Abell 1068 & 18650368 & $2006-12-27$ & 108 & 18638336 & $2006-12-08$ & 400, 320 \\
Abell 1795 & \nodata & \nodata & \nodata & 8788480 & $2004-07-11$  & 36, 42 \\
Abell 1835 & 4404480 & $2004-01-16$ & 3600 & 4764160(1), 4744448(2 \& 3) & $2004-02-20, 2005-06-28$ & 1800, 600, 150 \\
Abell 2597 & 13372160 & $2005-11-24$ & 3600 & 13371904 & $2005-06-18$ & 140, 150, 60 \\
Hydra A    & 26923008 & $2008-06-09$ & 3600 & 4707584 & $2004-05-04$ & 140, 120, 180 \\
MS0735.6+7421 & 7858688 & $2003-11-20$ & 500 &  \nodata &  \nodata &  \nodata \\
PKS 0745-19 & 18667776 & $2006-12-27$ & 108 & 18667520 & $2006-12-08$  & 400, 300 \\
\enddata
\tablenotetext{a}{Channel (1) is the 24-$\mu$m channel, Channels (2 \& 3) are the MIPS 70- and 160-$\mu$m channels respectively.}
\tablenotetext{b}{Two durations indicate total exposure time in seconds for 24- and 70-$\mu$m MIPS observations, respectively. Three durations indicate
exposure times for 24-, 70-, and 160-$\mu$m MIPS observation sequences, respectively.}
\end{deluxetable}

\subsection{Observations}
The {\it Spitzer} IRS 
observations took place in 2005 and 2006 (see Table \ref{obstable}), and the data were reprocessed in April 2009 (v18.7). 
Both short (SL) and long (LL) wavelength observations were obtained, at low spectral resolution, $R \sim 60-130$.
With two spectral orders each, we obtained a total of four spectral modules.
We have sparse spectral maps of nine cool-core BCGs, though we analyze 
only the central region here.  
We used IRSCLEAN v1.7 to apply the bad pixel mask supplied by the {\it Spitzer} pipeline and 
to find additional rogue pixels using a WCLEAN formula with an aggressive level of 0.5 (suitable for relatively faint targets such as these).
The LL pixels are $5.1\arcsec$ across, while the SL pixels are $1.8\arcsec$ across \citep{2004ApJS..154...18H}. 

To cross-check our flux calibration of the IRS spectroscopy, we also analyzed photometry 
data from the {\it Spitzer} Infrared Array Camera (IRAC) \citep{2004ApJS..154...10F} 
and the from the Multiband Imaging Photometer for Spitzer (MIPS) \citep{2004ApJS..154...25R}. 
We list the archival observations (known by their Astronomical Observation Requests or AORS)
in Table~\ref{obstable2}. The aperture photometry is discussed in \S~\ref{section:unc}.

%%%%%%%%%%%%%%%%%
% Update this table using spitzer_datared_table.pro.
\begin{deluxetable}{lccccccccc}
\tabletypesize{\scriptsize}
\tablecaption{\label{datared_table}Parameters used in data reduction.}
\tablehead{
\colhead{Cluster} & \colhead{Module} & \colhead{FWHM} & \colhead{FWHM} & \colhead{Type} & \multicolumn{2}{c}{Aperture center (J2000)} & \colhead{Aperture size} & \colhead{$\theta$\tablenotemark{a}} & \colhead{Factor\tablenotemark{b}} \\
\colhead{} & \colhead{} & \colhead{(\arcsec)} & \colhead{(kpc)} & \colhead{} & \colhead{RA} & \colhead{DEC} & \colhead{(\arcsec)} & \colhead{($^\circ$)} & \colhead{}
}
\startdata
 2A0335+096 & SL2 & $    $ & $    $ & Extended & $3:38:40.5$ & $9:58:12$ & $10.8 \times 10.8$ & $17$ & $ 1.16$ \\
            & SL1 & $ 8.0$ & $ 5.7$ &          & $3:38:40.6$ & $9:58:11$ & $10.8 \times 10.8$ & $17$ & $ 1.16$ \\
            & LL2 & $10.5$ & $ 7.5$ &          & $3:38:40.4$ & $9:58:9$ & $30.6 \times 15.3$ & $-80$ & $ 1.01$ \\
            & LL1 & $    $ & $    $ &          & $3:38:40.5$ & $9:58:10$ & $30.6 \times 15.3$ & $-80$ & $ 1.01$ \\
  Abell 478 & SL2 & $    $ & $    $ & Extended & $4:13:25.4$ & $10:27:55$ & $7.2 \times 7.2$ & $14$ & $ 2.26$ \\
            & SL1 & $ 7.0$ & $12.3$ &          & $4:13:25.4$ & $10:27:56$ & $7.2 \times 7.2$ & $14$ & $ 2.26$ \\
            & LL2 & $ 8.4$ & $14.7$ &          & $4:13:25.3$ & $10:27:57$ & $15.3 \times 10.2$ & $8$ & $ 1.71$ \\
            & LL1 & $    $ & $    $ &          & $4:13:25.3$ & $10:27:56$ & $15.3 \times 10.2$ & $8$ & $ 1.71$ \\
 Abell 1068 & SL2 & $    $ & $    $ &    Point & $10:40:44.5$ & $39:57:11$ & $12.6 \times 5.4$ & $-40$ & $ 0.65$ \\
            & SL1 & $ 3.7$ & $10.6$ &          & $10:40:44.4$ & $39:57:11$ & $19.8 \times 3.6$ & $-40$ & $ 1.00$ \\
            & LL2 & $ 9.0$ & $25.6$ &          & $10:40:44.7$ & $39:57:9$ & $35.7 \times 10.2$ & $43$ & $ 1.00$ \\
            & LL1 & $    $ & $    $ &          & $10:40:44.6$ & $39:57:10$ & $56.1 \times 10.2$ & $43$ & $ 1.00$ \\
 Abell 1795 & SL2 & $    $ & $    $ & Extended & $13:48:52.5$ & $26:35:33$ & $12.6 \times 3.6$ & $-6$ & $ 1.87$ \\
            & SL1 & $ 8.1$ & $10.5$ &          & $13:48:52.5$ & $26:35:34$ & $12.6 \times 3.6$ & $-6$ & $ 1.87$ \\
            & LL2 & $12.7$ & $16.5$ &          & $13:48:52.2$ & $26:35:35$ & $20.4 \times 10.2$ & $78$ & $ 1.33$ \\
            & LL1 & $    $ & $    $ &          & $13:48:52.3$ & $26:35:36$ & $20.4 \times 10.2$ & $78$ & $ 1.33$ \\
 Abell 1835 & SL2 & $    $ & $    $ &    Point & $14:1:2.1$ & $2:52:41$ & $16.2 \times 3.6$ & $-11$ & $ 1.20$ \\
            & SL1 & $ 3.8$ & $19.4$ &          & $14:1:2.1$ & $2:52:42$ & $16.2 \times 3.6$ & $-11$ & $ 1.20$ \\
            & LL2 & $ 9.5$ & $49.2$ &          & $14:1:1.9$ & $2:52:40$ & $45.9 \times 10.2$ & $72$ & $ 1.00$ \\
            & LL1 & $    $ & $    $ &          & $14:1:2.0$ & $2:52:39$ & $45.9 \times 10.2$ & $72$ & $ 1.00$ \\
 Abell 2597 & SL2 & $    $ & $    $ & Extended & $23:25:19.9$ & $-12:7:27$ & $9.0 \times 3.6$ & $27$ & $ 3.42$ \\
            & SL1 & $ 6.4$ & $10.7$ &          & $23:25:19.8$ & $-12:7:27$ & $9.0 \times 3.6$ & $27$ & $ 3.42$ \\
            & LL2 & $15.0$ & $25.3$ &          & $23:25:19.7$ & $-12:7:26$ & $15.3 \times 10.2$ & $21$ & $ 1.87$ \\
            & LL1 & $    $ & $    $ &          & $23:25:19.8$ & $-12:7:26$ & $10.2 \times 10.2$ & $21$ & $ 1.87$ \\
    Hydra A & SL2 & $    $ & $    $ &    Point & $9:18:5.7$ & $-12:5:43$ & $14.4 \times 3.6$ & $-20$ & $ 1.70$ \\
            & SL1 & $ 4.7$ & $ 5.3$ &          & $9:18:5.7$ & $-12:5:44$ & $18.0 \times 3.6$ & $-20$ & $ 1.70$ \\
            & LL2 & $ 8.7$ & $ 9.7$ &          & $9:18:5.4$ & $-12:5:43$ & $35.7 \times 10.2$ & $63$ & $ 1.70$ \\
            & LL1 & $    $ & $    $ &          & $9:18:5.5$ & $-12:5:44$ & $35.7 \times 10.2$ & $63$ & $ 1.00$ \\
    MS 0735 & SL2 & $    $ & $    $ & Extended & $7:41:44.6$ & $74:14:39$ & $5.4 \times 3.6$ & $-7$ & $ 3.58$ \\
            & SL1 & $ 6.2$ & $27.5$ &          & $7:41:44.6$ & $74:14:39$ & $3.6 \times 3.6$ & $-7$ & $ 3.58$ \\
            & LL2 & $10.4$ & $46.0$ &          & $7:41:44.2$ & $74:14:39$ & $10.2 \times 10.2$ & $-13$ & $ 1.72$ \\
            & LL1 & $    $ & $    $ &          & $7:41:43.8$ & $74:14:39$ & $10.2 \times 10.2$ & $-13$ & $ 1.72$ \\
PKS 0745-19 & SL2 & $    $ & $    $ &    Point & $7:47:31.4$ & $-19:17:39$ & $12.6 \times 3.6$ & $-11$ & $ 3.36$ \\
            & SL1 & $ 4.9$ & $10.3$ &          & $7:47:31.4$ & $-19:17:36$ & $27.0 \times 3.6$ & $-11$ & $ 1.29$ \\
            & LL2 & $ 9.4$ & $19.7$ &          & $7:47:31.3$ & $-19:17:37$ & $25.5 \times 10.2$ & $72$ & $ 1.00$ \\
            & LL1 & $    $ & $    $ &          & $7:47:31.2$ & $-19:17:36$ & $25.5 \times 10.2$ & $72$ & $ 1.00$
\enddata
\tablenotetext{{}}{(a) Angle of longer axis of aperture, in degrees east of north (CCW). (b) Factor applied to the extracted spectrum, to scale it up to the total light in a similar broadband aperture (only for extended-source spectra) and in a few cases to co-register modules and improve agreement with broadband measurements. }
\end{deluxetable}

%%%%%%%%%%%%%%%%%

\subsection{Spectral Data Filtering and Extraction \label{section:spectra}}

This section discusses our choices regarding the IRS spectral extraction
process, with particular attention to the treatment of extended sources compared
to point sources.
We used the software package 
CUBISM v1.7\footnote{\url{http://ssc.spitzer.caltech.edu/dataanalysistools/tools/cubism/}} 
\citep{2007PASP..119.1133S} to
combine the exposures into a data cube, or spectral map.  To reduce noise near 
the ends of the slit, we trimmed the exposures in the cross-dispersion 
direction by $3-5\%$.  We confirmed that the off-target, paired 
observations for each spectrum were indeed source-free. These blank sky spectra
were used for background subtraction.

To remove rogue pixels which were not caught by IRSCLEAN, we composed bad 
pixel lists using CUBISM's {\em autobadpix} algorithms, on both the global 
and individual record levels.
At the global level, we flagged any pixel which deviated by more than
$2.5\sigma$ from the median level in at least 50\% of its appearances  
in the cube.  This method flags only a few pixels, but each of these pixels
has a relatively large effect on the spectral map.  At the record
level, we conservatively flagged any pixel which is a $5\sigma$ outlier
in at least 75\% of its appearances in the cube.  This method flags 
more pixels, but only in individual exposures.  We also 
manually removed obvious rogue pixels.

The {\it Spitzer} pipeline is optimized for single-slit observations of point sources, 
including a correction for the light lost from the slit, which can be as much as 36\% \citep[Figure 4]{2007PASP..119.1133S}.  However, 
CUBISM includes an option to remove this slit-loss correction factor, in order to 
extract spectra of extended sources.  To determine which targets to treat as point sources, we
used CUBISM to create a map combining 
all SL1 wavelengths, and averaged the two rows which covered the peak source emission.  
Using this light profile, we measured the FWHM along the slit (see Table \ref{datared_table}).  We did the same for the LL2.
These modules were selected because their signal-to-noise is highest.
  {\it Spitzer's} PSF has an average FWHM of $2.6\arcsec$ in the SL and $6.6\arcsec$ in the LL 
\footnote{\url{http://ssc.spitzer.caltech.edu/irs/irsinstrumenthandbook/}}.
Four of our nine galaxies have a FWHM in the SL1 of $\lesssim 5.5\arcsec$ and were considered to be point sources.  
The other five sources do not uniformly fill the {\it Spitzer} slit, 
but they are not well characterized as point sources.

For the point sources, we extracted spectra using an aperture that just spans the slit (2 pixels) 
and a length that captures most of the light along the slit (see Table \ref{datared_table}).  
{\it Spitzer} data are calibrated for point sources and for this kind of aperture.  
Single-pointing software such as 
SMART\footnote{\url{http://ssc.spitzer.caltech.edu/dataanalysistools/tools/contributed/irs/smart/}} \citep{2004PASP..116..975H,2010PASP..122..231L} 
uses a similar aperture with its ``tapered column" extraction, increasing the length of the aperture as the 
PSF broadens with wavelength, and further optimizing the extraction by weighting each pixel by its 
signal-to-noise \citep{2010PASP..122..231L}.  In order to use the same software for all of our spectra, 
we used CUBISM to extract our point-source spectra, approximating the tapered column type of aperture. 
(For PKS 0745-19 SL2, we truncated the aperture to avoid a noisy region.  
For Abell 1068 SL2, CUBISM spreads the light from those two pixels across three rows.)  
To check our procedure, we also extracted the point-source spectra using SMART and found agreement to within about 10\%, sometimes to within 2\%.  

For the extended targets, we removed the pipeline slit-loss correction factor.  Our apertures include much of the available light with good signal-to-noise.  
However, our sparse spectral maps do not cover the full extent of the source.

After extraction, any noisy edges were trimmed from each order.
In the observed frame, the four low-resolution IRS modules span the following wavelength ranges:
SL2: $5.2 - 7.6 ~\mu$m; SL1: $7.5 - 14.5 ~\mu$m; LL2: $14.3 - 20.6 ~\mu$m; LL1: $20.5 - 37.5 ~\mu$m.
 Spectra extracted from CUBISM are reported in units of MJy sr$^{-1}$, so the size of the extraction 
aperture is used to convert spectra to units of flux density (mJy).  
We then corrected to the rest frame for each target (see Table \ref{obstable}) 
by dividing both the wavelength and fluxes by a factor of $(1+z)$.

\subsection{Aperture Photometry and Systematic Uncertainties\label{section:unc}}

The light collected by the narrow {\it Spitzer} slit and our sparse spectral maps represents only a portion of the MIR light.  
Therefore, to obtain meaningful luminosities, we rely on IRAC and MIPS photometry (see Table~\ref{table:scaling}) (as in \citet{2006ApJ...652L..21E}).  
The broadband aperture diameters are approximately three times the
source's FWHM measured using the IRS (or equal to the FWHM of {\it
Spitzer's} PSF for the 70 and 160 $\mu$m points) 
with subtraction of background computed from a larger annulus. 
We adjusted the aperture size to exclude unrelated 
foreground or background sources, and applied the suggested aperture
corrections given by the MIPS Instrument Handbook. At 24 $\mu$m, the
corrections are 1.17 (for apertures of $26\arcsec$ and $30\arcsec$) and 1.13 (for
apertures $50\arcsec$). For 70 and 160 $\mu$m the corrections are 1.22
and 1.752 respectively. 
The photometric uncertainty is 5\% for IRAC, and 10, 20 and 20\% for the
MIPS 24, 70 and 160 $\mu$m points respectively.

Because  the spectra within the cited IRAC and MIPS circular apertures may differ from the spectra obtained from within 
our smaller, rectangular IRS apertures, our analysis and conclusions rely most heavily on relative 
quantities, i.e., ratios, rather than absolute quantities.  
All correction factors for each module are listed in Table~\ref{datared_table}.  
The interested reader can recover the flux in the apertures listed in Table~\ref{datared_table} by dividing the fluxes published here by this factor.

In three cases (Hydra A, A1795, MS0735), 
we do not have complete IRAC and MIPS coverage in the IRS wavelength range, 
and we needed a robust, standalone scaling procedure. For this purpose, we developed a scaling procedure that did
not rely on IRAC or MIPS photometry. We validated this procedure, with scaled IRS spectra for sources with IRAC and MIPS photometry,
by comparing the resulting spectrophotometry to IRAC and MIPS photometry. We now describe our scaling procedure.

For the sources we identified as extended, we fit Gaussian, azimuthally symmetric light profiles 
(and neglected the fainter extended haloes) to estimate how much of the source was included within a circle of the same area as our rectangular aperture.    
The spatial profile from the SL1 map was used as the reference 
to determine the scale for the SL1 and SL2 spectra for light from
outside the rectangular aperture. The profile from the LL2 map was used for both LL orders.  
For point sources, this scaling has already been performed by the pipeline (See \S~\ref{section:spectra}).

An additional scale factor is needed to match the orders of the IRS spectrum.
Flux mismatches are expected between the 
spectral orders extracted with CUBISM, because it is impossible to perform an exact tapered column extraction using CUBISM.  
Even tapered column extractions with SMART sometimes show mismatches. 
In our sample, three targets had mismatches between the first and second orders (SL1/SL2 or LL1/LL2), 
and three had mismatches between the SL and LL.  
Some spectra are plagued by noise or decreased signal at the module interface (e.g. 2A0335 near 7.1 $\mu$m).
We selected the LL2 spectrum as the photometric reference point because of 
the good agreement between the LL spectrum and MIPS photometry in a similar aperture and 
because the LL is less vulnerable to slit loss due to pointing errors (because of its larger pixels) \citep{2007PASP..119.1133S}.
As mentioned above, the LL2 has superior signal-to-noise to the LL1. 
 We used low-order polynomial continuum fits to match the modules when major features did not interfere.

In three cases an additional overall factor was needed (Hydra A (1.70), 
A478 (1.55), MS0735 (1.12)), probably 
because of extended halo light contained within the broadband aperture 
but not represented by our Gaussian profile. 
This procedure worked well, because agreement between the IRS spectra 
and broadband photometry was relatively good, 
almost always within 10\%, and usually within 5\% (see Figures~\ref{figure:pahfit1}-\ref{figure:pahfit2}).  
Therefore, while the IRS spectrum appears to exceed the IRAC value at 
8 $\mu$m for Hydra A, the agreement between the integrated spectrum 
and the IRAC photometry point is actually excellent.
Note that the comparison here is between the MIPS and IRAC 
photometry and the observer-frame IRS spectra integrated over appropriate
bandpasses. These flux points should not be confused with 
  the integrated IRS photometry in the rest-frame 24-$\mu$m MIPS 
  bandpass reported in Table~\ref{table:fluxes}. 
 
For analysis and plots,
we combine the uncertainty calculated by the SSC pipeline in quadrature with a 15\% systematic uncertainty to all fluxes to account for 
the uncertainty in scaling the IRS photometry relative to {\em Spitzer} broadband photometry. The
systematic uncertainty dominates in almost all measurements.
An additional 5\% absolute photometric uncertainty is applied when making comparisons with data from 
other telescopes.\footnote{http://ssc.spitzer.caltech.edu/spitzermission/missionoverview/spitzertelescopehandbook/}.

\subsection{PAHFIT Spectral Decomposition}

\begin{figure}
\plotone{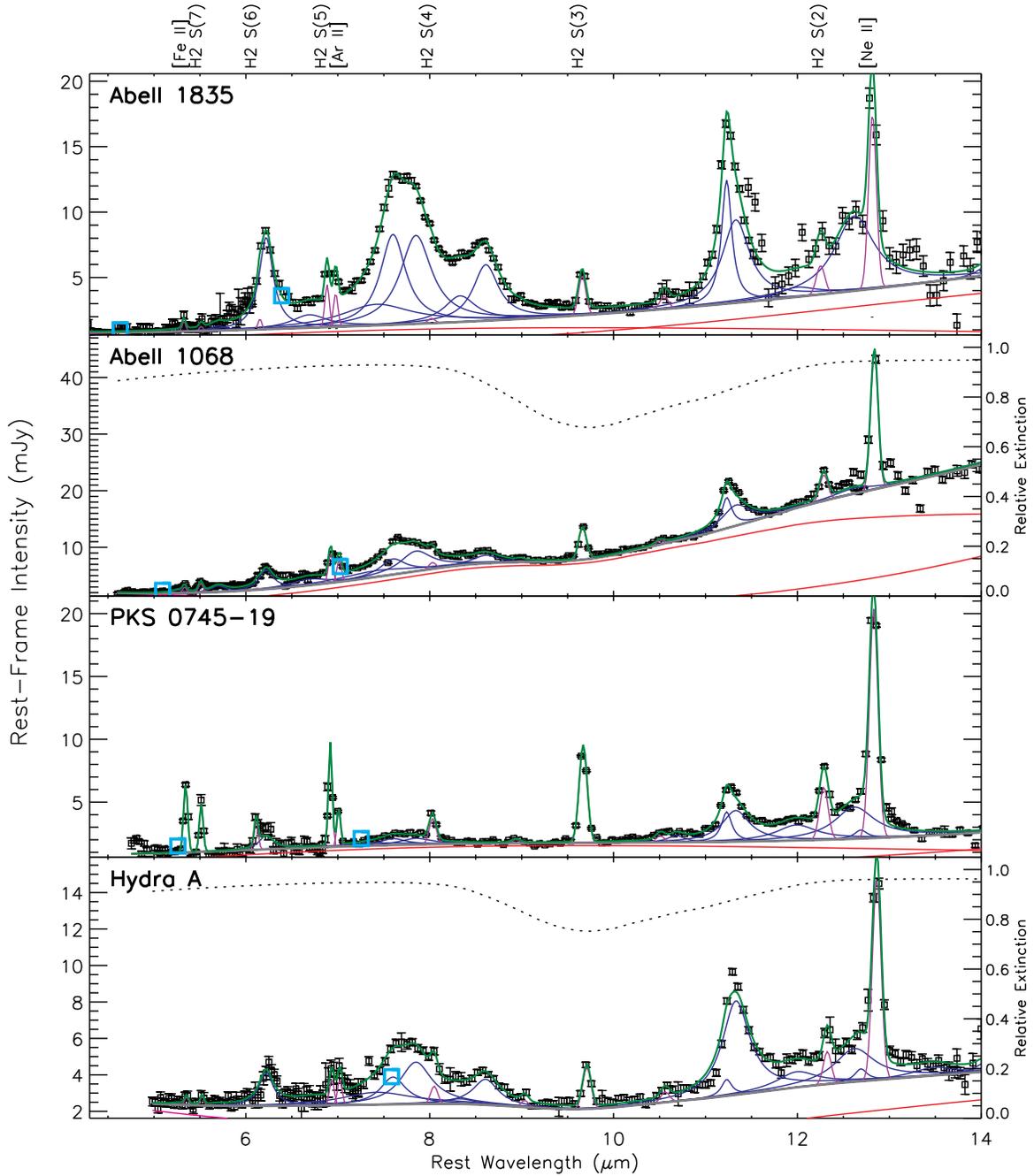}
\caption{\label{pahfit_plot_1}
Detailed decompositions of nine cool-core galaxy clusters from
4.3 to 14 $\mu$m, utilizing PAHFIT \citep{2007PASP..119.1133S}. 
Red lines represent thermal dust components; magenta, the
stellar continuum.  Their combination is a thick gray line. 
Broad PAH emission complexes are plotted in blue, and the
unresolved emission lines arising from low-ionization or
molecular hydrogen emission are plotted in violet and labeled at
the top.  The full spectral extraction is indicated by the green
line, plotted over the rest-frame flux intensities and
statistical uncertainties.  In the two cases where the empirical PAHFIT
detected silicate extinction, the extinction curve is
represented with a dotted line using the axis at right; all
components are diminished by the extinction.  Appropriately transformed 
IRAC photometry is indicated by cyan squares (see Table~\ref{table:fluxes}).
\label{figure:pahfit1}}
\end{figure}

\begin{figure}
\plotone{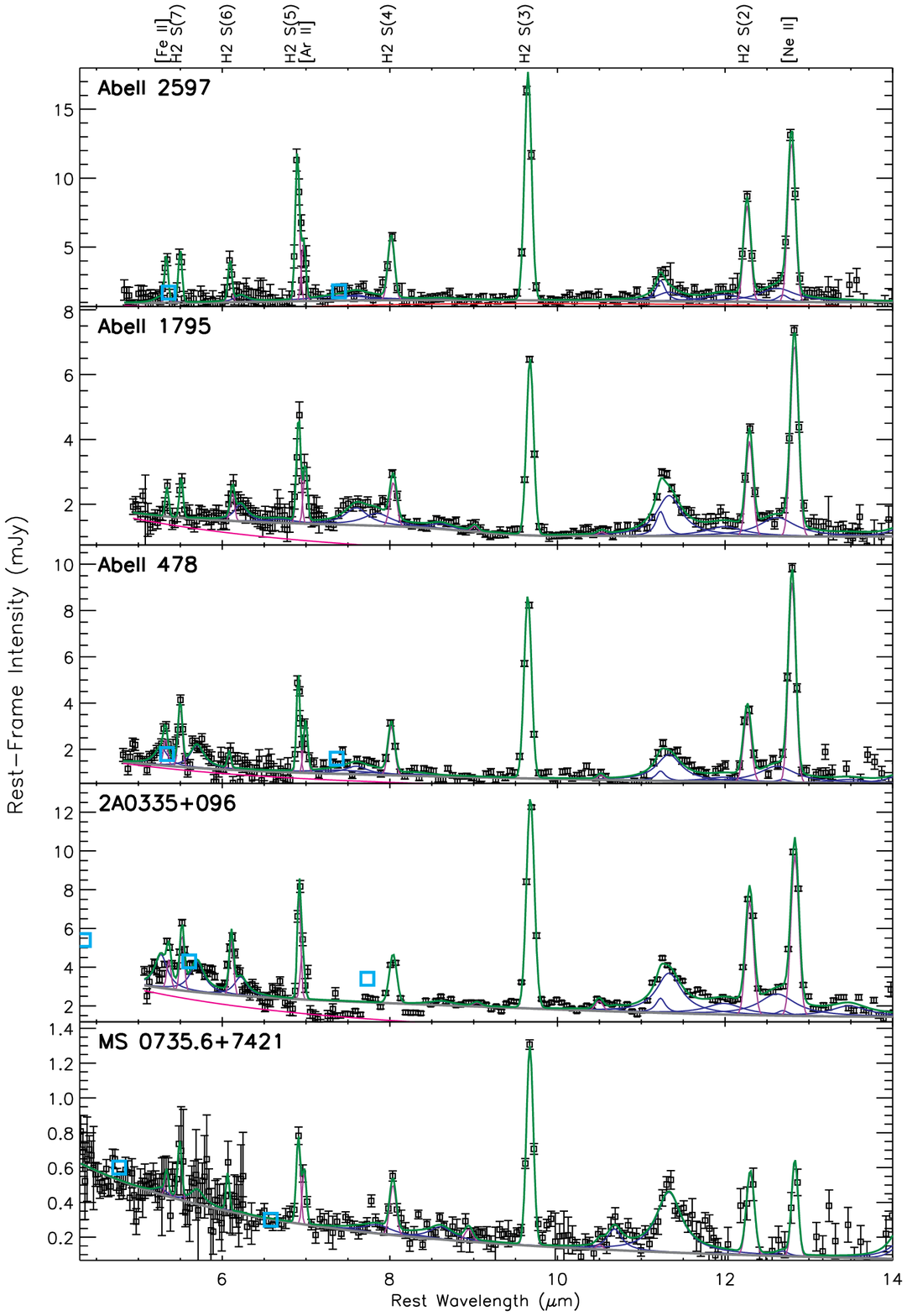}
\caption{Figure \ref{pahfit_plot_1} continued.
\label{figure:pahfit2}}
\end{figure}

\begin{figure}
\plotone{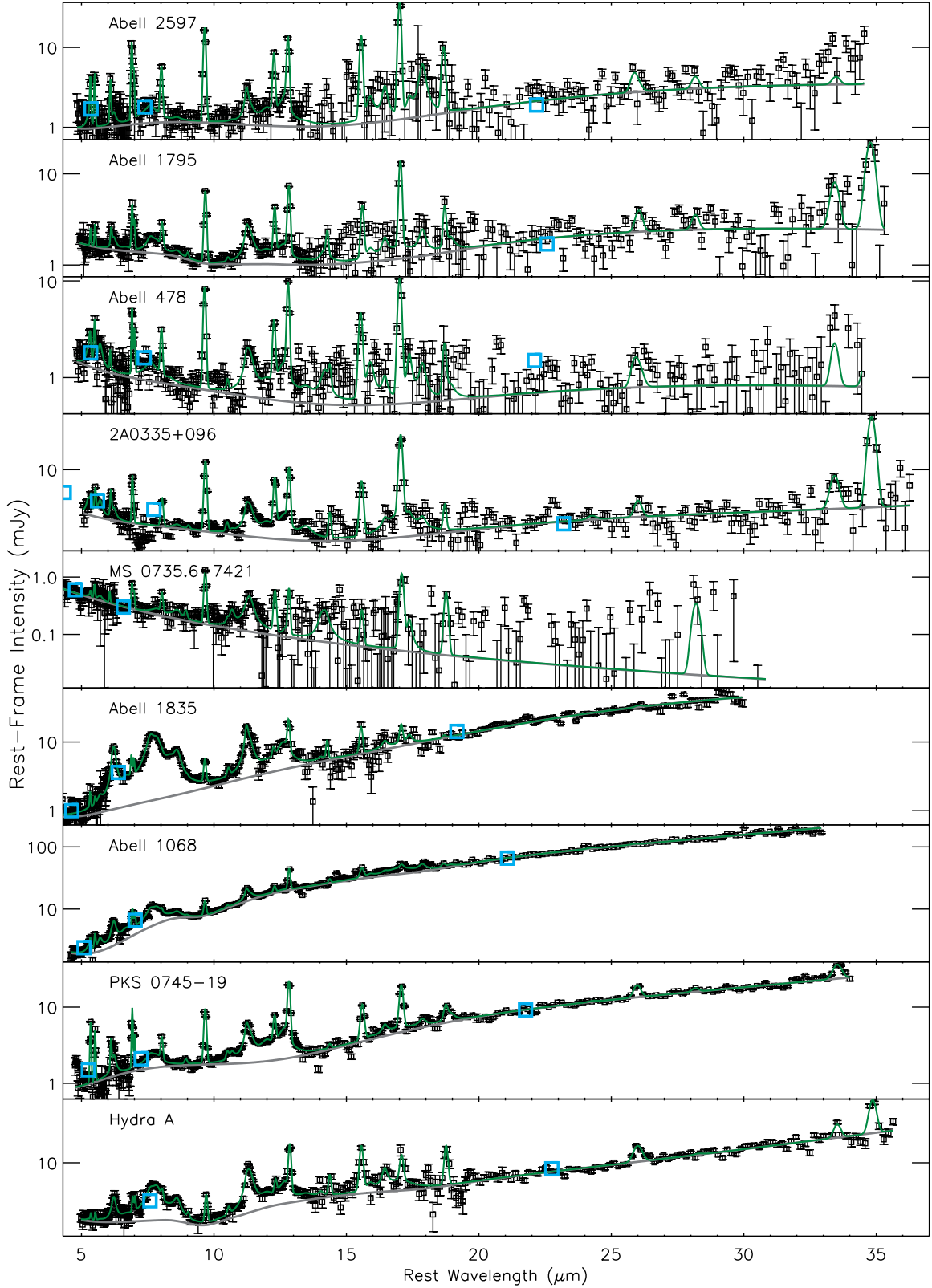}
\caption{PAHFIT results over the full wavelength coverage.  The
 continuum (stellar + thermal dust) is represented by a thick gray line, and
 the full spectral extraction by a green line.  The rest-frame flux
 intensities are plotted with statistical uncertainties.  Note the log
 scaling exaggerates the uncertainty of the faint, long-wavelength continua for
 several targets.  Appropriately transformed IRAC and MIPS photometry points
 are indicated by cyan squares (see Table~\ref{table:fluxes}).
\label{figure:pahfit3}}
\end{figure}

We used the spectral-decomposition package PAHFIT v1.2 \citep{2007ApJ...656..770S} 
 to make empirical fits to the IRS spectra and to facilitate direct comparison with results from
other workers using the same method.  
The short-wavelength PAHFIT results are plotted in Figures~\ref{figure:pahfit1}-\ref{figure:pahfit2} and
the full-wavelength results are shown in Figure~\ref{figure:pahfit3}.
PAHFIT fits the following components:
a starlight continuum, several thermal dust continuum components,
broad PAH emission bands, narrow atomic and molecular emission lines, and 
broad silicate absorption bands (Figures~\ref{figure:pahfit1}-\ref{figure:pahfit2}.) 
We customized the list of fitted emission features to a limited set,
excluding those that were very weak.
In the case of 2A0335, small parts of the spectrum (below 5.3$\mu$m and 
between 7.05 and 7.35$\mu$m) were excluded from the PAHFIT analysis because 
noise in those parts of the spectrum hindered a successful fit.

\begin{deluxetable}{rrrrrrrrrr}
\tablecaption{Broadband Spitzer Photometry \label{table:scaling}}
\tabletypesize{\scriptsize}
\tablehead{
\colhead{Band} & \colhead{2A0335+096} & \colhead{Abell 478} & \colhead{Abell 1068} & \colhead{Abell 1795} & \colhead{Abell 1835} & \colhead{Abell 2597} & \colhead{Hydra A} & \colhead{MS 0735} & \colhead{PKS 0745-19}
}
\startdata
  3.6   &   9.2 [12] &   4.2 [20] &   2.1 [10] &            &   2.5 [14] &   4.5 [20] &            &   1.4 [20] &   3.0 [14] \\
  4.5   &   5.6 [12] &   2.8 [20] &   2.0 [10] &            &   2.0 [14] &   2.9 [20] &   2.9 [14] &   1.0 [20] &   2.1 [14] \\
  5.8   &   4.4 [12] &   2.0 [20] &   2.7 [10] &            &   1.3 [14] &   1.8 [20] &            &   0.7 [20] &   1.7 [14] \\
  8.0   &   3.5 [12] &   1.7 [20] &   7.5 [10] &            &   4.5 [14] &   1.9 [20] &   4.1 [14] &   0.4 [20] &   2.3 [14] \\
 24.0   &   2.4 [26] &   1.6 [30] &  74.8 [30] &   1.8 [40] &  17.8 [30] &   2.1 [50] &   9.1 [30] &            &  10.2 [30] \\
 70.0   &  77.1 [70] &  62.8 [70] & 894.5 [70] &  37.2 [70] & 175.0 [70] &  89.0 [70] & 155.2 [70] &            & 154.3 [70] \\
160.0   &            &  56.4 [80] &            &            & 317.0 [80] &  42.0 [80] & 181.8 [80] &            &           
\enddata
\tablecomments{IRAC and MIPS waveband centers are in units of $\mu$m.  Observer-frame fluxes are in units of mJy, and aperture diameters in arcseconds are given in brackets.  Photometric uncertainties are $\sim 5\%$ for IRAC and, for MIPS, 10, 20, 20\% for 24, 70, and 160 $\mu$m respectively.}
\end{deluxetable}

%%%%%%%%%%

\section{Results}

Our BCG galaxies exhibit a number of
emission features from PAHs, ions, and H$_2$ molecules.  The H$_2$
features are unusually prominent.  The BCG spectra qualitatively fall into two 
general categories. Four galaxies (A1835, A1068, PKS0745, and Hydra A) exhibit the
strongly rising IR continuum at $>25\, \mu$m and distinct PAH features characteristic of galaxies with
strong signatures of star formation 
\citep[e.g., ][]{2006ApJ...653.1129B,2007ApJ...656..770S}.  In the remaining five
cases, the 5-7$\mu$m continuum is dominated by cool stars, in
contrast to the spectra of starbursts.

Several spectral features, which do not correspond to a known emission
feature, are artifacts of noise or the data reduction process.  For example, 
in the spectrum of A1835, there is noise on the red shoulder of the 
11.3 $\mu$m PAH band where the SL and LL modules do not perfectly
align; this is also responsible for noise near 13 $\mu$m in A478.  The
junction between the LL2 and LL1 accounts for some of the noise
near 17 $\mu$m for A1835, and near 20 $\mu$m for 2A0335.  
A feature near 24 $\mu$m in the spectrum of A1068 may be attributed to 
[Ne~V]24.3 $\mu$m, as discussed in \S~\ref{AGN}, but is probably spurious.  
An emission feature near 4.9 $\mu$m in A478, and possibly A1795,
A2597, and PKS0745, might be ascribed to [Ar V]4.93 $\mu$m, or to an
unidentified PAH. The feature at 20.7 $\mu$m in A478 is unidentified.
The noise at 15.0 $\mu$m in the spectra of A2597 and A1795, and 
18.0 $\mu$m in Hydra A, appears to be spurious.  
Note that noise increases dramatically past about 33 $\mu$m in the observed
frame.

The features we will examine most closely in this paper are the relatively bright forbidden emission lines of
[Ne II] at 12.8 $\mu$m, [Ne III] at 15.6 $\mu$m, and the PAH complexes at 7.7, 11.3, and 17 $\mu$m. 
We include an analysis of the correlation of the intensities of the 
brightest rotationally-excited molecular hydrogen transitions, S(2) and S(3), with those of 
other spectral features. A selection of line measurements and intrinsic luminosities 
including 1-$\sigma$ statistical errors from PAHFIT 
are presented in Table~\ref{table:fluxes}.
The continuum fluxes and
luminosities ($\nu L_\nu $) are found from the feature-free continuum (the
stellar blackbody, thermal dust components, and silicate
absorption).  The continuum measurements were determined by
averaging across bandpasses 1 $\mu$m in width at 6 and 15
$\mu$m, and weighting by the 24-$\mu$m MIPS response.    All
fluxes and luminosities in Table~\ref{table:fluxes} are presented in
the rest frame, at the rest wavelength.  To recover observed
fluxes, multiply by $(1+z)$. A more detailed analysis of the
molecular hydrogen line ratios and excitation diagrams is
deferred to a paper in preparation.

 \begin{deluxetable}{lccccccccc}
 \tablecaption{Line and Continuum Measurements\label{table:fluxes}}
 \tabletypesize{\scriptsize}
 \tablehead{ 
 \colhead{Line} & \colhead{2A0335} & \colhead{A478}  & \colhead{A1068} &  \colhead{A1795} &  \colhead{A1835}  & \colhead{A2597} &  \colhead{HydraA}  & \colhead{MS0735}  & \colhead{PKS0745}} 
 \startdata
Ne II 12.8 $\mu$m Flux &       17.7 &       17.4 &       47 &       13.0 &       26 &       23.2 &       24.6 &      0.95 &       38.9 \\ 
Ne II 12.8 $\mu$m Flux Error & $\pm$        0.3 & $\pm$        0.4 & $\pm$         2 & $\pm$        0.3 & $\pm$         4. & $\pm$        0.9 & $\pm$        0.6 & $\pm$        0.2 & $\pm$        0.3 \\ 
Ne II 12.8 $\mu$m Lum &     0.0499 &      0.319 &       2.4 &      0.126 &       5.0 &      0.386 &      0.176 &      0.129 &       1.05 \\ 
 & & & & & & & & & \\
Ne III 15.5 $\mu$m Flux &       11.6 &       8.2 &       28 &       5.63 &       16 &       21 &       22 &       $<0.5$ &       14.3 \\ 
Ne III 15.5 $\mu$m Flux Error & $\pm$        0.7 & $\pm$        0.6 & $\pm$         1 & $\pm$        0.7 & $\pm$         2. & $\pm$         2. & $\pm$         1. & \nodata & $\pm$        0.3 \\ 
Ne III 15.5 $\mu$m Lum &     0.0328 &      0.15 &       1.45 &     0.054 &       3.1 &      0.36 &      0.16 &       $<0.07$ &      0.385 \\ 
 & & & & & & & & & \\
PAH 7.7 $\mu$m Flux &       \nodata \tablenotemark{a} &       26 &       300. &       28 &       532. &       42 &       180 &       $<4$ &       46 \\ 
PAH 7.7 Flux Error &        \nodata   & $\pm$         6 & $\pm$         8 & $\pm$         9 & $\pm$         6 & $\pm$     10 & $\pm$     10 & \nodata & $\pm$         4 \\ 
PAH 7.7 Lum &       \nodata &      0.49 &       15.4 &      0.27 &       103. &      0.71 &       1.3 &       $<0.6$ &       1.2 \\ 
 & & & & & & & & & \\
PAH 11.3 $\mu$m Flux &       31.2 &       17 &       91 &       22 &       132. &       16 &       83 &       4.6 &       44.9 \\ 
PAH 11.3 Flux Error & $\pm$        0.7 & $\pm$         1 & $\pm$         2 & $\pm$         1 & $\pm$         2 & $\pm$         3 & $\pm$         3 & $\pm$        0.3 & $\pm$        0.8 \\ 
PAH 11.3 Lum &     0.0881 &      0.32 &       4.7 &      0.21 &       25.5 &      0.27 &      0.60 &      0.63 &       1.21 \\ 
 & & & & & & & & & \\
PAH 17 $\mu$m Flux &       32   &       $<14$  &       260. &     $<27$  &      $<36$  &   $<48$  &       42    &  $<0.89$   &       25   \\ 
PAH 17 Flux Error & $\pm$     10 &           \nodata    & $\pm$  80 &      \nodata      &       \nodata      &     \nodata     & $\pm$     20 &    \nodata      & $\pm$     10 \\ 
PAH 17 Lum &               0.090 &      $<0.25$ &     13    &     $<0.26$ &     $<7.0$   &   $<0.80$ &  0.30        &   $<0.12$ &   0.68        \\ 
 & & & & & & & & & \\
H2 S1 Flux       &       38.7 &       16.8 &       $<20$ &       18.0 &       $<11$ &       49 &       10. &       1.8 &       23.6 \\ 
H2 S1 Flux Error & $\pm$        0.9 & $\pm$        0.7 & \nodata & $\pm$         1. & \nodata  & $\pm$         2 & $\pm$         2 & $\pm$        0.2 & $\pm$        0.5 \\ 
H2 S1 Lum &      0.109 &      0.310 &      $<0.9$ &      0.174 &       $<2.2$ &      0.82 &     0.075 &      0.25 &      0.635 \\ 
 & & & & & & & & & \\
H2 S2 Flux  &       13.7 &       7.0 &       10.9 &       6.4 &       5.1 &       14.5 &       4.0 &       1.1 &       9.7 \\ 
H2 S2 Flux Error & $\pm$        0.3 & $\pm$        0.4 & $\pm$        0.7 & $\pm$        0.3 & $\pm$         2. & $\pm$        0.8 & $\pm$         1. & $\pm$        0.2 & $\pm$        0.3 \\ 
H2 S2 Lum &     0.0386 &      0.13 &      0.56 &     0.062 &      0.99 &      0.242 &     0.029 &      0.15 &      0.26 \\ 
 & & & & & & & & & \\
H2 S3 Flux &       40.8 &       27.8 &       28 &       22 &       9.6 &       56 &       11 &       3.73 &       27.0 \\ 
H2 S3 Flux Error & $\pm$        0.4 & $\pm$        0.5 & $\pm$         1 & $\pm$         2 & $\pm$        0.6 & $\pm$         1 & $\pm$         1 & $\pm$       0.09 & $\pm$        0.3 \\ 
H2 S3 Lum &      0.115 &      0.511 &       1.4 &      0.21 &       1.9 &      0.93 &     0.077 &      0.509 &      0.726 \\ 
 & & & & & & & & & \\
Cont 24 $\mu$m Flux (mJy) &       2.66 &      0.931 &       91.0 &       2.12 &       25.4 &       2.57 &       9.0 &     $<0.04$ &       11.6 \\ 
Cont 24 Flux Error & $\pm$       0.06 & $\pm$       0.05 & $\pm$        0.1 & $\pm$       0.07 & $\pm$        0.1 & $\pm$        0.1 & $\pm$       0.06 & \nodata & $\pm$       0.03 \\ 
Cont 24 Lum &      0.937 &       2.14 &       582. &       2.56 &       614. &       5.35 &       8.1 &      $<0.7$ &       38.9 \\ 
 & & & & & & & & & \\
Cont 15 $\mu$m Flux (mJy) &       1.45 &      0.520 &       28.4 &       1.04 &       6.13 &       1.07 &       4.49 &      $<0.1$ &       3.13 \\ 
Cont 15 Flux Error & $\pm$        0.1 & $\pm$        0.1 & $\pm$        0.2 & $\pm$        0.1 & $\pm$        0.3 & $\pm$        0.3 & $\pm$        0.2 & \nodata & $\pm$       0.06 \\ 
Cont 15 Lum &      0.817 &       1.91 &       291. &       2.01 &       237. &       3.57 &       6.43 &       $<2.73$ &       16.8 \\ 
 & & & & & & & & & \\
Cont 6 $\mu$m Flux (mJy) &       2.60 &       1.14 &       2.23 &       1.46 &      0.99 &       1.0 &       2.20 &      0.36 &       1.18 \\ 
Cont 6 Flux Error & $\pm$       0.03 & $\pm$       0.05 & $\pm$       0.05 & $\pm$       0.04 & $\pm$       0.09 & $\pm$        0.1 & $\pm$       0.05 & $\pm$       0.02 & $\pm$       0.05 \\ 
Cont 6 Lum &       3.66 &       10.5 &       57.1 &       7.04 &       96. &       8.4 &       7.8 &       25. &       15.9 \\ 
 \enddata
 \tablecomments{Emission line flux values are in rest-frame units of $10^{-18}$ W m$^{-2}$, while continuum ('Cont') fluxes are quoted in mJy. Luminosities are in units of 10$^{42}$ erg s$^{-1}$. Fluxes, luminosities, and wavelengths are corrected to their rest-frame values (To recover the observed flux, multiply by $(1+z)$.)  The $1\sigma$ flux errors are statistical only. A 15\% systematic error (0.06 dex) is included in the plots and in the analysis, as discussed in \S~\ref{section:unc}. Upper limits are $3\sigma$. }
 \tablenotetext{a}{The fit to 2A0335's 7.7$\mu$m PAH complex failed because of decreased signal near the expected location of the feature.  
We were unable to set an upper limit in this case.}
 \end{deluxetable}

\clearpage

\section{Best Fit Starburst and Old Stellar Population SED Models \label{groves}}

We fit the IRS spectra of the BCGs with an ensemble of simulated spectral energy distributions (SEDs). These
models allow us to estimate the total infrared luminosity and associated star formation rates and to
identify differences between the SED of a star-forming galaxy and the observed SEDs.
We use the suite of starburst models, including a wavelength-dependent attenuation
template, described in \citet{2008ApJS..176..438G} together with an SED of 
a 10 billion year old stellar population, derived using Starburst99 \citep{1999ApJS..123....3L}. 
\citet{2008ApJS..176..438G} model the starburst SED as the time-integrated sum of distinct HII regions and the 
the photodissociation regions (PDRs) surrounding them, over a range of cluster ages and cluster masses. These
models were used to reproduce the SEDs of typical template starbursts such as Arp 220
and NGC 6240. The models span 5 different metallicities (Z $=$ 0.05, 0.2, 0.4, 1.0, and 2.0 solar).
Metallicity affects the prominence of the PAH features and the dust-to-gas ratio. The models sample 6 
compactness parameters $\cal{C}$ which characterize the intensity of the stellar radiation at 
the HII region/PDR interface. More compact HII regions result in hotter grains.  $\cal{C}$ determines the location (the ``temperature")  
of the dust peak and thus controls the mid-IR emission peak, a feature that is not
well constrained by IRS spectral coverage.  

\begin{figure}
\plotone{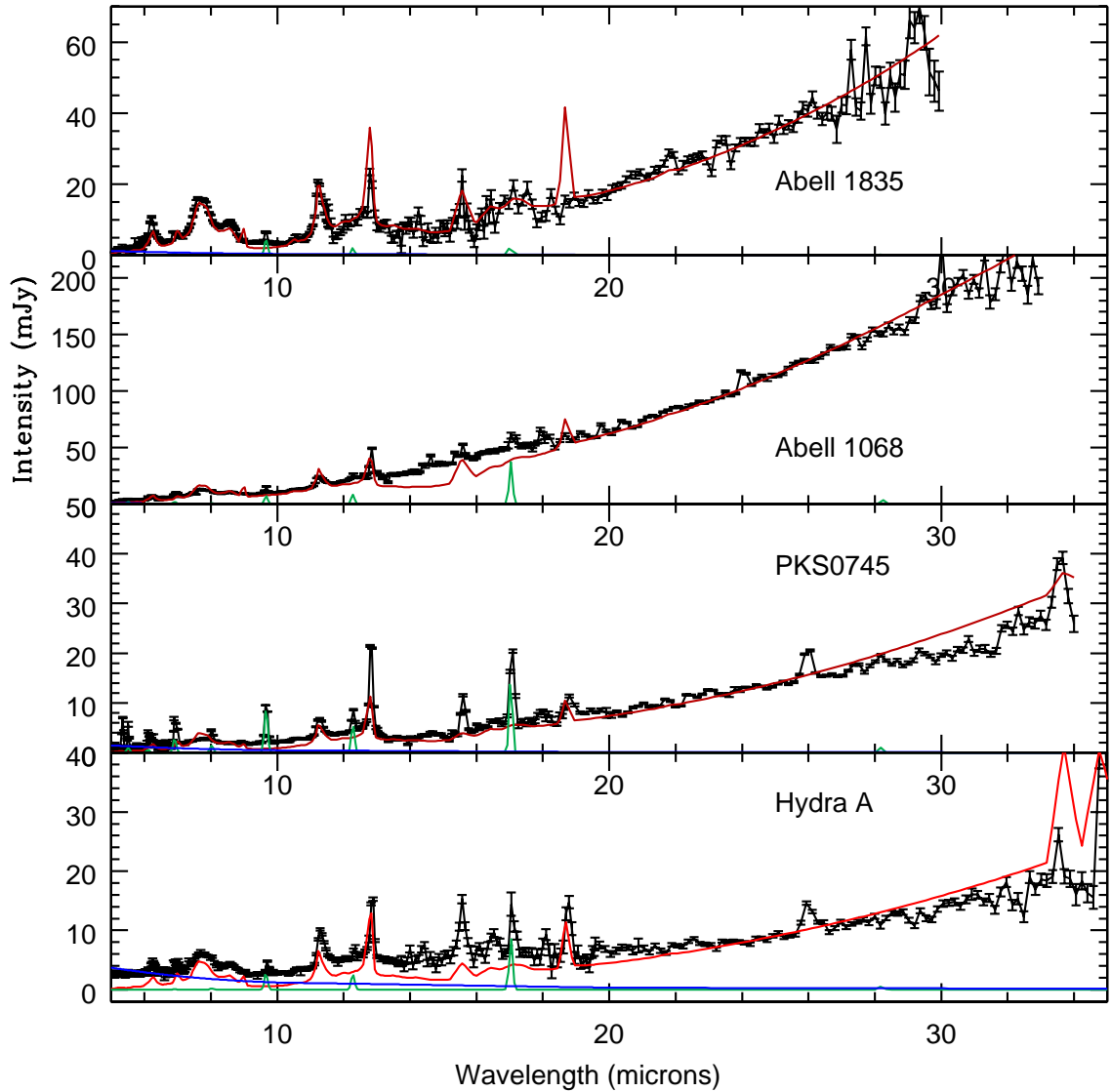}
\caption{Rest wavelength 5-35 $\mu$m spectra, observed fluxes. (To convert to rest-frame
flux, divide by (1+z).) Red line: Best fit constant star formation
model from \citet{2008ApJS..176..438G}. Blue line: Best fit old stellar population. Green line:
Molecular hydrogen, two-temperature LTE model.
\label{figure:Model1}}
\end{figure}

\begin{figure}
\plotone{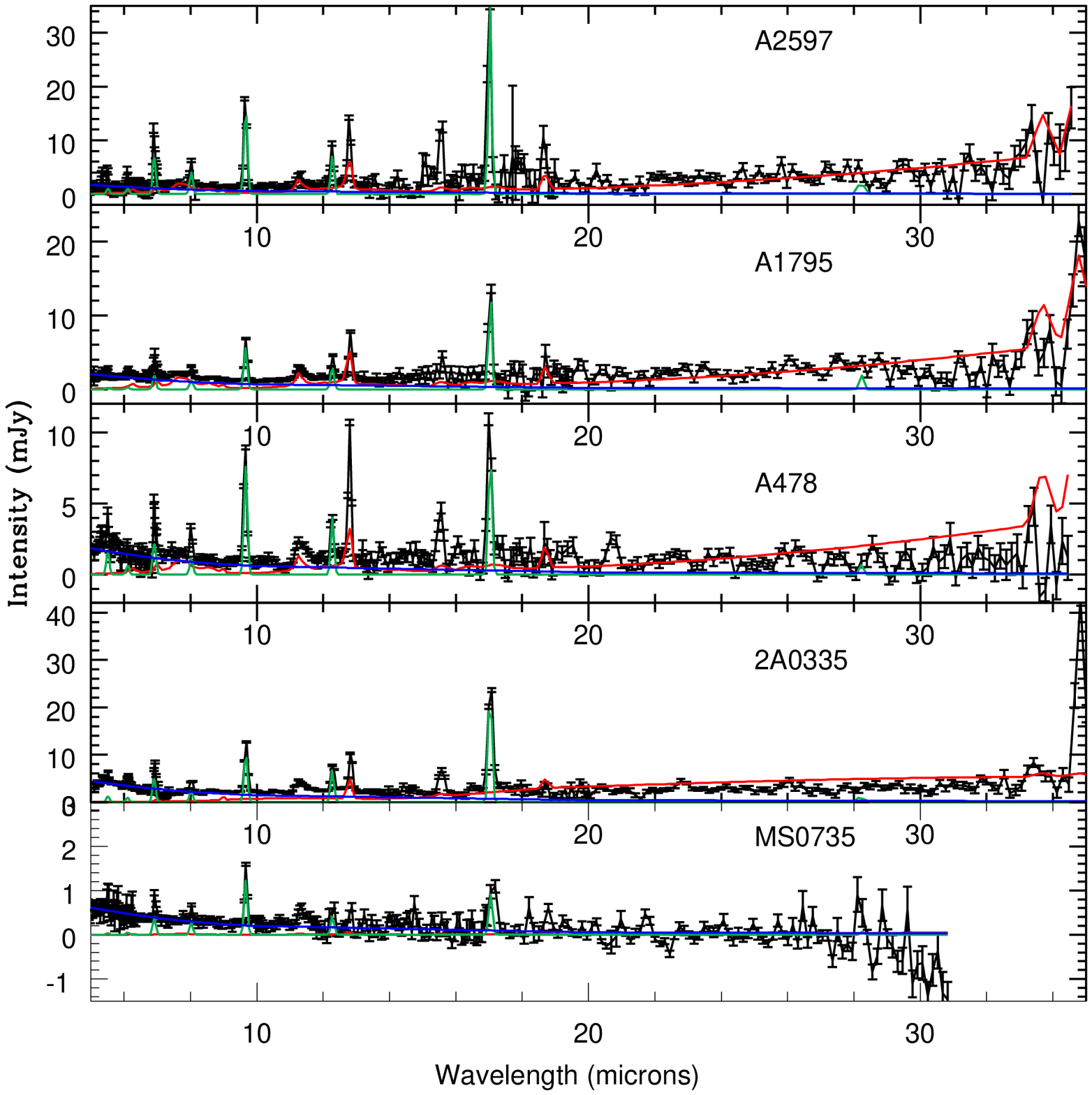}
\caption{Figure \ref{figure:Model1} continued.
\label{figure:Model2}}
\end{figure}

\citet{2008ApJS..176..438G} provide models for
five gas pressures ($P/k = 10^4, 10^5, 10^6, 10^7, 10^8$ K cm$^{-3}$),
spanning the range of lower pressures in  star-forming galaxies to the higher pressures in ULIRGs. 
However, the infrared SED between 5-25 $\mu$m is insensitive to the 
pressure, except, at a marginal level, the forbidden lines, and we checked that 
changing the gas pressure does not change the other parameters. 
Since the ICM pressure in cool-core clusters is $\sim10^6-10^7$ K cm$^{-3}$,
 the fits reported in Table~\ref{table:sbmodel} sample
the range of $10^6, 10^7, 10^8$  K cm$^{-3}$. 
Finally, the \citet{2008ApJS..176..438G} models were computed for unobscured HII regions and HII regions with PDRs. We opted here
to fit only the PDR models, since the models with only HII regions had no PAHs.
Models are normalized in luminosity to a star formation rate of $1 ~\rm{M}_\odot~\rm{yr}^{-1}$ 
sustained over the 10 Myr lifetime of the HII region. The spectra and the best-fit components are shown
in Figures~\ref{figure:Model1}-\ref{figure:Model2}.

We compared the observed spectra with the full suite of 
theoretical spectra and quantitatively identified the spectrum that best fit the data, in $L_\nu$ units,
by minimizing the sum of chi-squared. 
In addition to the quantized parameters described above, we fit the 
normalization for each starburst SED ($M_\odot~\rm{yr}^{-1}$), the attenuation ($A_V$), and 
the normalization of the old  stellar SED ($M_\odot$).
The \citet{2008ApJS..176..438G} models include forbidden lines and PAH emission, but not 
molecular emission, so we added a two-component local thermodynamic equilibrium (LTE) H$_2$ 
spectrum, each component specified by $T$ (K), and column density $N$ ($10^{18}$ cm$^{-2}$). We used
molecular properties and A-values from  \citet{1976ApJ...203..132B,Huber+Herzberg}.
The widths of the H$_2$ lines were fixed to 0.05 $\mu$m.

In this comparison, the compactness parameter ($\cal{C}$) seemed to divide BCGs with strong rising
IR continuum from those with relatively flat IR continuum in the IRS spectral range. 
The fit quality was sensitive to  $\cal{C}$, which mainly
affects the steepness of the long-wavelength IR continuum. The BCGs with the best-fit
${\cal C}=4$  are 2A0335, A1795, A2597, A478, and MS0735. The first four of these galaxies have the 
lowest four 24/70 $\mu$m MIPS  photometry ratios in the sample ($=0.03\pm 0.01$); 
MS0735 lacks MIPS photometry. The signal to noise in the IRS spectrum of MS0735 at long wavelengths 
is so low that there are no strong preferences for any starburst SED over another. 
The other 4 BCGs in the sample have higher 24/70 $\mu$m ratios 
($=0.08\pm0.02$). This trend in flux ratios is consistent
with the compactness parameter governing the peak in the dust spectrum, in the sense that
more compact HII regions have PDRs with hotter dust.

The metallicity also
affects the SED in this region, particularly the relative strength of the PAH feature to the
infrared continuum. The preferred starburst SED metallicities for the BCGs 
were at the high end ($Z/Z_\odot = 1$ or $2$). For all of the BCGs, the fit quality 
was very similar for $Z/Z_\odot = 1$ or $2$, except for that of A1068, which preferred 
$Z/Z_\odot = 0.5$ or $1$. That result is consistent with the fact that the BCG in A1068 has the
lowest PAH to mid-IR continuum ratio of the sample (see \S~\ref{section:pah}).
We note these high metallicities are consistent with the metallicity of the ICM in the centers of these clusters.
As expected, pressure did not affect the fit quality in
any significant way. The spectral features, including the 7.7 and 11.3 $\mu$m PAH features, 
were often well-matched by models including the PDRs at rest wavelengths shortward of 10-11 $\mu$m. 
The two-temperature molecular hydrogen template matched the H$_2$ spectrum very well in most cases.

\begin{figure}
\plottwo{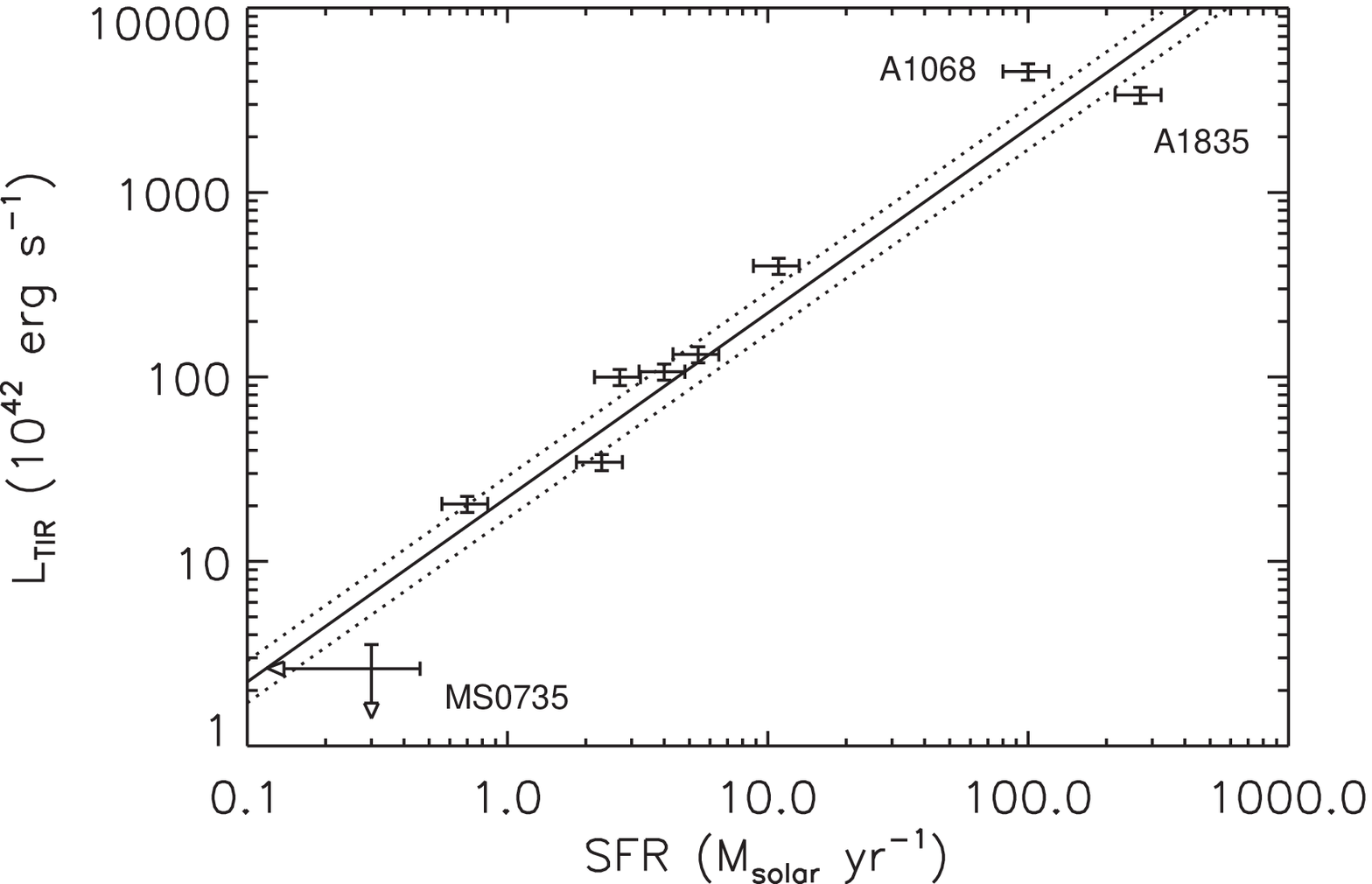}{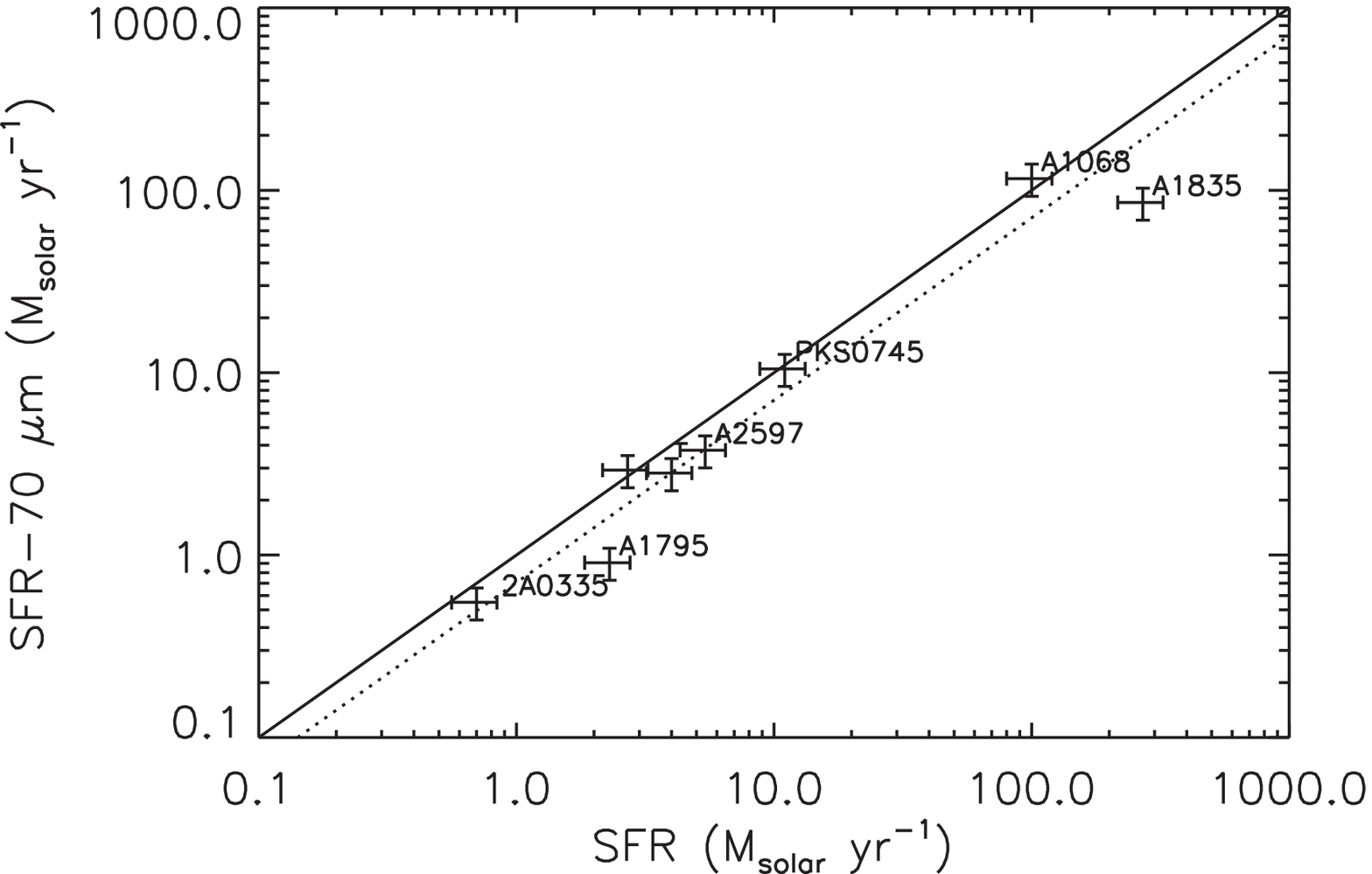}
\caption{A comparison of infrared Brightest Cluster Galaxy (BCG) 
star formation rates (SFRs) from the best-fit starburst models of \citet{2008ApJS..176..438G} 
and total infrared luminosity ($L_{TIR}$) and the SFRs inferred from $L_{TIR}$.
On left, we show these inferred SFRs are highly correlated with the total infrared luminosity 
($L_{TIR}$ DH; Table~\ref{table:TIR}).
For comparison, the solid line shows the Kennicutt (1998) relation for starbursts 
( $\rm{SFR}(\rm{M}_\odot ~\rm{yr}^{-1}) = 0.045 (L_{TIR}/10^{42}) ~\rm{erg~ s}^{-1}$ ), 
along with $\pm30\%$ typical calibration scatter (dotted lines). 
On right, the same SFRs are plotted against the SFRs based on 70 $\mu$m luminosities, using
the mean relation in  
\citet{2010ApJ...714.1256C}  
($\rm{SFR}(\rm{M}_\odot) = 0.059 (L_{70}/10^{42}) ~\rm{erg s}^{-1}$). The solid line shows the line of equality; 
the dotted line shows how the predicted SFR from $L_{70}$ would change, 
from \citet{2010ApJ...714.1256C}, if a different H$\alpha$-based SFR  from Kennicutt (1998) is used for calibration. 
\label{figure:MdotIR}}
\end{figure}

To check this process, we estimated the total infrared luminosity from the expression 
$L_{TIR} = 1.559 L_{24} + 2.1156 L_{70}$, adapted from \citet{2002ApJ...576..159D}. $L_{24}$
is $\nu L_{\nu}$ at 24 $\mu$m and $L_{70}$ is $\nu L_{\nu}$ at 70 $\mu$m.  To adapt their equation 4 to  
this expression, we assume that $\nu L_{\nu}$ at 70 $\mu$m $\sim \nu L_{\nu}$ at 160 $\mu$m. Some assumption was necessary
since the 160 $\mu$m flux is only available for 4 objects. This approximation may underestimate the TIR luminosity by
about 30\% compared to estimates using the 160 micron luminosities, but these 160 $\mu$m luminosity estimates are subject to large
systematic uncertainties, and the detected sources may not be representative of the sources lacking
photometry.  (For MS0735+74, we have no MIPS photometry and only a 3-$\sigma$ flux limit at 24 $\mu$m
from the IRS spectrum. We do not include this source in plots of TIR.) 
We compared $L_{TIR}$ estimated from this adaptation of \citet{2002ApJ...576..159D} to $L_{TIR}$
derived using the calibration of $L_{24}$ in Equation 2 of \citet{2010ApJ...723..895W}. To get rest-frame
70/24 $\mu$m ratios used in this calibration, we convert the observed MIPS luminosity
70 to 24 $\mu$m ratios to rest-frame ratios for 8 of  the 9 galaxies 
in the sample with k-correction factors ($k_{70/24}$) based on the best-fit SEDs ($<3-25\%$; Table~\ref{table:TIR}). 
The derived total IR luminosities were consistent to better than $20\%$ (Table~\ref{table:TIR}).
The relationship between either estimate of $L_{TIR}$ and the SED-inferred SFR is consistent 
with the Kennicutt (1998) SFR relation for starbursts (Figure~\ref{figure:MdotIR}a). This consistency indicates the 24 and 70 $\mu$m 
MIPS data, not included in the fits, are 
consistent with the star formation rates inferred from the IRS data alone.

\begin{deluxetable}{lccccc}
\tabletypesize{\scriptsize}
%\rotate
\tablecaption{Total Infrared Luminosity Estimates\label{table:TIR}}
\tablewidth{0pt}
\tablehead{
	\colhead{Name} & \colhead{$L_{24}$ } & \colhead{$L_{TIR}$ (DH)} & \colhead{$L_{TIR}$ (W)} & \colhead{$k_{70/24}$} & \colhead{$\log L_{70}/L_{24}$} \\
	\colhead{}     & \colhead{$10^{42}$ erg s$^{-1}$} & \colhead{$10^{42}$ erg s$^{-1}$} & \colhead{$10^{42}$ erg s$^{-1}$} & \colhead{} & \colhead{(rest)}  \\ }
\startdata
2A0335 & 0.94    & 21.2          &  25.8          &   0.97  & 1.07 \\
A478   & 2.14    & 108.          &  76.2          &   0.92  & 1.09 \\
A1068  & 582.    & 5060.         &  4120          &   0.73  & 0.48 \\
A1795  & 2.56    & 36.6          &  40.9          &   0.94  & 0.82 \\
A1835  & 614.    & 4028.         &  4210.         &   0.83  & 0.45 \\
A2597  & 5.35    & 143.          &  177.          &   0.92  & 1.13 \\
HydraA & 7.32    & 112.          &  94.3          &   0.94  & 0.74 \\
PKS0745 & 38.9   & 437.          &  389.          &   0.84  & 0.64 \\
\enddata
\tablecomments{ All luminosities are in units $10^{42}$ erg s$^{-1}$. $L_{24}$ is rest-frame $\nu L_\nu$ at 24 $\mu$m from
the IRS spectra. $L_{TIR}$ (DH) is the TIR derived from a relation adapted from \citep{2002ApJ...576..159D}. $L_{TIR}$ (W) is 
based on $L_{24}$ and the observed ratio $L_{70}/L_{24}$ luminosities based on MIPS photometry, from Equation 2
in \citet{2010ApJ...723..895W}. The k-correction for the 70/24 $\mu$m luminosity ratio 
is based on rest- and observer-frame MIPS response functions convolved with the best-fit SED shapes. The last
column is the log$_10$ ratio of the k-corrected, rest-frame $L_{70}/L_{24}$ luminosities.}
\end{deluxetable}

In Figure~\ref{figure:MdotIR}b, we plot the star formation rates based on the SED fits versus 
the \citet{2010ApJ...714.1256C} mean relation for SFRs derived from the observed MIPS 70 $\mu$m luminosities ($\nu L_\nu$) 
($\rm{SFR}(\rm{M}_\odot) = 0.059 (L_{70}/10^{42}) ~\rm{erg s}^{-1}$). The dotted line shows the same relation, also
from \citet{2010ApJ...714.1256C}, if a different H$\alpha$-based SFR  from Kennicutt (1998) is used for calibration. 
Figure~\ref{figure:MdotIR} demonstrates reasonable consistency between SFRs estimated based on SED fits to the IRS
spectra and SFRs derived from MIPS photometry. One caveat to this comparison is that the IR continuum of galaxies with a significant
old stellar population may have contributions from dust heated by these cool stars. 
 The observed 70-$\mu$m luminosities of the BCGs are somewhat higher than what is
predicted by the best-fit SEDs, by a factor of 1.3-2.4, which, if taken literally may indicate that dust heated by evolved stars may be 
contributing between 30-60\% of the 70 $\mu$m luminosity.
The 70 $\mu$m luminosities of these BCGs are in the valid domain for
applying the relationships in \citet{2010ApJ...714.1256C}; the 160 $\mu$m luminosities are likely 
to be even more contaminated from cool dust emission unrelated to star formation.

\begin{figure}
\epsscale{.80}
\plotone{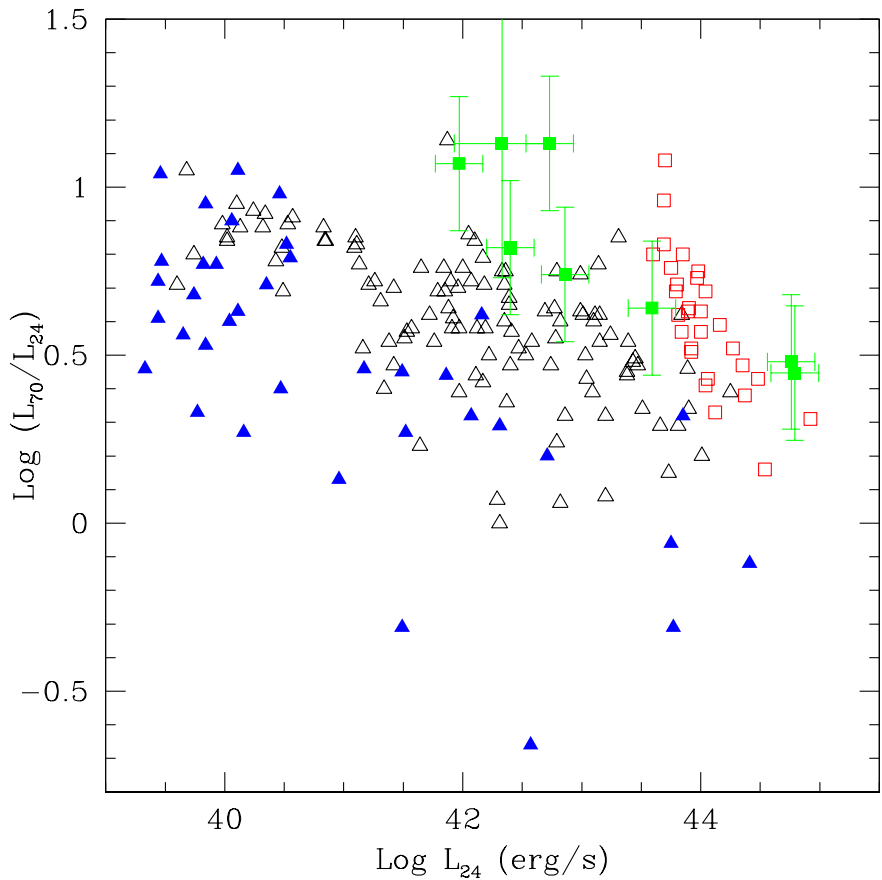}
\caption{ The $L_{70}/L_{24}$ ratio as a function of the 24 $\mu$m luminosity $L_{24}$.The open black 
triangles are higher-metallicity star forming galaxies; the closed blue triangles are lower-metallicity
star-forming galaxies; and the open red squares are luminous infrared galaxies (LIRGs), all from Calzetti et al. (2010). The green 
symbols with error bars are the rest-frame 24 $\mu$m luminosities and rest-frame (k-corrected)
70 to 24 $\mu$m luminosity ratios for the 8 BCGs in our sample with MIPS photometry.
\label{figure:Calzetti17}}
\end{figure}

It is interesting to note that the SFR estimates based on 24-$\mu$m luminosities, from Calzetti et al. (2010),
are systematically lower compared to SFRs based on these SED fits or the 70-$\mu$m luminosities for
the 4 galaxies with high 70/24 $\mu$m flux ratios (A478, 2A0335, A2597, and A1795) and low best-fit 
${\cal C} = 4$, as well as MS0735 (which has no MIPS photometry).
In Figure~\ref{figure:Calzetti17}, we plot the rest-frame 70/24 $\mu$m luminosity ratio of the BCGs
together with the local sample studied by Calzetti et al. (2010; Figure 17). The rest-frame 24
$\mu$m luminosity is calculated from the IRS spectra, but it is consistent with MIPS photometry.
The rest-frame ratios are based on MIPS photometry alone, with small k-corrections ($<10\%$ for $z<0.1$,
and $10-25\%$ for $z>0.1$), based on the best-fit
starburst SEDs. The 70/24 $\mu$m ratio varies by about factor of 10 for a given 24 $\mu$m luminosity in
the full sample of Calzetti galaxies, and 
some of this scatter is due to metallicity, with the highest metallicity systems having the highest 70/24
$\mu$m luminosity ratios. The 70/24 $\mu$m ratios of BCGs are similar to, but somewhat higher than, those of 
high-metallicity star-forming galaxies. The two BCGs most like starbursts, A1835 and A1068, have
ratios similar to the luminous infrared galaxies (LIRGs) of the Calzetti et al. (2010) sample.

We briefly discuss the results of the comparison to theoretical SEDs of star-forming galaxies 
for each source, ordered approximately by decreasing mid-IR luminosity. Note that all fluxes and
luminosities in this paper are derived from PAHFIT, not these fits.

{\bf Abell 1835}. The IRS spectrum of Abell 1835 shows a strong red continuum and PAH features 
whose shapes and intensities are well-fit by the model in Table~\ref{table:sbmodel}. The most prominent
residual is the overestimate of the flux of the [S~III] 18.7 $\mu$m line.

{\bf Abell 1068}. The IRS spectrum of Abell 1068 is similar to that of Abell 1835 in that it has
a strong red continuum, but its emission features are weaker. All features are well matched except
for a 24 $\mu$m feature that is likely spurious but could be associated with [Ne~V]24.3 $\mu$m. 
(The deficiency in the model spectrum between 13-18 $\mu$m is a shortcoming of the \citet{2008ApJS..176..438G}
models compared to many starburst spectra.)

{\bf PKS0745-19}. The IRS spectrum of PKS0745-19 also has a strong red continuum but fainter PAHs 
than Abell 1835.  The observed long wavelength slope is flatter than the model. 
The observed blend of [Fe II] + [O IV] at 26 $\mu$m is bright. For this source, the [Ne II]12.8 $\mu$m
and [Ne III]15.6 $\mu$m features are underpredicted by our simulated SED. The H$_2$ S(1) transition
is quite bright. 

{\bf Hydra A}. The fit to the IRS spectrum of Hydra A overpredicts the [S~III] 33 $\mu$m
line somewhat, while the  [S~III] 18.7 $\mu$m line is underpredicted.
The ratio of this line pair is set by gas density. Shortward of 20 $\mu$m, the 
spectrum resembles the best fit starburst model fairly well, except for 
the previously noted deficiency in the model spectra between 10-19 $\mu$m. 
As in other systems, the [FeII] + [OIV] blend is not modeled.

{\bf Abell 2597}. The IRS spectrum of Abell 2597 is fairly flat. Most features seem to be represented. There
are small excesses of [Ne II] and [Ne III] emission visible in the residuals.

{\bf Abell 1795}. The IRS spectrum of Abell 1795 is fairly well represented by the starburst models.

{\bf Abell 478}. The IRS spectrum of Abell 478 is faint. The data exhibit stronger lines of 
 [Ne II] and [Ne III] and a flatter long-wave spectral slope.

{\bf 2A0335+096}. The IRS spectrum of 2A0335+096 has a flat red continuum, and molecular hydrogen
and neon lines are well fit by the model. The most prominent feature
in the residual spectrum is the strong [Si II] 34.8 $\mu$m line. 2A0335 is the lowest redshift
source, so we have no other sources for context, and it does sit near the edge of the spectrum where
noise spikes are not uncommon.  The [Ne III]/[Ne II] ratio is higher than predicted in the best fit model.

{\bf MS0735+74}. The IRS spectrum of MS0735+74 has a flat red continuum with very weak emission line
features, and to the limits of the data, fit but not tightly constrained in this exercise.

In summary, the starburst models of time-averaged HII regions and PDRs, based on fits to the IRS data, 
do a surprisingly good job at qualitatively representing the continuum, PAH, and nebular
features of the IRS spectra of BCGs, but far from perfectly. On the other hand,  
the star formation rates derived are consistent with estimates based on the 70$\mu$m continuum or PAH features. We
will discuss this further in \S~\ref{section:H2}, where we show that the H$_2$ and [Ne~II] luminosities are
significantly higher than those of star-forming galaxies with similar infrared luminosities.
While the PAHs and the IR continuum are usually well represented, the models do not match
the nebular [Ne~II] (and [Ne~III]) emission relative to the continuum, and the slope of the
continuum through the longest wavelengths of the IRS spectra is not consistently fit for 
spectra with faint IR continuua. We note that the metric for a best-fit 
for a given starburst model is dominated by the continuum since most of the points are continuum-dominated.

\begin{deluxetable}{rccccccccccc}
\tabletypesize{\scriptsize}
%\rotate
\tablecaption{Best Fit Starburst PDR + HII Model Parameters and Stellar Masses\label{table:sbmodel}}
\tablewidth{0pt}
\tablehead{
\colhead{Name} & \colhead{$Z/Z_\odot$} & \colhead{$\cal{C}$} & \colhead{Log P/k} & \colhead{HII} &
\colhead{$A_V$} & \colhead{Old Star} & \colhead{SFR } &
\colhead{T1} & \colhead{N1 (H$_2$)} &
\colhead{T2} & \colhead{N2 (H$_2$)} \\
\colhead{} & \colhead{} & \colhead{} & \colhead{(K cm$^{-3}$)} & \colhead{or PDR} &
\colhead{(mag)} & \colhead{$10^{11}~\rm{M}_\odot$} & \colhead{M$_\odot$ yr$^{-1}$} &
\colhead{(K)} & \colhead{ ($10^{18}~\rm{cm}^{-2}$) } &
\colhead{(K)} & \colhead{ ($10^{18}~\rm{cm}^{-2}$) } }
\startdata
2A0335  & 1 & 4   & 8 & PDR & 0 & 1.5 & 0.7 & 350 (5) & 52 (0.2) & 1200 (30) & 0.8 (0.1) \\
A478    & 2 & 4   & 8 & PDR & 0 & 4.4 & 2.7 & 460 (10) & 13 (0.7) & 1920 (200) & 0.12 (0.02) \\
A1068   & 1 & 6.5 & 8 & PDR & 2.5 & 8.6 & 100 & 320 (4) & 98 (7) & 2000* & 0.120 (0.007) \\
A1795   & 2 & 4   & 8 & PDR & 0 & 2.6 & 2.3 & 260 (13) & 48 (0.6) & 760 (40) & 1.7 (0.2) \\
A1835   & 2 & 5.5 & 6 & PDR & 0 & 26  & 270 & 530 (40) & 4.7 (1.5) & 1300 (750) & 0.03 (0.02) \\
A2597   & 2 & 4   & 8 & PDR & 0 & 4.1 & 5.4 & 240 (4) & 160 (8) & 810 (17) & 4.9 (0.3) \\
HydraA  & 2 & 5   & 7 & PDR & 0 & 3.3 & 4.3 & 380 (25) & 15 (0.3) & -- & -- \\
MS0735  & 2 & 4   & 6 & PDR & 0 & 9.7  & 0.3 (0.12) & 300 (240) & 4.4 (7) & 730 (70) & 0.5 (0.2) \\
PKS0745 & 2 & 6.0 & 8 & PDR & 0 & 5 &  11 & 390 (20) & 26 (1) & 1200 (65) & 0.48 (0.05) \\
\enddata
%% Text for table notes should follow after the \enddata but before
%% the \end{deluxetable}. Make sure there is at least one \tablenotemark
%% in the table for each \tablenotetext.
\tablecomments{*Value pegged at extreme temperature. Uncertainties are quoted in parentheses. 
Component normalizations for the old stellar population SED and the starburst SED had statistical  
uncertainties of less than 1-3\%, with the exception of MS0735. However,
the photometric calibration and  scaling uncertainties were $\sim15-20\%$, so at minimum,  
uncertainties at that level apply to the stellar masses and SFRs estimated here. }
\end{deluxetable}

\section{Discussion}

In typical star-forming galaxies, the luminosities of dust, PAHs, [Ne~II] lines, and even the
rotationally-excited molecular hydrogen lines are linearly correlated with each other and with the
star formation rate (SFR). There is a significant uncertainty in the SFR  inferred for any individual galaxy, 
a factor of 5-10, because of the dispersion, but these quantities are highly correlated in star-forming
galaxies. Therefore we compare the correlations we see for the BCGs in our sample with those
of star-forming galaxies.
We will show here, based on the correlations and ratios that we observe for BCGs, 
that the infrared continuum and PAH features are consistent with being powered
primarily by star formation in BCGs. In contrast, the emission lines from rotational transitions 
of hydrogen are uncorrelated with the dust and PAH features, and are primarily powered by a 
second process. The forbidden lines of neon are correlated
with the IR emission, but not linearly. This pattern is consistent with these lines being powered
by star formation and a second process that does not provide much heat to the PAHs and dust but
is very effective at producing H$_2$ emission. This second heating mechanism is consistent with
heating by a population of suprathermal electrons, either from the hot gas or perhaps associated with the
radio source. 

All linear correlation coefficients ($r$) in the analysis below are based on the measurements in {\em logarithmic} 
quantities  unless stated otherwise. For reference, a correlation
of $r>0.66 (r>0.86)$ might be considered significant at the $2\sigma$ ($3\sigma$) 
level  for $N = 9$ points (e.g. Bevington 1969; 
Bevington \& Robinson 2003). In general, if two quantities are correlated linearly (i.e., the log-log slope is unity), 
they may have common origins. But if they are correlated, but not linearly, there may be something more
interesting going on. For that reason, we plot dotted lines with 
unity slope in our correlation graphs.

\subsection{Dust and PAH Luminosity Correlations \label{section:pah}}

The smallest dust grains are the polycyclic aromatic hydrocarbons (PAHs), composed of only 
a hundred atoms or so. These structures generate emission from C-H or C-C-C bending modes
\citep{1984A&A...137L...5L,1998A&A...339..194B,2000A&A...357.1013V}, excited by the absorption 
of UV photons \citep{1985ApJ...290L..25A,1999ApJ...513L..65S}. 
Such photons can heat these tiny grains stochastically, causing them to suddenly
increase in temperature then cool \citep[e.g.,][]{2001ApJ...554..778L}. PAH features at 3.3, 6.2,
7.7, 8.6, and 11.3 $\mu$m in spectra
are thought to be an excellent tracer of B stars, or of relatively recent star formation
\citep{2004ApJ...613..986P,2006ApJ...653.1129B,2004A&A...419..501F}. 
The continuum IR luminosity, which traces star formation, and PAH luminosity is strongly correlated in normal
star-forming galaxies \citep{2010ApJ...723..895W}.  Studies of
low metallicity star-forming dwarf galaxies by \citet{2008ApJ...674..814R} and of
star forming regions in irregular galaxies by \citet{2007AJ....134..721H}
show the PAH emission decreases as metallicity decreases, so metallicity is one factor that can lead to 
scatter in the correlation between PAH emission and IR luminosity from dust.

\begin{figure}
\plottwo{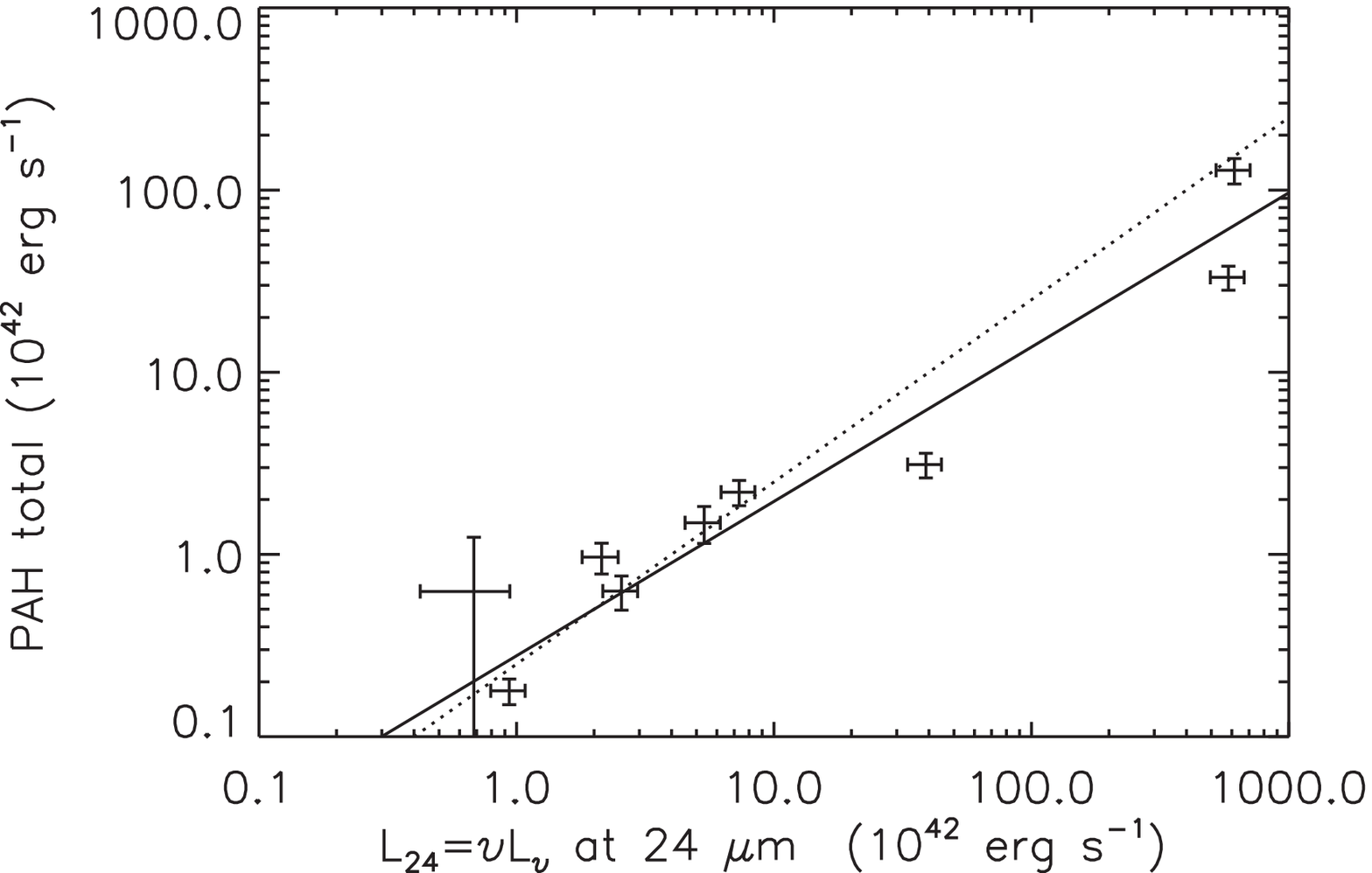}{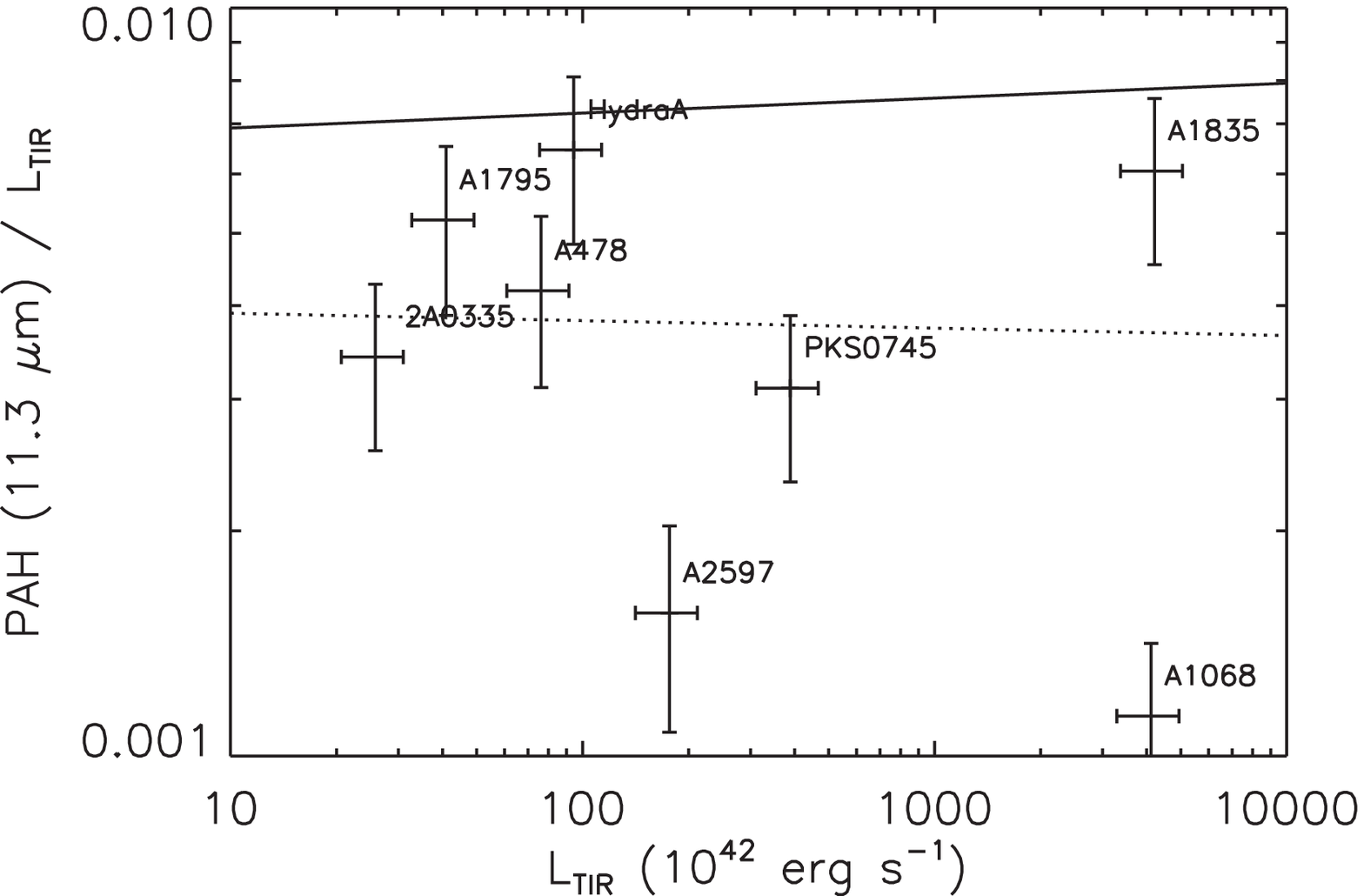}
\caption{PAH and infrared continuum  properties compared. 
The 24 $\mu$m continuum luminosity ($L_{24}$) is strongly correlated with the sum of the 
luminosities of the PAH complexes at  11.3 $\mu$m and 7.7 $\mu$m. The slope  of the best-fit power
law (plotted as a solid line) is $0.90 \pm 0.03$. For comparison, a dotted line of slope unity is shown.
On the right, we show that the ratio of PAH (11.3 $\mu$m) to $L_{TIR}$ (W; Table~\ref{table:TIR}) is $\sim 0.0039 \pm 0.0020$. 
The solid line is the mean $\sim 0.0066^{+0.0045}_{-0.0042}$ 
from \citet{2010ApJ...723..895W}; the dashed line is the lower limit of the range of their best-fit to the mean
for starburst galaxies. Given the uncertainties
in converting from $L_{70}$ and $L_{24}$ to $L_{TIR}$, this plot shows that these galaxies have 
PAH/IR luminosity ratios that are only somewhat lower than  normal star-forming galaxies, 
with the exception of A1068 and A2597.
\label{fnupah}}
\end{figure}

The brightest features from the polyaromatic hydrocarbons (PAHs) are the complexes at 11.3 $\mu$m and 7.7
$\mu$m. The sum of those lines in our BCG sample is strongly correlated with the 24 $\mu$m continuum
 in both flux and luminosity  (Figure~\ref{fnupah}). The correlation coefficient 
$r=0.93$ for both. Excluding MS0735 does not affect the correlation. 
The relationship is very close to linear: 
$L_{11.3+7.7} \propto L_{24}^{0.90\pm0.03}$, where $L_{24}=\nu L_{\nu}$ at 24 $\mu$m rest frame.
($L_{11.3} \propto L_{TIR}^{0.96\pm0.05}$.)

The ratio of the 11.3 $\mu$m luminosity to $L_{TIR}$ is $\sim 0.0039 \pm 0.0020$ (omitting MS0735 from
the sample for lack of MIPS data), plotted in Figure~\ref{fnupah}b, to compare to the mean of 
$\sim 0.0066^{+0.0045}_{-0.0042} L_{11.3 \mu \rm{m}}^{0.02\pm 0.03}$ 
from 123 starburst dominated galaxies from a 24-$\mu$m flux-limited sample of 330 galaxies in \citet{2010ApJ...723..895W}. 
The best-fit power law relating $L_{11.3 \mu \rm{m}}$ and $L_{TIR}^\alpha$ is $\alpha = 1.05 \pm 0.05$, also similar
to that seen for starburst galaxies \citep{2010ApJ...723..895W}.
Given the uncertainties in converting from $L_{70}$ and $L_{24}$ to $L_{TIR}$, this comparison shows that these galaxies have 
PAH/IR luminosity ratios that are only somewhat lower than  normal star-forming galaxies, 
with the exception of A1068 and possibly A2597. The PAHs in Abell 1835 are about 4 times 
brighter compared to $L_{24}$ than the nearly equally IR luminous Abell 1068, so we detect
significant intrinsic scatter in this ratio. There is no
correlation in the ratio of PAH/IR luminosities to IR luminosity (see Figure~\ref{fnupah}b).

The 11.3 $\mu$m PAH luminosity is highly correlated with the 24 $\mu$m and the TIR luminosity, 
which makes sense if the dust and PAHs are heated
by the same process. The 15 and 24 $\mu$m continuum luminosities 
are strongly correlated with each other ($r=0.96$) and nearly 
linearly correlated ($L_{24} \propto L_{15}^{1.042\pm0.006}$), 
as expected since both quantities are usually produced by dust grains.

\begin{figure}
\plotone{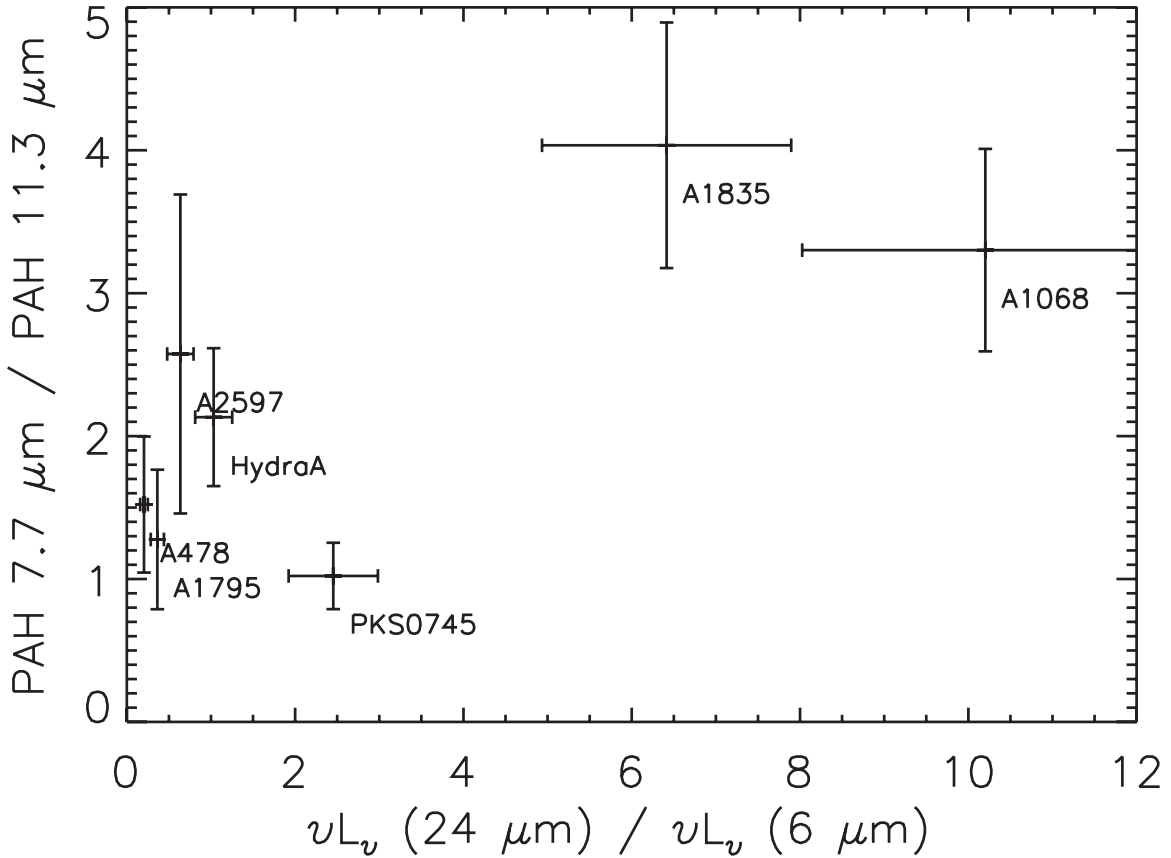}
\caption{A plot of the PAH 7.7 to 11.3 $\mu$m ratio vs. the ratio of the 24/6 $\mu$m continuum luminosities, 
The PAH ratio indicates the ionization level of PAHs, and this ratio is higher in systems in which the 24 $\mu$m
luminosity from dust exceeds the 6 $\mu$m stellar continuum. 
This trend is consistent with the PAHs in the BCG systems most like starbursts experiencing a
harder incident radiation spectrum. \label{starlighttoSFR}}
\end{figure}

In contrast,  24 $\mu$m continuum and 11.3  $\mu$m PAH luminosities are 
not correlated with 6 $\mu$m continuum luminosity. In fact, when
 MS0735 is excluded, there is no correlation between 24 $\mu$m and
6 $\mu$m luminosities, $r=0.17$.
Similarly excluding MS0735, there is no correlation between the 6 $\mu$m
and PAH flux at 11.3 $\mu$m ($r=0.23$) or [Ne~II] ($r=0.004$). (Including MS0735
in the tests increases the correlations to $\sim0.6$, under the $2\sigma$ threshold, but because
the computed significance relies on the inclusion of a single source, it must be considered spurious.)
The 6 $\mu$m light in these BCGs is produced primarily by old (cool) stars, and therefore is a
metric for the stellar mass. The lack of correlation between dust and stellar continua luminosities
suggests that dust, PAH, and gas heating is not determined by  
cool stars. The systems where the dust luminosity well exceeds 
the 6 $\mu$m luminosity from stars, A1068 and A1835, exhibit higher PAH 7.7 to 11.3 $\mu$m ratios,
consistent with the hypothesis that these are like starburst galaxies with levels of 
PAH ionization similar to those seen in starbursts (Figure~\ref{starlighttoSFR}). 
We will consider these ratios more fully in \S~\ref{section:PAHratios}.

For the rest of this discussion, we will assume that the long wavelength IR continuum is powered primarily
by obscured recent star formation. However, even though dust heating by evolved stars does not seem to 
dominate these systems at 24-70 $\mu$m, evolved stars may be the dominant source of heat for cooler dust emitting
at longer wavelengths, and this dust therefore could {\em contribute} to emission at shorter wavelengths (see 
\S~\ref{GROVES}). Furthermore, 
processes of interest such as suprathermal electron heating and weak AGN may also supply energy to these systems. 
Since the observed global quantities are galaxy-wide averages, they suffer from the same 
interpretation ambiguitiy as high redshift, unresolved 
sources. Even if we could interpret these spectra in the context of 
star formation alone, it is impossible to unambiguously distinguish between a star formation episode
of a single age and stellar mass and a time-averaged star formation history of ``constant" star formation.
Physically, signatures of star formation in normal galaxies include 
cold, dusty molecular gas, excess UV continuum, H$\alpha$, PAH emission, and infrared
dust emission. We will discuss these data in a framework where obscured star formation is tracked by
the infrared and PAH emission. 
However, we will show that star formation alone is inadequate to explain the full set of 
infrared spectral features in these systems.

\subsection{Forbidden Neon Line Correlations\label{section:neii}}

The luminosities of forbidden lines of neon, which are channels for radiative cooling, are sensitive to the thermal energy input into 
the ionized gas. They therefore have also been shown to be good tracers of star formation rates in normal star-forming galaxies.
\citet{2007ApJ...658..314H} showed that the sum of the fine structure lines of Ne~II (12.8 $\mu$m)
and Ne~III (15.6 $\mu$m) correlates strongly with IR luminosity in normal star-forming galaxies
over 5 orders of magnitude in luminosity. 
The sums of [Ne II] and [Ne III] luminosities in our BCG sample also correlate strongly with $L_{24}$ 
($r=0.95$, fluxes correlate with $r=0.90$). [Ne II] alone is just as correlated ($r=0.94$, fluxes correlate with $r=0.91$). The relationship, however,
deviates even more from linearity than the PAH-IR luminosity relationship,
with $L([Ne~II])$ scaling as $L_{24}^{0.58\pm0.03}$ (or $L_{TIR}^{0.79\pm0.04}$). We suspect that while the dust and the PAHs are heated primarily by 
star formation, this lack of linearity in the [Ne~II]-IR correlation suggests that 
star formation may not be the sole process producing [Ne~II] emission.

\begin{figure}
\plottwo{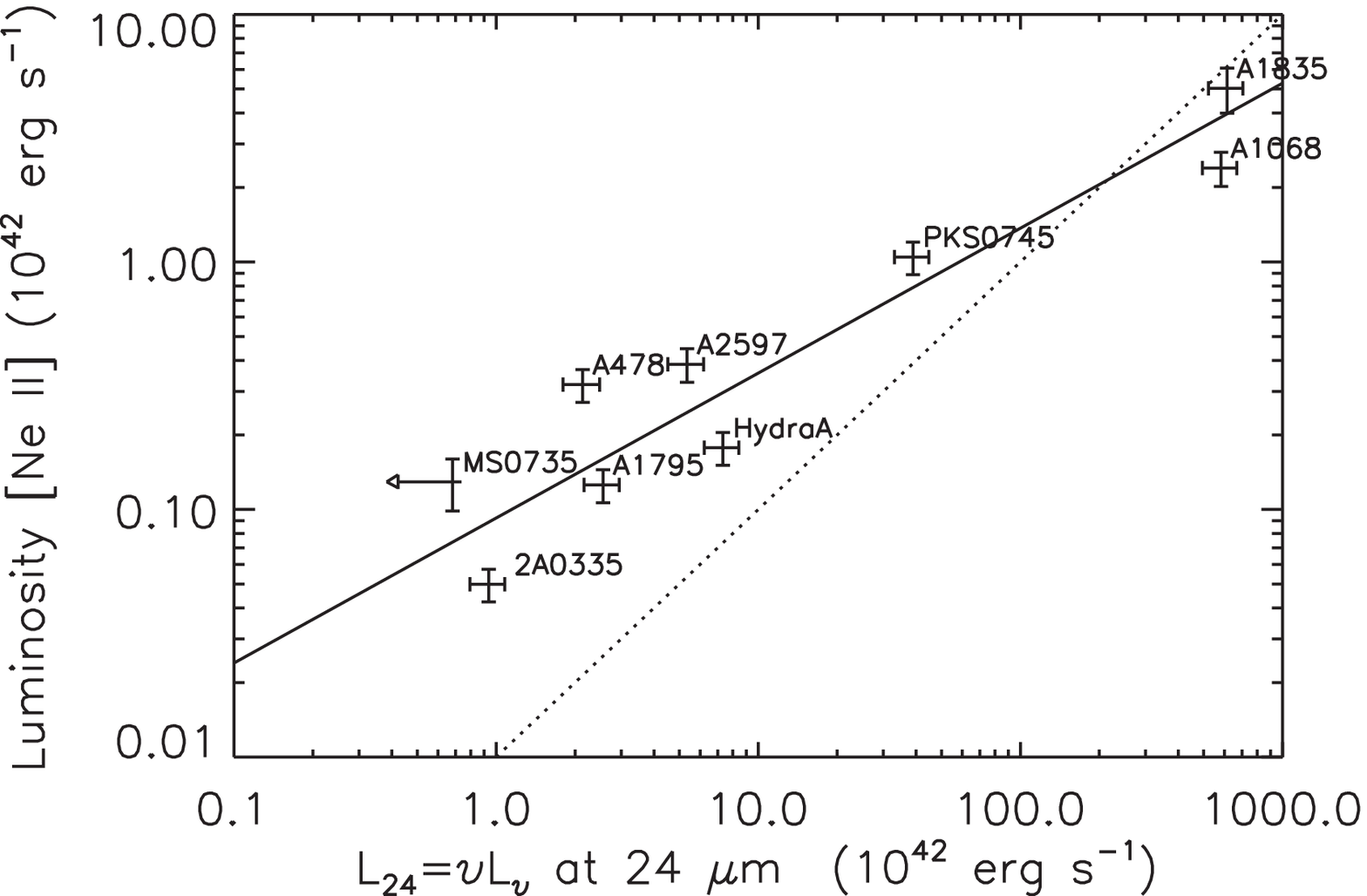}{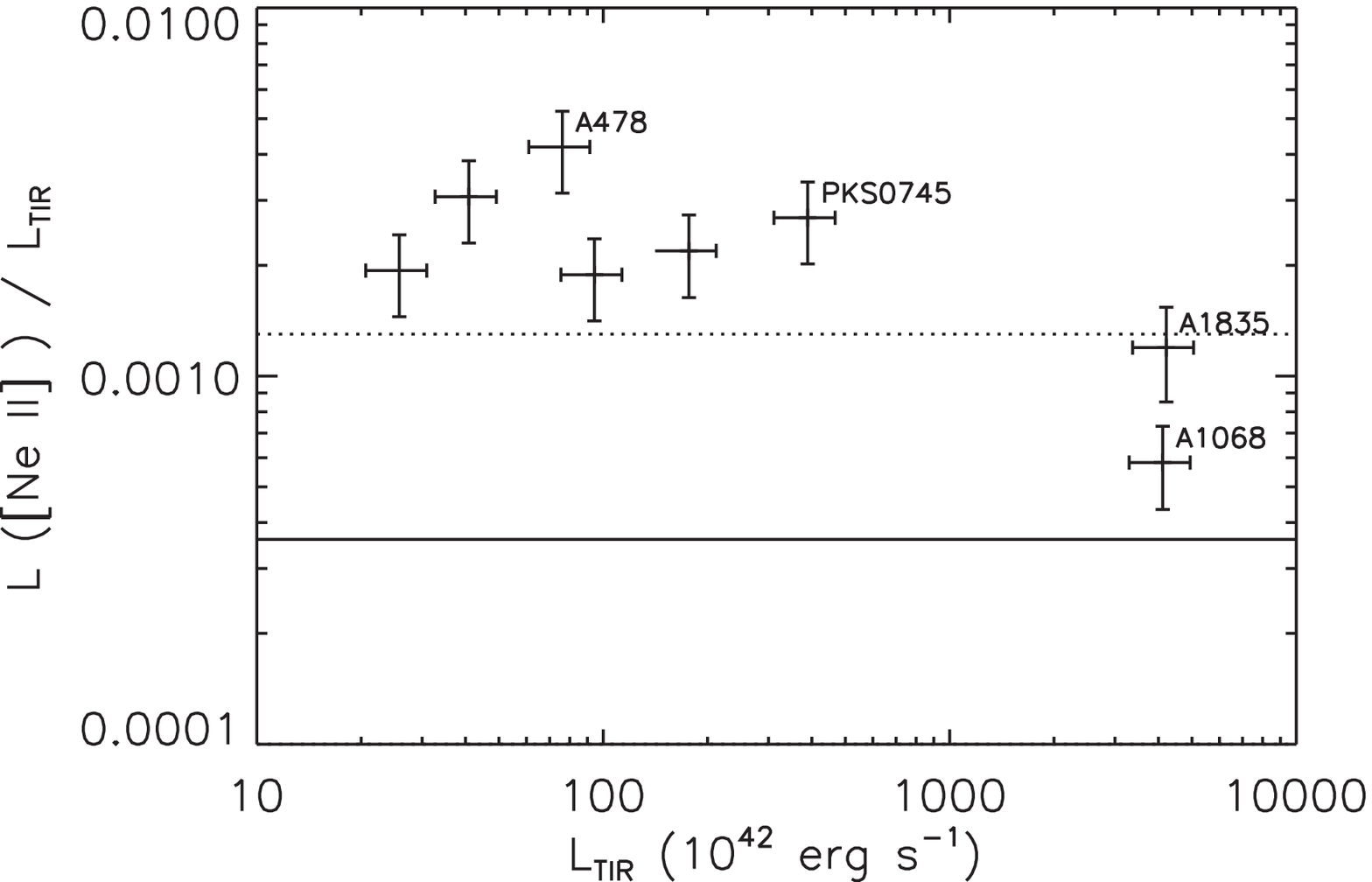}
\caption{[Ne~II] and infrared continuum  properties compared. On the left, [Ne~II] luminosity is  strongly
correlated with $L_{24}$ luminosity, but the best-fit power-law (solid) is flatter than linear, with a index of $0.59 \pm 0.03$.
A dotted line with slope unity is plotted for comparison.  On the right, the ratio of [Ne II] luminosity to 
$L_{TIR}$ ($L_{TIR}$ (W); Table~\ref{table:TIR}) continuum luminosity decreases with increasing $L_{TIR}$. The ratio for typical star-forming
galaxies, $\log({\rm [Ne~II]/L_{TIR}}) = {-3.44 \pm 0.56}$, from Ho \& Keto (2007) is plotted.
The dotted line is the approximate upper bound of the intrinsic scatter.  \label{LnuNeII}}
\end{figure}

The ratio of [Ne II] to total infrared luminosity ($L_{TIR}$ (W); Table~\ref{table:TIR}) 
decreases somewhat with increasing IR luminosity (Figure~\ref{LnuNeII}). 
The \citet{2007ApJ...658..314H}  mean relationship between [Ne~II] and $L_{TIR}$ for star-forming
galaxies is $\log({\rm [Ne~II]/L_{TIR}}) = {-3.44 \pm 0.56}$, nearly independent of $L_{TIR}$.
Excluding MS0735, the [Ne~II] luminosity in BCGs is about $1.6-12$ times higher than the mean [Ne~II] luminosities of
normal star-forming galaxies of similar infrared luminosities. The largest differences are found 
for the BCGs with lower IR luminosities ($<11^{10} ~\rm{L}_\odot$).  MS0735+74, for which there is only an upper limit 
continuum estimate, is particularly bright in [Ne~II] compared
to its infrared luminosity ($\gtrsim 0.009$), $>25 \times$ the mean. 
The observed scatter of this ratio for normal star-forming galaxies in \citet{2007ApJ...658..314H} is large, $\pm0.5$ dex;
nevertheless, the BCG ratios sit consistently on the high side of the scatter for normal star-forming galaxies, indicating that 
another process beyond star formation is also contributing to the heating of the ionized gas, particularly in the low-luminosity systems.
In summary, the [Ne~II] luminosities seen in the low IR-luminosity BCGs exceed what would be expected from a 
star-forming galaxy with the same IR luminosity, but the two quantities are strongly correlated. 

\begin{figure}
\plotone{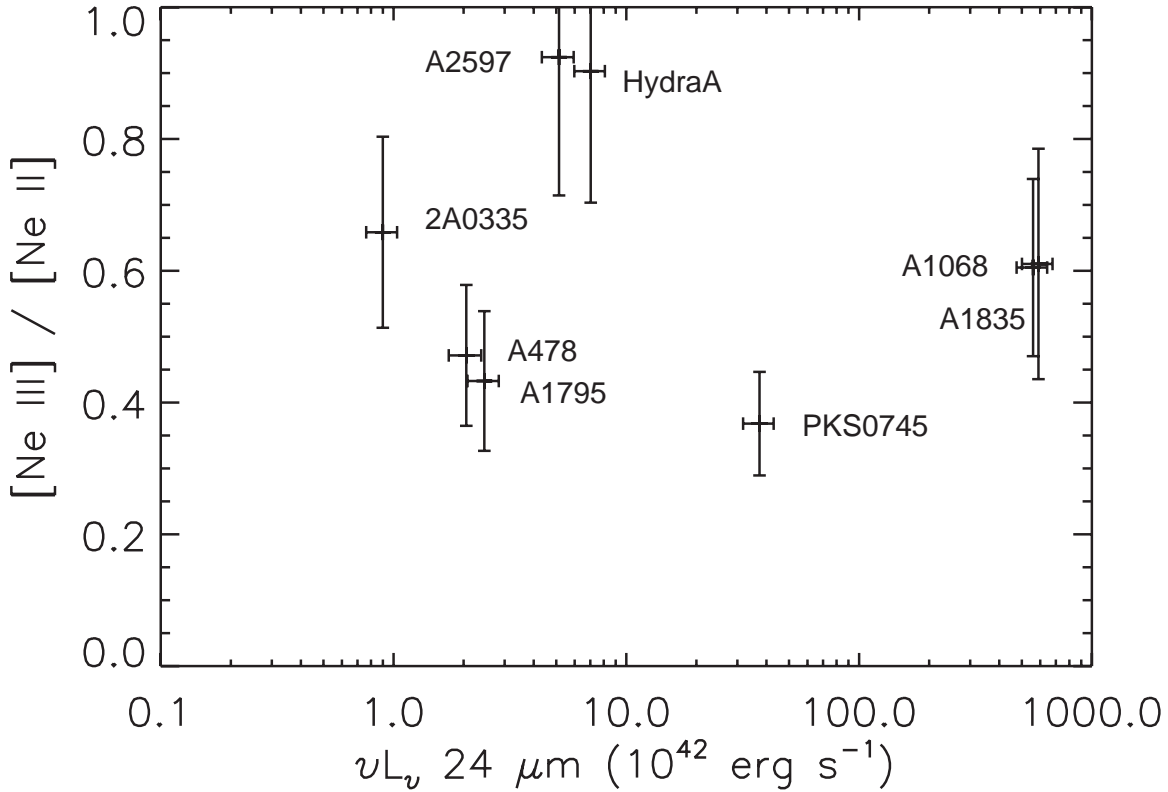}
\caption{[Ne III] / [Ne II] ratios, indicators of the hardness of the incident UV spectrum for the ionized gas, 
are plotted against the 24 $\mu$m luminosity, a surrogate for star formation
rate. The ionization level of the gas shows no correlation with the mid-IR continuum. The neon
line ratios exhibit intrinsic scatter, possibly evidence that either the ages of the youngest stars vary
from system to system (which would mean a constant star formation rate is not a good assumption), 
or the mix of heating mechanisms (between star formation and energetic particles) 
differs from galaxy to galaxy. \label{neonratioIRlum}}
\end{figure}

The [Ne III]/[Ne II] ratio is not at all correlated with the 
mid-IR luminosity (Figure~\ref{neonratioIRlum}).  
It is possible that the [Ne III]/[Ne II]  ratio indicates an approximate starburst age, with BCGs having
the highest [Ne III]/[Ne II]  also having the youngest starburst populations, but not necessarily the
largest numbers of young stars \citep{2000ApJ...539..641T,2004ApJ...606..237R,2007ApJ...669..269S}.

\subsection{Molecular Hydrogen Correlations \label{section:H2}}

Extremely luminous pure rotational H$_2$ lines, usually 
S(3) 9.67 $\mu$m, S(2) 12.28 $\mu$m, S(1) 17.04 $\mu$m, 
 are detected in all nine galaxies.   S(0) 28.22 $\mu$m was not detected in
 any of these sources. Rotational transitions from S(5) - S(7) (5.51 $\mu$m) 
are seen in a majority of these spectra. 
While the luminosities of rotationally-excited molecular hydrogen lines are correlated
with IR luminosities of star-forming galaxies \citep{2010ApJ...719.1191T}, 
that is certainly not the case with our BCG sample.
The line luminosities from rotational molecular hydrogen transitions from
these BCGs are much greater than expected from the level of star formation heating
the warm dust. H$_2$ emission is also uncorrelated with the continuum 
at 15 or 24 $\mu$m, $r=0.4-0.5$.  

Rotational emission from 
molecular hydrogen is commonly detected in ULIRGs \citep{2006ApJ...648..323H}
and in star-forming normal galaxies \citep{2007ApJ...669..959R}. In such
galaxies, the luminosities of these lines are only about $4 \times 10^{-4}$ of the
total infrared power between 8-1000 $\mu$m. 
However, the ratio of H$_2$ luminosity to $L_{24}$ 
for the BCGs in this sample, ranging from 0.004-0.3, is about 5-100 times more
than one would expect from a photodissociation region. The most extreme
object is MS0735, owing to its faint (and uncertain) IR continuum. The large H$_2$
luminosity from off-nuclear regions in the BCG 
NGC1275 \citep{2007MNRAS.382.1246J} led \citet{2008MNRAS.386L..72F,2009MNRAS.392.1475F} to propose
that much of the H$_2$ luminosity in BCGs located in X-ray cool-core clusters 
can be generated by cosmic ray heating, or
by non-radiative processes such as plasma waves.

Brightest cluster galaxies are not the only galaxies to exhibit unusually large luminosities of rotational molecular hydrogen.
\citet{2007ApJ...668..699O} find the FR~II radio galaxies have strong H$_2$ lines, but these galaxies are dissimilar to
the BCGs in our sample. For example, 3C~326 exhibits
high ionization [Ne V] and [O IV] emission, indicating AGN or LINER-like lines, 
very tiny star formation rates ($<0.1 \rm{M}_\odot~\rm{yr}^{-1}$), and the H$_2$ line transitions 
are primarily S(0) and S(1), indicative of cooler molecular 
gas than in our sample. These transitions are also seen in  
IRS mapping of the nearby group of galaxies, Stephan's Quintet, which exhibits bright H$_2$ 
\citep{2010ApJ...710..248C}. Similarly, H$_2$ S(0) and S(1) emission lines 
have been reported from IRS mapping of edge-on spiral galaxies \citep{2010AJ....140..753L}.
An archival study of ULIRGs by \citet{2010Natur.465...60Z} suggests that their H$_2$
emission is not associated with star formation.
 While many of these studies speculate that
shocks might be a source of energy \citep[e.g.][]{2010arXiv1009.4533O}, and might be quite common, the 
unifying thread to all of their discussions is that the molecular hydrogen rotational lines are 
surprisingly bright and their source of energy is still unidentified.
The situation is not much different here, except the BCGs tend to also exhibit rotational lines characteristic
of warmer molecular gas than the groups or radio galaxies (S(2), S(3), to S(7)).

\begin{figure}
\plottwo{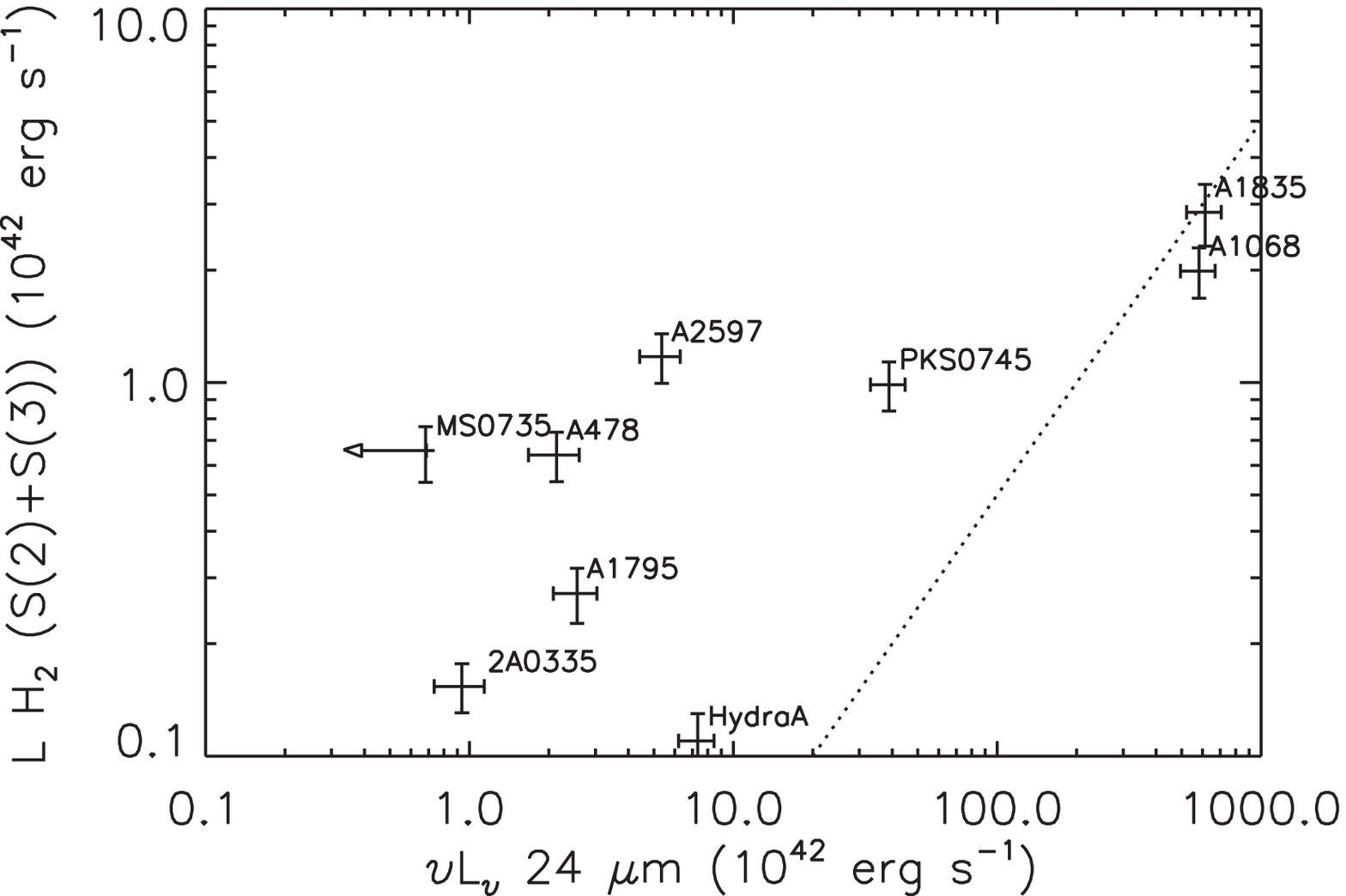}{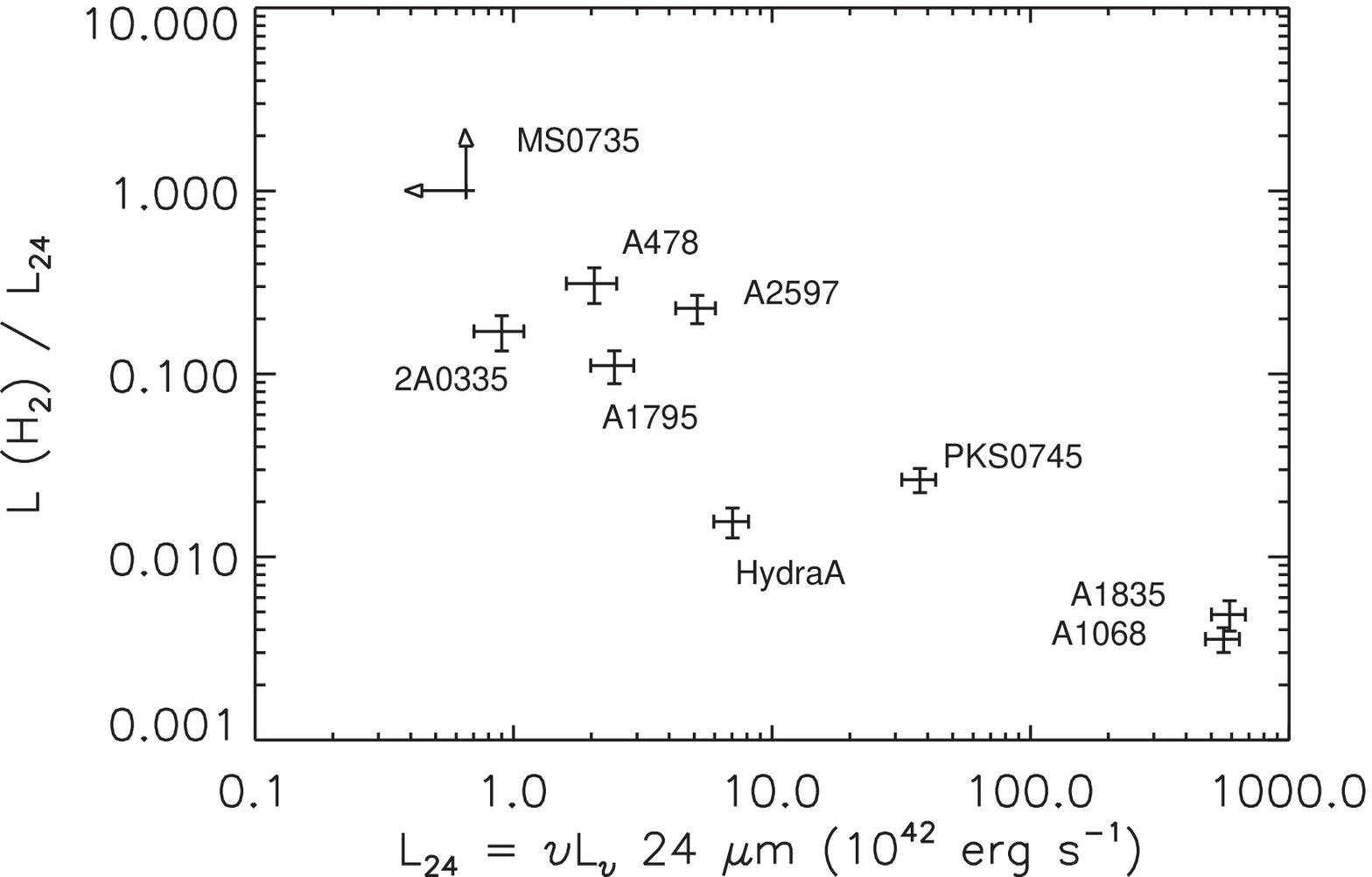}
\caption{Rotational H$_2$ and infrared continuum  properties compared.
On left, the summed luminosity of the two most prominent rotational hydrogen
lines (S(2), S(3)) is very weakly correlated ($<2\sigma$ significance) 
with the $L_{24}$  as might
be expected if the sample is (approximately) flux-limited; if plotted
as a flux-flux diagram all correlation disappears. A dotted line of slope unity is shown. 
On right, the ratio of molecular hydrogen to $L_{24}$ 
decreases with increasing $L_{24}$, which suggests that the source of energy powering 
H$_2$ is unrelated to star formation in most of this sample. At the highest IR luminosities (A1068 and A1835), the H$_2$ might be generated
by star formation processes.
\label{H2L15corr}}
\end{figure}

The mid-IR luminosity is not significantly correlated with the 
summed luminosity  of the molecular hydrogen lines, here represented by the sum of S(2) and
S(3) lines, which were reliably detected in all 9 systems (Figure~\ref{H2L15corr}a). 
While we showed in \S~\ref{section:neii} that [Ne~II] emission is correlated with dust continuum emission, 
here we see that dust continuum is not significantly correlated with molecular 
hydrogen emission (for fluxes, $r=0.5$; for luminosities $r=0.68$).
We plot
the ratio of H$_2$ sum to mid-IR luminosity (Figure~\ref{H2L15corr}b). These ratios 
decrease for the systems with the highest mid-IR luminosities. 
This trend appears because the H$_2$ luminosities are limited
in range (factor of 20) while the IR luminosity spans a large range ($>1000$).
We interpret the trend to mean that
the H$_2$ heat source is more important and in fact dominates the IR emission features 
in systems with low mid-IR luminosities.

\begin{figure}
\plotone{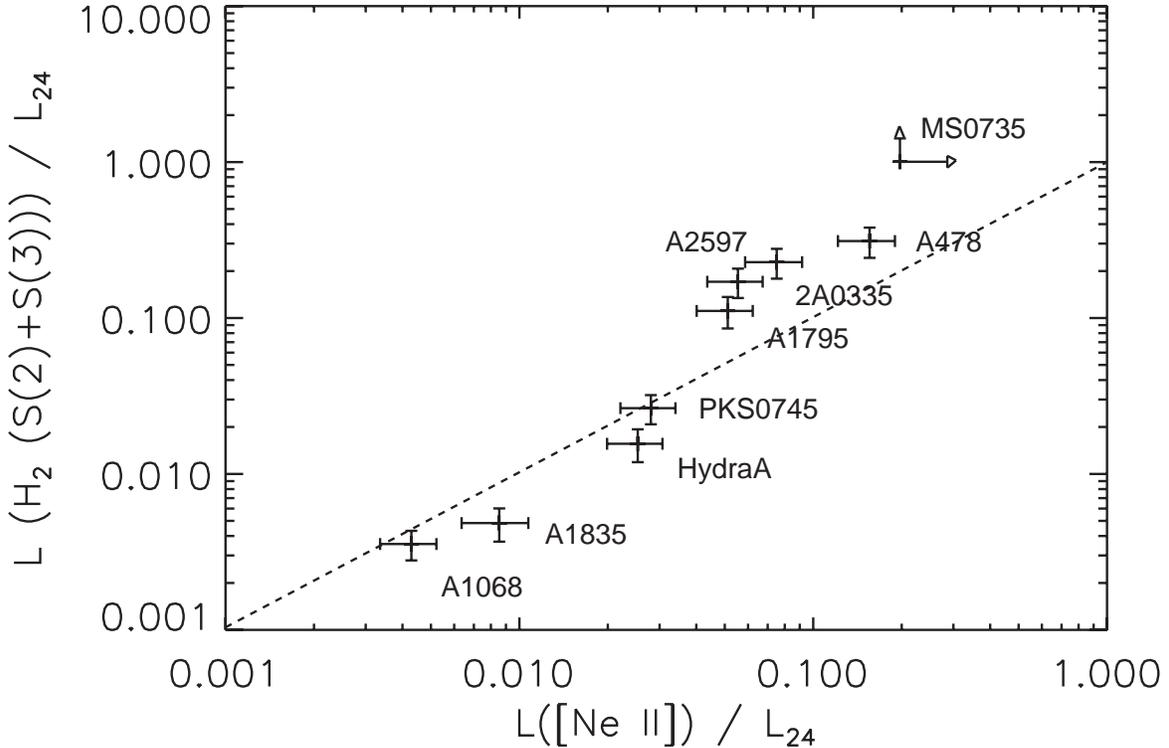}
\caption{[NeII]/IR ratio correlates with H$_2$/IR, which might be expected if the heating source that 
powers the molecular hydrogen well over that expected from star formation also 
elevates the [Ne~II] emissivity of the ionized gas over that expected from star formation. This 
additional heating source might simultaneously explain the H$_2$ luminosity 
and the excess [Ne~II] luminosity. The dashed line indicates a slope of unity.  
\label{Ne2H2IR}}
\end{figure}

 The rather insignificant correlation between H$_2$ and $L_{24}$ 
utterly vanishes once MS0735 is omitted from the sample. A similar
effect happens when [Ne~II] flux is compared with H$_2$ flux: dropping MS0735 from the sample
causes a very weak (less than $2\sigma$) correlation to completely vanish. On the other hand,
the luminosity of [Ne~II] is correlated with the luminosity of the H$_2$ S(2) + S(3) lines
($r=0.92$). The presence or absence of MS0735 has little effect on the inferred 
strong correlations between Ne II, PAHs, and mid-IR continuum flux and luminosity correlations.
This correlation analysis suggests that while there is some relationship between the heat sources
for the ionized gas and the dust, there appears to be a much weaker relationship between
the heating sources for the molecular hydrogen and the dust. The
ratio of molecular hydrogen to IR luminosity (Figure~\ref{H2L15corr}b) decreases with
increasing IR luminosity, however, and suggests that the H$_2$ heating process becomes less
important to the total luminosity budget as star formation increases. 

The strong luminosity correlation between [Ne~II] and H$_2$ required further investigation,
since the lack of correlations in the flux quantities suggested the luminosity correlation may be 
a simple ``bigger is bigger" luminosity-luminosity comparison. (See
\citet{1990ASSL..161..405K} for an infamous description of this type of error, involving a cigar.) 
Intriguingly, we find that the ratio of [Ne II] to $L_{24}$
correlates even more strongly with the ratio of H$_2$ S(2)+S(3)  summed luminosity  to $L_{24}$  ($r=0.98$). 
The best power-law fit to this relationship is H$_2$/IR $\sim$ ([Ne II]/IR)$^{1.49\pm0.12}$ 
(Figure~\ref{Ne2H2IR}). 
The correlation of these ratios suggests that whatever process heats
the molecular hydrogen is likely to be the culprit that boosts the forbidden line luminosity (heating the ionized gas)
as well. 

\begin{figure}
\plottwo{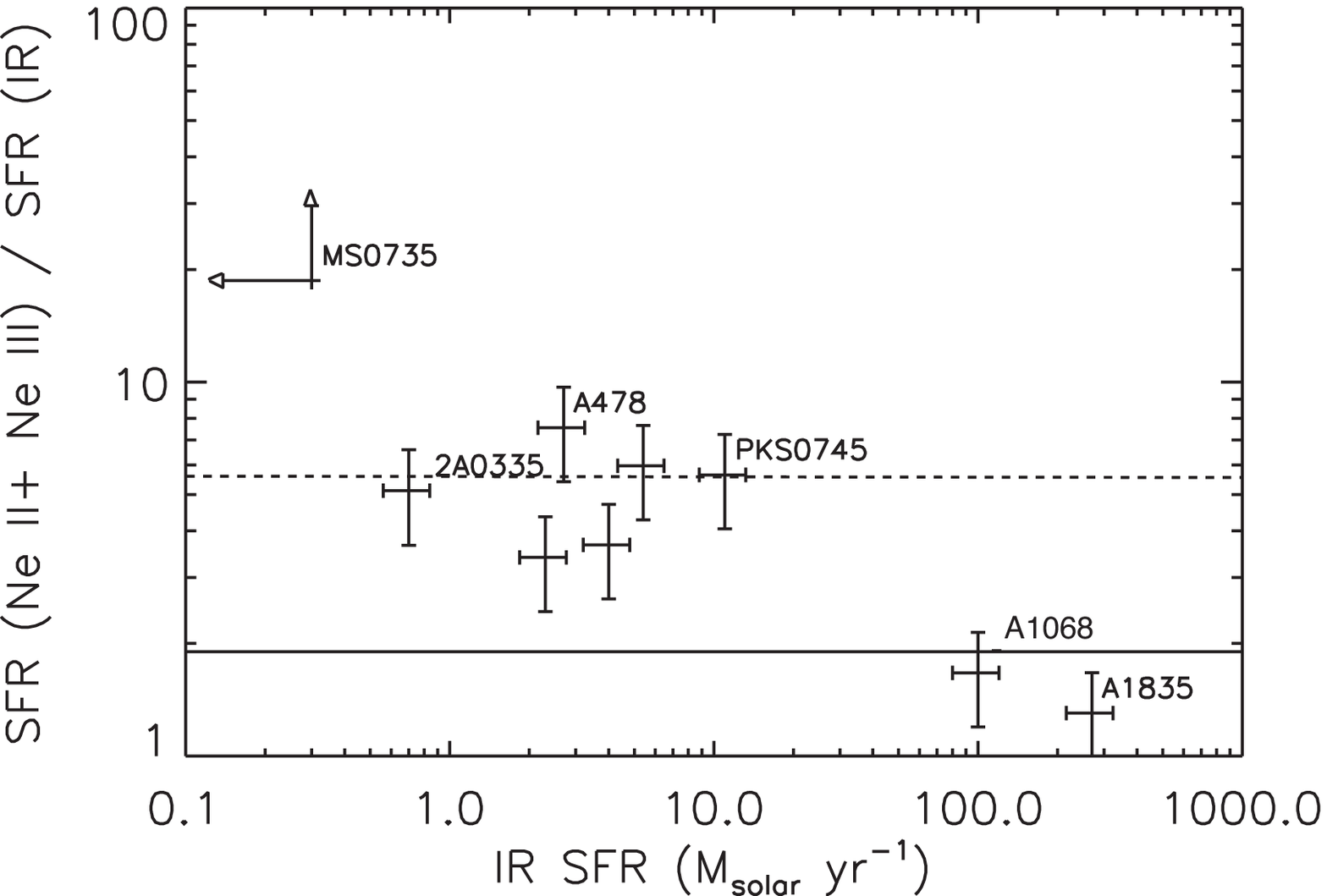}{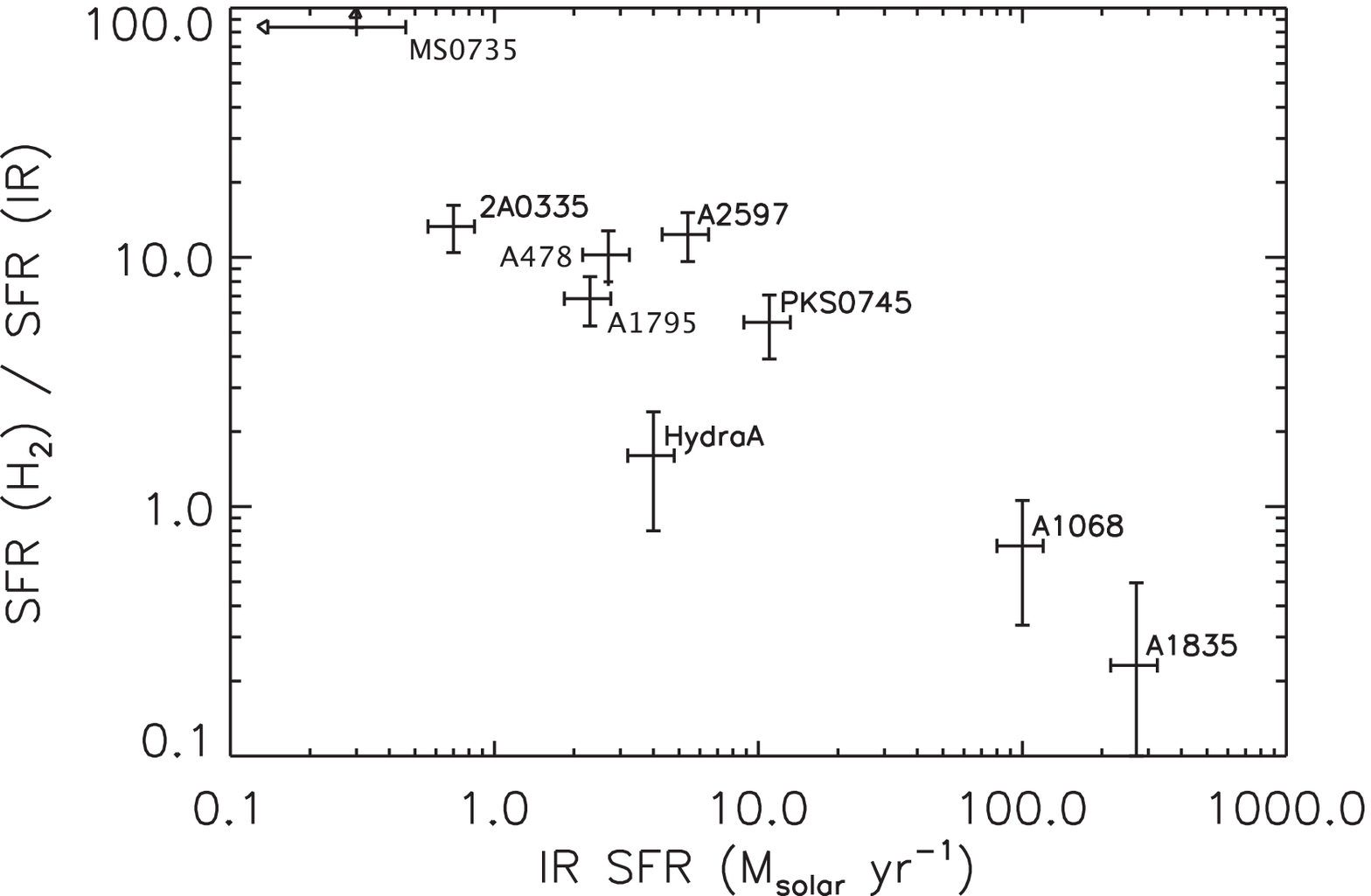}
\caption{ Ratios of inferred star formation rates demonstrate that both the forbidden neon lines and the molecular rotational lines are being emitted in excess of
what would be expected of a star-forming galaxy where the star formation rate is tracked by the IR luminosity. The left plot shows the ratio of the star formation
rates from the sum of [Ne II]+[Ne III] (Ho \& Keto 2007) compared to our estimated IR star formation rates, based on fits to the infrared continuum. The lower line
shows the average of the Ho \& Keto rate compared to the IR rate based on Kennicutt (1998) quantities for their sample. The higher line shows the upper limit
of their scatter. Therefore [Ne~II] in BCGs is moderately over-luminous, by a factor of $\sim 3$ over the mean for star-forming galaxies. 
The rotational transitions in H$_2$ are overluminous
by a factor of 5-15. In both these relations, MS0735 is an extreme example. 
\label{SFR_IR_Ne_H2}}
\end{figure}

To explore this idea further, we investigated how much more luminous  the neon and molecular
hydrogen lines are than one would expect from a star forming galaxy, if the dust luminosity were
a reliable indicator of the level of star formation. In Figure~\ref{SFR_IR_Ne_H2}, we 
estimated the star formation rates inferred from the IR continuum (the best fits to the Groves et al. 2008 models, which
are consistent with Kennicutt (1998) estimates), from 
the [Ne~II]+[Ne~III] luminosity \citep{2007ApJ...658..314H}, and from the H$_2$ luminosity \citep{2010ApJ...719.1191T}. 
For the latter estimate, we assumed that the $Ne^+/Ne = 0.75$ and $Ne^{++}/Ne = 0.15$.  
The H$_2$-based star formation rate in Treyer et al. (2010) 
relies on the sum of the S(0), S(1), and S(2) transitions, which were not all detected in our systems. The 
sums plotted are based only on luminosities of the detected lines. These plots show that for most of the BCGs, 
while [Ne~II] is moderately over-luminous for the inferred IR-based star formation rate (a factor of 2-5 above the
upper end of the scatter exhibited by the galaxy sample of Ho \& Keto 2007, and a factor of $\sim 10$ over the mean), 
the H$_2$ luminosity is a factor of 5-15 over-luminous based on the IR-based star formation rates. The BCGs in A1068,
A1835, and Hydra A have ratios typical of starbursts.

It is interesting that the best fit for the points in Figure~\ref{Ne2H2IR} is nearly linear. 
The slightly steeper than linear fit might be explained in the context of heating by suprathermal particles 
if the luminosity of the ionized gas ([Ne~II]) is limited by the finite column density of ionized gas, while
the luminosity of the rotational line emission from the 
molecular gas is limited by the penetration depth of the suprathermal particles into 
the molecular gas, not the total column density of molecular hydrogen.

We defer a full discussion of the excitation analysis of the individual molecular hydrogen lines 
to a future paper. The more approximate dual temperature fit that we have done here, however,
shows very similar trends to those seen in NGC1275 filaments:
the H$_2$ rotational line intensities cannot be fit by a single temperature. 
This trend is consistent with any model with a non-radiative
energy source \citep{2008MNRAS.386L..72F}.

In summary, as we examined correlations of continuum, PAH, and emission lines of [Ne~II] and H$_2$, 
and compared them to correlations and
infrared line ratios in other types of galaxies, it emerged 
that a single heating process cannot explain
the range of infrared properties we see in these BCGs. Star formation seems to play a role, albeit with
varying levels of dominance, in  producing the emission from these systems, but other processes unrelated
to star formation must
also contribute, particularly in systems with apparently low rates of star formation but high fluxes
from rotationally-excited transitions of molecular hydrogen.

\subsection{ PAHs and Dust Grain Survival and Processing \label{section:PAHratios}} 

If the dust in these BCGs spent much time in contact with the hot, X-ray emitting gas (or more generally,
suprathermal electrons), one
might expect the dust properties, such as its size distribution or ionization fraction, to be different from
dust that has not undergone such a traumatic
experience.  PAH survival alone is problematic if suprathermal particles alone
provide heat: radiation and collisions make PAH lifetimes
in the harsh environment of the center of a cool-core cluster of galaxies quite short.
Using order of magnitude cross sections from \citet{1992MNRAS.258..841V}, and 0.5 keV photon fluxes of
about $10^6$ cm$^{-2}$ s$^{-1}$ we estimate lifetimes of order one million years. 
The damage from particle collisions may be
even more dire.  From the analysis of \citet{2010A&A...510A..37M}, the
lifetime of PAH molecules embedded in $\sim~1$ keV gas with a density of $\sim~0.1$
cm$^{-3}$ is limited to hundreds of years by collisions with the hot
electrons and ions.  Any processing along these lines causes the PAHs and
small grains to evaporate preferentially compared to large grains. The presence of PAHs 
requires the dusty gas to be shielded from the hot gas and its
radiation. 

As in other galaxies observed with the {\em Spitzer} IRS 
\citep{2007ApJ...656..770S,2008ApJ...684..270K}, 
the fluxes and luminosities of the PAH complex at  7.7 $\mu$m and  at 11.3 $\mu$m are strongly
and linearly correlated. The  7.7 and 11.3 $\mu$m PAH complex luminosities are
strongly correlated  ($r=0.98$, fluxes at $r=0.94$) for all 7 systems in which both are detected. 2A0335 and MS0735 lack 
PAH 7.7 $\mu$m detections. 

\begin{figure}
\plotone{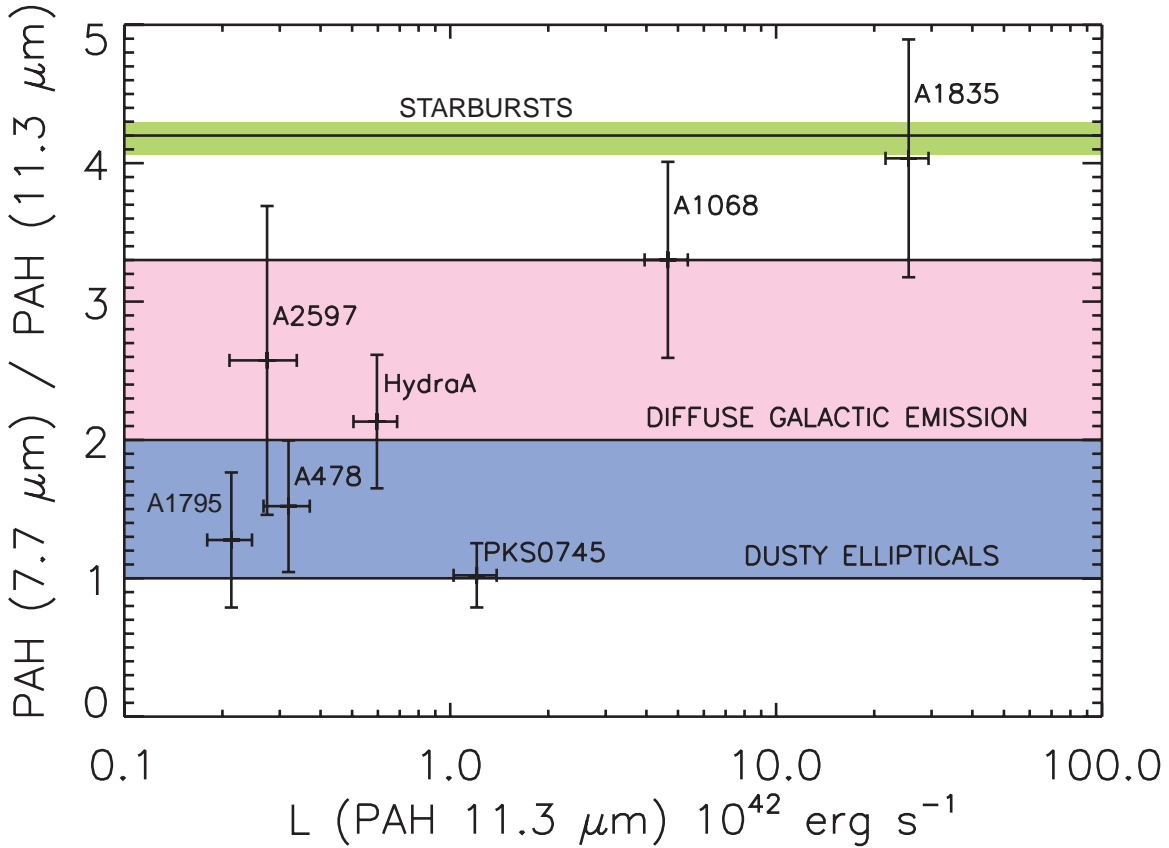}
\caption{The PAH complex at 7.7 $\mu$m is strongly correlated with the PAH complex at 11.3 $\mu$m.
The mean ratio for starbursts, indicated with a single horizontal line at the top of the plot,  is $\sim 4$. 
Diffuse Galactic ISM generates ratios from $2-3.3$, shaded pink (light grey), 
and the ratio in dusty ellipticals ranges from  $1-2$, shaded blue  (darker grey). 
The weighted mean for seven BCGs with PAH 7.7 $\mu$m detections in our sample is $2.7\pm 0.2$, 
indicating radiation fields intermediate in
hardness between dusty elliptical and normal star-forming galaxies.
\label{PAHcorr}}
\end{figure}

The ratio of 7.7 to 11.3 $\mu$m PAH complexes is relatively insensitive to the
sizes of the PAHs \citep{1993ApJ...415..397S}, but sensitive to the ratio of ionized
to neutral PAHs \citep{1989ApJS...71..733A,2007ApJ...657..810D}. For example,
the ratio of the PAH complex at 7.7 to the PAH complex at 11.3 
$\mu$m is lower in galaxy centers that are AGN-dominated compared
with those which are HII-dominated \citep{2007ApJ...656..770S}. 
The mean range for seven ratios of the PAH complexes at 7.7 and 11.3 $\mu$m 
in our sample is $2.7\pm 0.2$ (Figure~\ref{PAHcorr}).
The mean sample ratio is intermediate between that of  HII/starburst galaxies (4.2, \citet{2007ApJ...656..770S})
and  diffuse Galactic emission (2-3.3, \citet{2004ApJ...609..203S}). The BCGs with classic 
indicators of starburst activity (Abell 1068 and Abell 1835) have ratios
typical of starbursts; the others are closer to that of Galactic ISM.
In elliptical galaxies, the PAH 7.7/11.3 ratio is unusually
weak ($\sim 1-2$) \citep{2008ApJ...684..270K}; the lowest ratios seen in dusty ellipticals are
lower than the ratios detected in our BCG sample. 
For BCGs with a low PAH 7.7 $\mu$m to 11.3 $\mu$m 
ratio, the incident spectrum on the PAHs may be dominated
by evolved stars, compared to a harder, ionizing spectrum with contributions 
from young massive stars.  Our result suggests that
the radiation fields in BCGs are intermediate in their hardness 
between those of dusty elliptical galaxies and star-forming galaxies, which is consistent
with what one might expect in the ISM of galaxies with enormous old stellar populations, together with small numbers of recently
formed stars.

\begin{figure}
\plotone{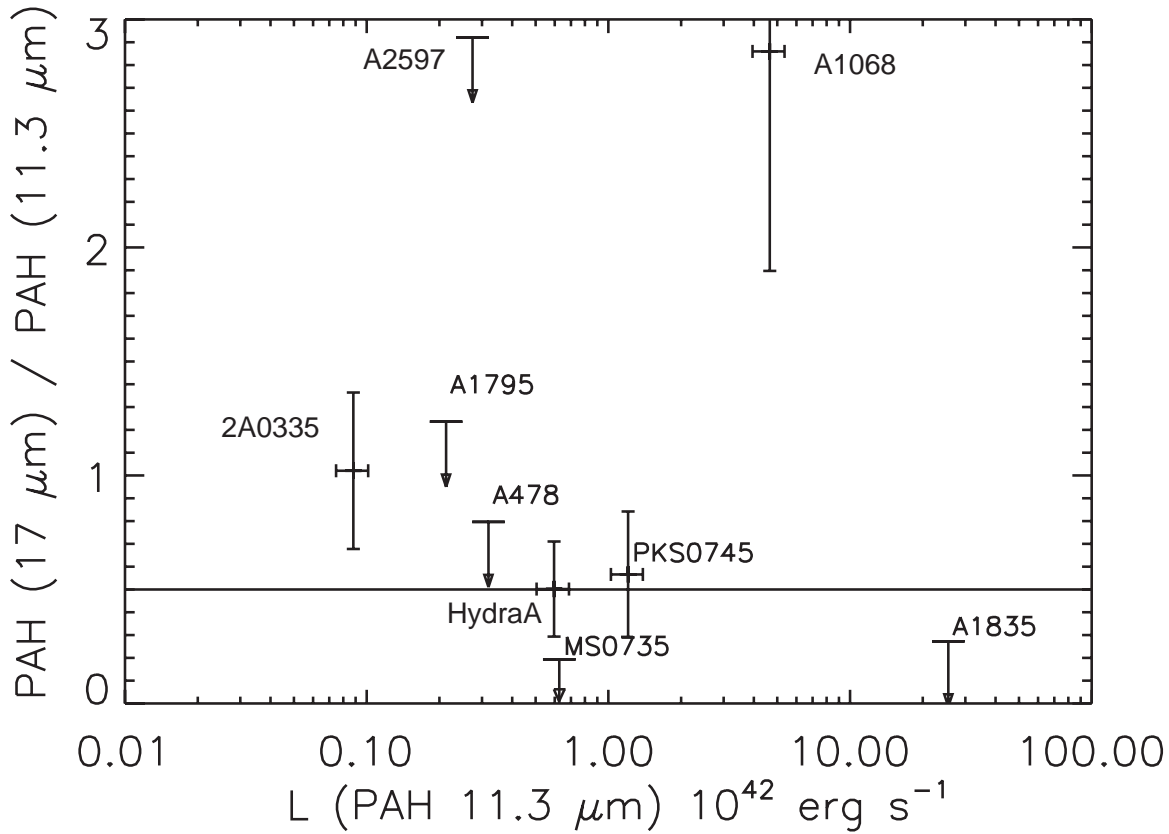}
\caption{The luminosity ratios and upper limits for the PAH complexes at 17 and 11.3 $\mu$m.
The line indicates the mean value of the ratio between the 17 and 11.3 $\mu$m PAH complexes 
(0.5) for normal star-forming galaxies \citep{2007ApJ...656..770S}. Only A1068 has a ratio that is distinctly atypical, suggesting
some processing of the smaller PAHs have occurred in that system. \label{PAHcorr2}}
\end{figure}

The ratio of PAH complexes at 17 and 11.3 $\mu$m is thought to be regulated by PAH sizes with large PAHs contributing
more to the longer wavelength complex \citep[e.g.,][]{2007ApJ...657..810D}. 
The broad PAH feature at 17 $\mu$m was detected with confidence in 4 BCGs (2A0335, A1068, Hydra A, and PKS0745), while
the others have $3\sigma$ upper limits (Figure~\ref{PAHcorr2}). With the single exception of Abell 1068, 
the detections and upper limits are consistent with the ratio of PAH 17 $\mu$m to PAH 11.3 $\mu$m fluxes 
typical of normal star-forming galaxies ($\sim0.5$) \citep{2007ApJ...656..770S}.  
The consistency of the observed ratios and limits in nearly all of these BCGs compared with those  
with normal galaxies suggests that the PAH size distribution may be normal.
 The exception of Abell 1068, with a ratio $>2$, indicates it may have 
deficit of small PAHs compared to large PAHs. This BCG also has a smaller
PAH to $L_{TIR}$ ratio. In the BCG of Abell 1068, PAHs, and particularly the small PAHs, may have been
 preferentially destroyed by collisions with particles or photons.

\begin{figure}
\plotone{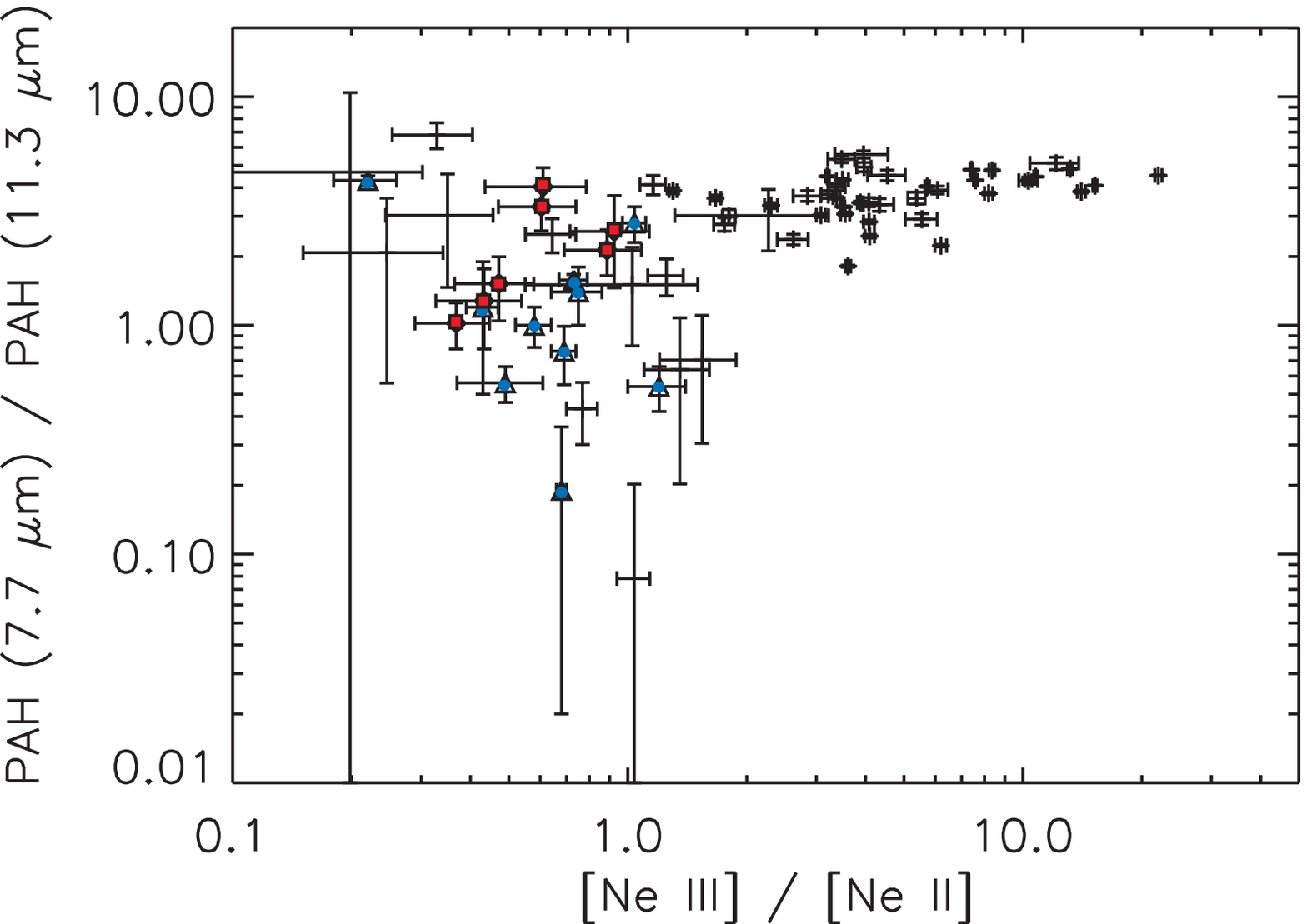}
\caption{The [Ne~III]/[Ne~II] and PAH 7.7 $\mu$m and 11.3 $\mu$m ratios for 55 SINGS galaxies \citep[][using on-line data from 
their Table 4]{2007ApJ...656..770S} with black 
points, elliptical galaxies \citep{2008ApJ...684..270K} with blue filled circles inside triangles, 
and our sample (red, filled square points). 
The [Ne III]/[Ne II] line ratios of BCGs are similar to those seen in star-forming galaxies with low [Ne~III]/[Ne~II] ratios,
LINERS, and dusty ellipticals. Three of the red (filled  square) points (A478, A1795, PKS0745) have somewhat lower PAH 7.7 to 11.3 $\mu$m 
ratios than seen in SINGS galaxies with similarly low [Ne III]/[Ne II] ratios, but are rather similar
to the ratios seen in ellipticals in \citet{2008ApJ...684..270K}. 2A0335 and MS0735 are not plotted here.
\label{fig14Smith}}
\end{figure}

Since both [Ne III]/[Ne II] and the ratio of PAH 7.7 $\mu$m to 11.3 $\mu$m flux are related to 
the hardness of radiation, we compared these quantities to see whether they are correlated 
(Figure~\ref{fig14Smith}) 
and whether the BCG points cover similar parameter space as other galaxies.
We see no correlation between
the PAH ratios and the neon forbidden line ratios, but that is not surprising since
the gas producing the forbidden lines is different from the gas hosting the PAHs; 
similarly, the photons ionizing neon are not the same photons setting the ionization
level of PAHs. As shown in Figure~\ref{fig14Smith}, the seven BCGs with detections in
all 4 quantities exhibit ratios rather similar to SINGS
galaxies \citep{2007ApJ...656..770S}.  There are spiral galaxies in the SINGS sample in Figure~\ref{fig14Smith} 
with very high [Ne~III]/[Ne~II] ratios ($>10$); the explanation may be that these galaxies (about 15\% of the
SINGS sample) are dominated by very recent star formation, and therefore hotter 
O stars \citep{2000ApJ...539..641T,2004ApJ...606..237R,2007ApJ...669..269S}, compared to the BCGs.
 LINERs  in the \citet{2007ApJ...656..770S} sample do not exhibit  [Ne III]/[Ne II] $<0.7-0.8$ 
together with low PAH 7.7/11.3 $\mu$m fractions ($<2$), so a few of the BCGs in our sample with
low [Ne~III]/[Ne~II] ratios (A1795, A478,
and PKS0745) also have lower PAH 7.7/11.3 $\mu$m ratios than seen in the SINGS sample,
more similar to those of dusty ellipticals 
\citep{2008ApJ...684..270K} or diffuse Galactic ISM \citep{2004ApJ...609..203S}.     
The two systems with the highest IR and PAH luminosities of the sample 
(Abell 1068 and Abell 1835) also exhibit
the largest PAH 7.7/11.3 $\mu$m ratios, indicating the PAHs in the most luminous BCG systems
are more ionized than those in the other BCGs.

We arrive at the following conclusions about the PAHs and dust in these systems.

\begin{itemize}

\item The presence of PAHs, and the similarity of the PAH emission ratios to those in star-forming galaxies, 
mean that these tiny grains must be protected from the ICM and shocks. 
PAHs are easily destroyed by ionizing UV and X-rays, by collisions with hot thermal particles, and by shocks. 

\item The emissivities of the PAHs and the ionized gas are correlated with the mid-IR continuum emitted by
dust grains. Therefore there is a common source of heat for these components, consistent with being star formation. However,
the excess luminosity of [Ne~II] and H$_2$ in the less luminous systems suggests that another component is contributing to the heating of
the ionized gas in addition to star formation.

\item The emissivity of the PAHs is not correlated with the stellar (6 $\mu$m) IR continuum. Correlation is expected if the
stars were the main agent heating the dust.

\end{itemize}

We speculate that the evolved stars in the BCGs are the main production source of dust. The dust, except in a couple of
cases, shows little indication that it may have been processed by hot, X-ray emitting plasma or AGN. 
The source of the cool BCG gas may be the hot intracluster medium, but it seems
unlikely that the dust came with the gas. Dust, however, is an extremely effective coolant, and if mixed with the
hot gas, could precipitate the cooling necessary to fuel the formation of dusty molecular star-forming clouds.

\subsection{AGN Contributions \label{AGN}}

Synchrotron radiation from a jet could contribute some of the infrared continuum. 
\citet{2007ApJ...660..117C} find
in their survey of an unbiased set of 3CR and quasars that only the quasars have a nonthermal contribution at
15 $\mu$m of $>20\%$. About half of the 15-$\mu$m emission from nucleus of M87 is synchrotron
emission, based on analysis of its IRS spectrum \citep{2009ApJ...705..356B}. Of the 9 sources in
our sample, only Hydra A has a radio synchrotron source that is luminous enough to contribute more than $\sim10\%$
to the IRS continuum at 24 $\mu$m, if its power law extends unbroken into the infrared. 

High ionization lines in the infrared could indicate a buried AGN or the presence of very hot stars.
[Ne V] 14.3 $\mu$m and 24.3 $\mu$m, with an ionization potential of 97.1 eV, would be an unambiguous AGN indicator 
\citep{1992ApJ...399..495V}. 
We did not detect this emission line in any of our sources at a level 
$\gtrsim 10^{-17}$ W m$^{-2}$), although a 
a weak [Ne V]14.3 $\mu$m feature is detected in Abell 1835 and Hydra A, and
a suspicious (possibly spurious) feature appears near 24 $\mu$m 
in the spectrum of Abell 1068, which is known to have some AGN contribution in the infrared
\citep{2008ApJS..176...39Q}. The lack of [Ne~V]
does not rule out some AGN contribution, since this line can be faint compared to [Ne~II].  
[O IV] at 25.9 $\mu$m would also be an unambiguous sign of AGN photoionization, since
O III has an ionization potential just above
that of He I (to He II), 54.4 eV, and is therefore rare in regions ionized by stars. However,
[O IV] 25.9 $\mu$m is blended with [Fe II] at 26 $\mu$m. We detected this
blend unambiguously only in PKS0745-19 and Hydra A. Puzzlingly (if this blend is an indication
of the presence of O IV), the [Ne III]/[Ne II]
ratio in PKS0745-19 is one of the lowest in the sample.
The [O~IV]-[Fe II] blend does not appear at all in the spectrum of Abell 1068,
however, nor does [Ne V] 14.3 $\mu$m. The lack of these features 
casts further doubt on the candidate [Ne~V] feature in that spectrum.

The presence of high ionization lines such as [S IV] (10.51 $\mu$m) and [Ne III] indicate 
the presence of young hot stars in these galaxies. 
The ionization parameters consistent with the [NeIII]/[NeII] ratios 
in the ionized gas are similar to those seen in star-forming galaxies.
The energy required to ionize neon once to Ne II is 21.6 eV, compared to 41 eV 
for reaching Ne III. 
[Ne II] was easily detected in every galaxy in our sample, and only MS0735+74 lacked detectable [Ne III].
Although S III to S IV has an ionization potential (34.8 eV) similar to Ne  II to Ne III, it has a very low
photoionization cross-section. So a significant detection of [S~IV] not only indicates
hot stars, but a high density of them. In contrast to [Ne III], [S IV] was not detected in any of the systems.
PAHFIT results for 2A0335+096, 
Abell 1835, Abell 1068, and Hydra A indicate faint formal detections at $\sim 3-7 \sigma$, but  
inspection of the fits and data causes us to regard these detections as extremely marginal.

Based on the lack of [Ne~V] emission lines from the gas, we have no conclusive evidence in favor of the gas being 
photoionized by an AGN, consistent with conclusions based on spatially-resolved 
optical emission-line studies of these and similar galaxies \citep[e.g.,][]{1989ApJ...338...48H}.

The detection of spatially extended UV continuum in a number of cases
is unambiguous evidence of the importance of recent star formation over AGN contributions in
this sample of BCGs 
\citep{1993AJ....105..417M, 2002AJ....123.1357M, 2004ApJ...612..131O, 2005ApJ...635L...9H, 2010ApJ...719.1844H,2010ApJ...715..881D}.
Since lack of evidence is not the same as evidence of lack, we  
keep in mind that some of the spectra may have contributions from a low-luminosity AGN since none of these spectra
exclude the nucleus. However, Occam's Razor prefers the simplest explanation for heating these
systems, and as such, no need for AGN excitation or heating is required by these observations.

\section{Conclusions}

In our {\em Spitzer} IRS spectroscopic 
study of 9 brightest cluster galaxies residing in cool-core X-ray clusters, we have detected
very bright molecular hydrogen rotational transitions, PAH features, and forbidden
lines from ionized gas (Ne II, Ne III), in addition to dust continuum at 15-25 $\mu$m. 
These galaxies were known previously to have prominent optical forbidden
and Balmer emission line nebulae \citep[e.g.][]{1989ApJ...338...48H}.  
Photometric MIPS and IRAC studies of similar BCGs have
shown prominent mid-infrared dust emission, together with UV and blue signatures of star formation
\citep[e.g.][]{2008ApJ...681.1035O}. 
The low resolution {\em Spitzer} 5-25 $\mu$m IRS spectra
reveal that BCGs host PAHs, and the ratios of these features indicate that the PAHs in such BCGs are similar to PAHs in
other types of galaxies. The emission from PAHs and dust are highly correlated, nearly linearly, as might be expected
if the PAHs and dust are heated by star formation. The ratios of the 11.3 $\mu$m PAH luminosities to total IR luminosities
are similar to or slightly lower than those of normal star-forming galaxies, with the exception of MS0735 whose
PAH 11.3 $\mu$m feature is well-detected, and its IR continuum is very weak.
Fits of simulated starburst models published in Groves et al. (2008) show that the star formation
rates inferred from the 5-25 $\mu$m spectra are consistent with star formation rates inferred from our 70 $\mu$m MIPS
photometry using relations from Kennicutt (1998) and Calzetti et al (2010). These simulated starburst SEDs also provide reasonable
fits to the PAH features in the spectra, again consistent with the strong, nearly linear correlation between
PAH luminosity and dust continuum luminosities.

The luminosity of the warm ionized gas, traced by the [Ne II]12.6 $\mu$m emission line, is also
correlated with the dust luminosity, so it is likely that the star formation powering the dust
luminosity is also contributing to the heating of the warm ISM. However, 
the relation is distinctly non-linear, in the sense that the systems with
lower IR luminosities have larger [Ne II]/IR ratios. All of the systems are over-luminous
in [Ne II] compared with PAH or dust emission, as most lie at or above the top end of the scatter of this
luminosity relation as seen in star-forming galaxies \citep{2007ApJ...656..770S}.

Even more strikingly, the molecular hydrogen luminosities of BCGs are very high compared to that expected from
star-forming galaxies of similar infrared luminosities. The H$_2$ luminosities are only weakly 
correlated with the mid IR or PAH luminosities, suggesting that the H$_2$ emission line 
power source is nearly independent of that powering the dust. 
The strong correlation of the molecular hydrogen line luminosity scaled by $L_{24}$ with 
[Ne II] scaled in the same way suggests that the warm gas has a second heat source that may be related to 
the primary power source for the molecular hydrogen emission lines, a scenario consistent with excess 
heating by energetic particles and/or conduction from the hot intracluster gas 
\citep{1989ApJ...345..153S,2008MNRAS.386L..72F,2009MNRAS.392.1475F}.

The existence of dusty gas and, in particular, PAHs, along with the fact that PAH feature ratios  
are similar to those found in star-forming galaxies, suggests that (a) the dust and PAHs are
similar in ionization and size distributions as in spiral galaxies and that (b) 
the PAHs and dust are shielded from the destructive X-ray radiation and fast-moving thermal particles
in the ICM. On the other hand, the ionized and molecular gas may indeed receive a noticeable energy dose
from thermal particles from the hot ICM, boosting the forbidden line radiation from the ionized gas
and the rotational line emission from molecular hydrogen.

The fits with starburst SEDs from \citet{2008ApJS..176..438G} give
a consistent story to our empirical comparisons: we can simultaneously fit the mid-IR continuum
and PAH features and some of the ISM emission lines, and the fits give SFRs which
are consistent with Kennicutt (1998) relations for the 24-70 $\mu$m photometry measured
with {\em Spitzer} MIPS and with Calzetti et al. (2010) relations for 70 $\mu$m. However, the models do not
reliably predict the neon line spectrum, which we suggest may be partially generated by the
same physical mechanism that generates the majority of  the molecular H$_2$ 
emission. 

Furthermore, our results are in agreement with the \citet{2007MNRAS.382.1246J}  
analysis of IRS spectra of NGC1275 and NGC4696. Our sample further demonstrates  
that there is no correlation between the star-formation signatures in
these galaxies and the strength of the molecular hydrogen luminosities, and that molecular hydrogen
luminosities are more extreme than produced by typical star formation-related
processes. The neon forbidden lines and rotational lines of molecular hydrogen 
can be enhanced without significantly modifying the luminosities of dust 
if the ionized gas and some of the molecular gas is heated by non-radiative 
agents that can penetrate the molecular gas clouds, such as cosmic rays, hot
electrons, MHD waves.

\acknowledgments
The authors are grateful for tabulated plot data from Daniela Calzetti and for
discussions with Brent Groves and Michael Dopita. We acknowledge the referee for
a careful review of our manuscript. MD, GMV, and AH acknowledge support from a 
Jet Propulsion Lab contract JPL 1353923 and a NASA LTSA grant NASA NNG-05GD82G.
GMV acknowledges partial support from NSF grant AST-0908819.
MD would like to thank the Aspen Center of Physics where she began this work 
in the summer of 2008, and the Kavli Institute of Theoretical
Physics in Santa Barbara, CA, where she finished the paper in February 2011.
This research was supported in part by the National Science Foundation under
grant NSF PHY05-51164.
This work is based on observations made with the {\it Spitzer} Space Telescope, 
which is operated by the Jet Propulsion Laboratory, California Institute of Technology 
under a contract with NASA. Support for GDM, BRM, and RWO was provided by NASA through 
two awards issued by JPL/Caltech.


%% in parentheses, after the keyword, but they will not be verified.

{\it Facilities:} \facility{Spitzer (IRS, MIPS, IRAC)}

\bibliography{spitzer}

\begin{thebibliography}{110}
\expandafter\ifx\csname natexlab\endcsname\relax\def\natexlab#1{#1}\fi

\bibitem[{{Allamandola} {et~al.}(1985){Allamandola}, {Tielens}, \&
  {Barker}}]{1985ApJ...290L..25A}
{Allamandola}, L.~J., {Tielens}, A.~G.~G.~M., \& {Barker}, J.~R. 1985, \apjl,
  290, L25

\bibitem[{{Allamandola} {et~al.}(1989){Allamandola}, {Tielens}, \&
  {Barker}}]{1989ApJS...71..733A}
---. 1989, \apjs, 71, 733

\bibitem[{{Allen} {et~al.}(1992){Allen}, {Edge}, {Fabian}, {Boehringer},
  {Crawford}, {Ebeling}, {Johnstone}, {Naylor}, \&
  {Schwarz}}]{1992MNRAS.259...67A}
{Allen}, S.~W., {Edge}, A.~C., {Fabian}, A.~C., {Boehringer}, H., {Crawford},
  C.~S., {Ebeling}, H., {Johnstone}, R.~M., {Naylor}, T., \& {Schwarz}, R.~A.
  1992, \mnras, 259, 67

\bibitem[{{Armus} {et~al.}(2009){Armus}, {Mazzarella}, {Evans}, {Surace},
  {Sanders}, {Iwasawa}, {Frayer}, {Howell}, {Chan}, {Petric}, {Vavilkin},
  {Kim}, {Haan}, {Inami}, {Murphy}, {Appleton}, {Barnes}, {Bothun}, {Bridge},
  {Charmandaris}, {Jensen}, {Kewley}, {Lord}, {Madore}, {Marshall},
  {Melbourne}, {Rich}, {Satyapal}, {Schulz}, {Spoon}, {Sturm}, {U}, {Veilleux},
  \& {Xu}}]{2009PASP..121..559A}
{Armus}, L., {Mazzarella}, J.~M., {Evans}, A.~S., {Surace}, J.~A., {Sanders},
  D.~B., {Iwasawa}, K., {Frayer}, D.~T., {Howell}, J.~H., {Chan}, B., {Petric},
  A., {Vavilkin}, T., {Kim}, D.~C., {Haan}, S., {Inami}, H., {Murphy}, E.~J.,
  {Appleton}, P.~N., {Barnes}, J.~E., {Bothun}, G., {Bridge}, C.~R.,
  {Charmandaris}, V., {Jensen}, J.~B., {Kewley}, L.~J., {Lord}, S., {Madore},
  B.~F., {Marshall}, J.~A., {Melbourne}, J.~E., {Rich}, J., {Satyapal}, S.,
  {Schulz}, B., {Spoon}, H.~W.~W., {Sturm}, E., {U}, V., {Veilleux}, S., \&
  {Xu}, K. 2009, \pasp, 121, 559

\bibitem[{{Balogh} {et~al.}(2001){Balogh}, {Pearce}, {Bower}, \&
  {Kay}}]{2001MNRAS.326.1228B}
{Balogh}, M.~L., {Pearce}, F.~R., {Bower}, R.~G., \& {Kay}, S.~T. 2001, \mnras,
  326, 1228

\bibitem[{{Bildfell} {et~al.}(2008){Bildfell}, {Hoekstra}, {Babul}, \&
  {Mahdavi}}]{2008MNRAS.389.1637B}
{Bildfell}, C., {Hoekstra}, H., {Babul}, A., \& {Mahdavi}, A. 2008, \mnras,
  389, 1637

\bibitem[{{B{\^i}rzan} {et~al.}(2004){B{\^i}rzan}, {Rafferty}, {McNamara},
  {Wise}, \& {Nulsen}}]{2004ApJ...607..800B}
{B{\^i}rzan}, L., {Rafferty}, D.~A., {McNamara}, B.~R., {Wise}, M.~W., \&
  {Nulsen}, P.~E.~J. 2004, \apj, 607, 800

\bibitem[{{Black} \& {Dalgarno}(1976)}]{1976ApJ...203..132B}
{Black}, J.~H., \& {Dalgarno}, A. 1976, \apj, 203, 132

\bibitem[{{Boulanger} {et~al.}(1998){Boulanger}, {Boisssel}, {Cesarsky}, \&
  {Ryter}}]{1998A&A...339..194B}
{Boulanger}, F., {Boisssel}, P., {Cesarsky}, D., \& {Ryter}, C. 1998, \aap,
  339, 194

\bibitem[{{Bower} {et~al.}(2006){Bower}, {Benson}, {Malbon}, {Helly}, {Frenk},
  {Baugh}, {Cole}, \& {Lacey}}]{2006MNRAS.370..645B}
{Bower}, R.~G., {Benson}, A.~J., {Malbon}, R., {Helly}, J.~C., {Frenk}, C.~S.,
  {Baugh}, C.~M., {Cole}, S., \& {Lacey}, C.~G. 2006, \mnras, 370, 645

\bibitem[{{Brandl} {et~al.}(2006)}]{2006ApJ...653.1129B}
{Brandl}, B.~R., {et~al.} 2006, \apj, 653, 1129

\bibitem[{{Burns}(1990)}]{1990AJ.....99...14B}
{Burns}, J.~O. 1990, \aj, 99, 14

\bibitem[{{Buson} {et~al.}(2009){Buson}, {Bressan}, {Panuzzo}, {Rampazzo},
  {Vald{\'e}s}, {Clemens}, {Marino}, {Chavez}, {Granato}, \&
  {Silva}}]{2009ApJ...705..356B}
{Buson}, L., {Bressan}, A., {Panuzzo}, P., {Rampazzo}, R., {Vald{\'e}s}, J.~R.,
  {Clemens}, M., {Marino}, A., {Chavez}, M., {Granato}, G.~L., \& {Silva}, L.
  2009, \apj, 705, 356

\bibitem[{{Calzetti} {et~al.}(2010)}]{2010ApJ...714.1256C}
{Calzetti}, D., {et~al.} 2010, \apj, 714, 1256

\bibitem[{{Cavagnolo} {et~al.}(2008){Cavagnolo}, {Donahue}, {Voit}, \&
  {Sun}}]{2008ApJ...683L.107C}
{Cavagnolo}, K.~W., {Donahue}, M., {Voit}, G.~M., \& {Sun}, M. 2008, \apjl,
  683, L107

\bibitem[{{Churazov} {et~al.}(2001){Churazov}, {Br{\"u}ggen}, {Kaiser},
  {B{\"o}hringer}, \& {Forman}}]{2001ApJ...554..261C}
{Churazov}, E., {Br{\"u}ggen}, M., {Kaiser}, C.~R., {B{\"o}hringer}, H., \&
  {Forman}, W. 2001, \apj, 554, 261

\bibitem[{{Cleary} {et~al.}(2007){Cleary}, {Lawrence}, {Marshall}, {Hao}, \&
  {Meier}}]{2007ApJ...660..117C}
{Cleary}, K., {Lawrence}, C.~R., {Marshall}, J.~A., {Hao}, L., \& {Meier}, D.
  2007, \apj, 660, 117

\bibitem[{{Cluver} {et~al.}(2010)}]{2010ApJ...710..248C}
{Cluver}, M.~E., {et~al.} 2010, \apj, 710, 248

\bibitem[{{Colless} {et~al.}(2003)}]{2003astro.ph..6581C}
{Colless}, M., {et~al.} 2003, ArXiv Astrophysics e-prints

\bibitem[{{Crawford} {et~al.}(1999){Crawford}, {Allen}, {Ebeling}, {Edge}, \&
  {Fabian}}]{1999MNRAS.306..857C}
{Crawford}, C.~S., {Allen}, S.~W., {Ebeling}, H., {Edge}, A.~C., \& {Fabian},
  A.~C. 1999, \mnras, 306, 857

\bibitem[{{Croton} {et~al.}(2006){Croton}, {Springel}, {White}, {De Lucia},
  {Frenk}, {Gao}, {Jenkins}, {Kauffmann}, {Navarro}, \&
  {Yoshida}}]{2006MNRAS.365...11C}
{Croton}, D.~J., {Springel}, V., {White}, S.~D.~M., {De Lucia}, G., {Frenk},
  C.~S., {Gao}, L., {Jenkins}, A., {Kauffmann}, G., {Navarro}, J.~F., \&
  {Yoshida}, N. 2006, \mnras, 365, 11

\bibitem[{{Dale} \& {Helou}(2002)}]{2002ApJ...576..159D}
{Dale}, D.~A., \& {Helou}, G. 2002, \apj, 576, 159

\bibitem[{{Dale} {et~al.}(2006)}]{2006ApJ...646..161D}
{Dale}, D.~A., {et~al.} 2006, \apj, 646, 161

\bibitem[{{Dale} {et~al.}(2009)}]{2009ApJ...693.1821D}
---. 2009, \apj, 693, 1821

\bibitem[{{Donahue} {et~al.}(2000){Donahue}, {Mack}, {Voit}, {Sparks},
  {Elston}, \& {Maloney}}]{2000ApJ...545..670D}
{Donahue}, M., {Mack}, J., {Voit}, G.~M., {Sparks}, W., {Elston}, R., \&
  {Maloney}, P.~R. 2000, \apj, 545, 670

\bibitem[{{Donahue} {et~al.}(2007{\natexlab{a}}){Donahue}, {Sun}, {O'Dea},
  {Voit}, \& {Cavagnolo}}]{2007AJ....134...14D}
{Donahue}, M., {Sun}, M., {O'Dea}, C.~P., {Voit}, G.~M., \& {Cavagnolo}, K.~W.
  2007{\natexlab{a}}, \aj, 134, 14

\bibitem[{{Donahue} {et~al.}(2007{\natexlab{b}})}]{2007ApJ...670..231D}
{Donahue}, M., {et~al.} 2007{\natexlab{b}}, \apj, 670, 231

\bibitem[{{Donahue} {et~al.}(2010)}]{2010ApJ...715..881D}
---. 2010, \apj, 715, 881

\bibitem[{{Draine} \& {Li}(2007)}]{2007ApJ...657..810D}
{Draine}, B.~T., \& {Li}, A. 2007, \apj, 657, 810

\bibitem[{{Dunn} \& {Fabian}(2006)}]{2006MNRAS.373..959D}
{Dunn}, R.~J.~H., \& {Fabian}, A.~C. 2006, \mnras, 373, 959

\bibitem[{{Edge}(2001)}]{2001MNRAS.328..762E}
{Edge}, A.~C. 2001, \mnras, 328, 762

\bibitem[{{Edge} {et~al.}(2010{\natexlab{a}})}]{2010A&A...518L..46E}
{Edge}, A.~C., {et~al.} 2010{\natexlab{a}}, \aap, 518, L46+

\bibitem[{{Edge} {et~al.}(2010{\natexlab{b}})}]{2010A&A...518L..47E}
---. 2010{\natexlab{b}}, \aap, 518, L47+

\bibitem[{{Egami} {et~al.}(2006{\natexlab{a}}){Egami}, {Rieke}, {Fadda}, \&
  {Hines}}]{2006ApJ...652L..21E}
{Egami}, E., {Rieke}, G.~H., {Fadda}, D., \& {Hines}, D.~C. 2006{\natexlab{a}},
  \apjl, 652, L21

\bibitem[{{Egami} {et~al.}(2006{\natexlab{b}})}]{2006ApJ...647..922E}
{Egami}, E., {et~al.} 2006{\natexlab{b}}, \apj, 647, 922

\bibitem[{{Elston} \& {Maloney}(1994)}]{1994ASSL..190..169E}
{Elston}, R., \& {Maloney}, P. 1994, in Astrophysics and Space Science Library,
  Vol. 190, Astronomy with Arrays, The Next Generation, ed. {I.~S.~McLean},
  169--+

\bibitem[{{Fabian}(1994)}]{1994ARA&A..32..277F}
{Fabian}, A.~C. 1994, \araa, 32, 277

\bibitem[{{Farrah} {et~al.}(2007)}]{2007ApJ...667..149F}
{Farrah}, D., {et~al.} 2007, \apj, 667, 149

\bibitem[{{Fazio} {et~al.}(2004)}]{2004ApJS..154...10F}
{Fazio}, G.~G., {et~al.} 2004, \apjs, 154, 10

\bibitem[{{Ferland} {et~al.}(2008){Ferland}, {Fabian}, {Hatch}, {Johnstone},
  {Porter}, {van Hoof}, \& {Williams}}]{2008MNRAS.386L..72F}
{Ferland}, G.~J., {Fabian}, A.~C., {Hatch}, N.~A., {Johnstone}, R.~M.,
  {Porter}, R.~L., {van Hoof}, P.~A.~M., \& {Williams}, R.~J.~R. 2008, \mnras,
  386, L72

\bibitem[{{Ferland} {et~al.}(2009){Ferland}, {Fabian}, {Hatch}, {Johnstone},
  {Porter}, {van Hoof}, \& {Williams}}]{2009MNRAS.392.1475F}
---. 2009, \mnras, 392, 1475

\bibitem[{{F{\"o}rster Schreiber} {et~al.}(2004){F{\"o}rster Schreiber},
  {Roussel}, {Sauvage}, \& {Charmandaris}}]{2004A&A...419..501F}
{F{\"o}rster Schreiber}, N.~M., {Roussel}, H., {Sauvage}, M., \&
  {Charmandaris}, V. 2004, \aap, 419, 501

\bibitem[{{Genzel} {et~al.}(1998)}]{1998ApJ...498..579G}
{Genzel}, R., {et~al.} 1998, \apj, 498, 579

\bibitem[{{Groves} {et~al.}(2008){Groves}, {Dopita}, {Sutherland}, {Kewley},
  {Fischera}, {Leitherer}, {Brandl}, \& {van Breugel}}]{2008ApJS..176..438G}
{Groves}, B., {Dopita}, M.~A., {Sutherland}, R.~S., {Kewley}, L.~J.,
  {Fischera}, J., {Leitherer}, C., {Brandl}, B., \& {van Breugel}, W. 2008,
  \apjs, 176, 438

\bibitem[{{Heckman} {et~al.}(1989){Heckman}, {Baum}, {van Breugel}, \&
  {McCarthy}}]{1989ApJ...338...48H}
{Heckman}, T.~M., {Baum}, S.~A., {van Breugel}, W.~J.~M., \& {McCarthy}, P.
  1989, \apj, 338, 48

\bibitem[{{Hicks} \& {Mushotzky}(2005)}]{2005ApJ...635L...9H}
{Hicks}, A.~K., \& {Mushotzky}, R. 2005, \apjl, 635, L9

\bibitem[{{Hicks} {et~al.}(2010){Hicks}, {Mushotzky}, \&
  {Donahue}}]{2010ApJ...719.1844H}
{Hicks}, A.~K., {Mushotzky}, R., \& {Donahue}, M. 2010, \apj, 719, 1844

\bibitem[{{Higdon} {et~al.}(2006){Higdon}, {Armus}, {Higdon}, {Soifer}, \&
  {Spoon}}]{2006ApJ...648..323H}
{Higdon}, S.~J.~U., {Armus}, L., {Higdon}, J.~L., {Soifer}, B.~T., \& {Spoon},
  H.~W.~W. 2006, \apj, 648, 323

\bibitem[{{Higdon} {et~al.}(2004)}]{2004PASP..116..975H}
{Higdon}, S.~J.~U., {et~al.} 2004, \pasp, 116, 975

\bibitem[{{Hill} \& {Oegerle}(1993)}]{1993AJ....106..831H}
{Hill}, J.~M., \& {Oegerle}, W.~R. 1993, \aj, 106, 831

\bibitem[{{Ho} \& {Keto}(2007)}]{2007ApJ...658..314H}
{Ho}, L.~C., \& {Keto}, E. 2007, \apj, 658, 314

\bibitem[{{Houck} {et~al.}(2004)}]{2004ApJS..154...18H}
{Houck}, J.~R., {et~al.} 2004, \apjs, 154, 18

\bibitem[{{Hu} {et~al.}(1985){Hu}, {Cowie}, \& {Wang}}]{1985ApJS...59..447H}
{Hu}, E.~M., {Cowie}, L.~L., \& {Wang}, Z. 1985, \apjs, 59, 447

\bibitem[{{Huber} \& {Herzberg}(1979)}]{Huber+Herzberg}
{Huber}, K.~P., \& {Herzberg}, G. 1979, {Molecular Spectra and Molecular
  Structure IV. Constants of Diatomic Molecules} (New York: Van Nostrand:
  Reinhold)

\bibitem[{{Hudson} {et~al.}(2010){Hudson}, {Mittal}, {Reiprich}, {Nulsen},
  {Andernach}, \& {Sarazin}}]{2010A&A...513A..37H}
{Hudson}, D.~S., {Mittal}, R., {Reiprich}, T.~H., {Nulsen}, P.~E.~J.,
  {Andernach}, H., \& {Sarazin}, C.~L. 2010, \aap, 513, A37+

\bibitem[{{Hunstead} {et~al.}(1978){Hunstead}, {Murdoch}, \&
  {Shobbrook}}]{1978MNRAS.185..149H}
{Hunstead}, R.~W., {Murdoch}, H.~S., \& {Shobbrook}, R.~R. 1978, \mnras, 185,
  149

\bibitem[{{Hunter} \& {Kaufman}(2007)}]{2007AJ....134..721H}
{Hunter}, D.~A., \& {Kaufman}, M. 2007, \aj, 134, 721

\bibitem[{{Jaffe} \& {Bremer}(1997)}]{1997MNRAS.284L...1J}
{Jaffe}, W., \& {Bremer}, M.~N. 1997, \mnras, 284, L1

\bibitem[{{Johnstone} {et~al.}(1987){Johnstone}, {Fabian}, \&
  {Nulsen}}]{1987MNRAS.224...75J}
{Johnstone}, R.~M., {Fabian}, A.~C., \& {Nulsen}, P.~E.~J. 1987, \mnras, 224,
  75

\bibitem[{{Johnstone} {et~al.}(2007){Johnstone}, {Hatch}, {Ferland}, {Fabian},
  {Crawford}, \& {Wilman}}]{2007MNRAS.382.1246J}
{Johnstone}, R.~M., {Hatch}, N.~A., {Ferland}, G.~J., {Fabian}, A.~C.,
  {Crawford}, C.~S., \& {Wilman}, R.~J. 2007, \mnras, 382, 1246

\bibitem[{{Kaneda} {et~al.}(2008){Kaneda}, {Onaka}, {Sakon}, {Kitayama},
  {Okada}, \& {Suzuki}}]{2008ApJ...684..270K}
{Kaneda}, H., {Onaka}, T., {Sakon}, I., {Kitayama}, T., {Okada}, Y., \&
  {Suzuki}, T. 2008, \apj, 684, 270

\bibitem[{{Kennicutt}(1990)}]{1990ASSL..161..405K}
{Kennicutt}, Jr., R.~C. 1990, in Astrophysics and Space Science Library, Vol.
  161, The Interstellar Medium in Galaxies, ed. {H.~A.~Thronson Jr.~\&
  J.~M.~Shull}, 405--435

\bibitem[{{Kennicutt}(1998)}]{1998ARA&A..36..189K}
{Kennicutt}, Jr., R.~C. 1998, \araa, 36, 189

\bibitem[{{Kennicutt} {et~al.}(2003)}]{2003PASP..115..928K}
{Kennicutt}, Jr., R.~C., {et~al.} 2003, \pasp, 115, 928

\bibitem[{{Kere{\v s}} {et~al.}(2005){Kere{\v s}}, {Katz}, {Weinberg}, \&
  {Dav{\'e}}}]{2005MNRAS.363....2K}
{Kere{\v s}}, D., {Katz}, N., {Weinberg}, D.~H., \& {Dav{\'e}}, R. 2005,
  \mnras, 363, 2

\bibitem[{{Laine} {et~al.}(2010){Laine}, {Appleton}, {Gottesman}, {Ashby}, \&
  {Garland}}]{2010AJ....140..753L}
{Laine}, S., {Appleton}, P.~N., {Gottesman}, S.~T., {Ashby}, M.~L.~N., \&
  {Garland}, C.~A. 2010, \aj, 140, 753

\bibitem[{{Laurent} {et~al.}(2000){Laurent}, {Mirabel}, {Charmandaris},
  {Gallais}, {Madden}, {Sauvage}, {Vigroux}, \&
  {Cesarsky}}]{2000A&A...359..887L}
{Laurent}, O., {Mirabel}, I.~F., {Charmandaris}, V., {Gallais}, P., {Madden},
  S.~C., {Sauvage}, M., {Vigroux}, L., \& {Cesarsky}, C. 2000, \aap, 359, 887

\bibitem[{{Lebouteiller} {et~al.}(2010){Lebouteiller}, {Bernard-Salas},
  {Sloan}, \& {Barry}}]{2010PASP..122..231L}
{Lebouteiller}, V., {Bernard-Salas}, J., {Sloan}, G.~C., \& {Barry}, D.~J.
  2010, \pasp, 122, 231

\bibitem[{{Leger} \& {Puget}(1984)}]{1984A&A...137L...5L}
{Leger}, A., \& {Puget}, J.~L. 1984, \aap, 137, L5

\bibitem[{{Leitherer} {et~al.}(1999){Leitherer}, {Schaerer}, {Goldader},
  {Gonz{\'a}lez Delgado}, {Robert}, {Kune}, {de Mello}, {Devost}, \&
  {Heckman}}]{1999ApJS..123....3L}
{Leitherer}, C., {Schaerer}, D., {Goldader}, J.~D., {Gonz{\'a}lez Delgado},
  R.~M., {Robert}, C., {Kune}, D.~F., {de Mello}, D.~F., {Devost}, D., \&
  {Heckman}, T.~M. 1999, \apjs, 123, 3

\bibitem[{{Li} \& {Draine}(2001)}]{2001ApJ...554..778L}
{Li}, A., \& {Draine}, B.~T. 2001, \apj, 554, 778

\bibitem[{{Martel} {et~al.}(2002){Martel}, {Sparks}, {Allen}, {Koekemoer}, \&
  {Baum}}]{2002AJ....123.1357M}
{Martel}, A.~R., {Sparks}, W.~B., {Allen}, M.~G., {Koekemoer}, A.~M., \&
  {Baum}, S.~A. 2002, \aj, 123, 1357

\bibitem[{{McNamara} \& {Nulsen}(2007)}]{2007ARA&A..45..117M}
{McNamara}, B.~R., \& {Nulsen}, P.~E.~J. 2007, \araa, 45, 117

\bibitem[{{McNamara} \& {O'Connell}(1989)}]{1989AJ.....98.2018M}
{McNamara}, B.~R., \& {O'Connell}, R.~W. 1989, \aj, 98, 2018

\bibitem[{{McNamara} \& {O'Connell}(1993)}]{1993AJ....105..417M}
---. 1993, \aj, 105, 417

\bibitem[{{Micelotta} {et~al.}(2010){Micelotta}, {Jones}, \&
  {Tielens}}]{2010A&A...510A..37M}
{Micelotta}, E.~R., {Jones}, A.~P., \& {Tielens}, A.~G.~G.~M. 2010, \aap, 510,
  A37+

\bibitem[{{O'Dea} {et~al.}(2004){O'Dea}, {Baum}, {Mack}, {Koekemoer}, \&
  {Laor}}]{2004ApJ...612..131O}
{O'Dea}, C.~P., {Baum}, S.~A., {Mack}, J., {Koekemoer}, A.~M., \& {Laor}, A.
  2004, \apj, 612, 131

\bibitem[{{O'Dea} {et~al.}(2008)}]{2008ApJ...681.1035O}
{O'Dea}, C.~P., {et~al.} 2008, \apj, 681, 1035

\bibitem[{{O'Dea} {et~al.}(2010)}]{2010ApJ...719.1619O}
{O'Dea}, K.~P., {et~al.} 2010, \apj, 719, 1619

\bibitem[{{Ogle} {et~al.}(2007){Ogle}, {Antonucci}, {Appleton}, \&
  {Whysong}}]{2007ApJ...668..699O}
{Ogle}, P., {Antonucci}, R., {Appleton}, P.~N., \& {Whysong}, D. 2007, \apj,
  668, 699

\bibitem[{{Ogle} {et~al.}(2010){Ogle}, {Boulanger}, {Guillard}, {Evans},
  {Antonucci}, {Appleton}, {Nesvadba}, \& {Leipski}}]{2010arXiv1009.4533O}
{Ogle}, P., {Boulanger}, F., {Guillard}, P., {Evans}, D.~A., {Antonucci}, R.,
  {Appleton}, P.~N., {Nesvadba}, N., \& {Leipski}, C. 2010, ArXiv e-prints

\bibitem[{{Peeters} {et~al.}(2004){Peeters}, {Spoon}, \&
  {Tielens}}]{2004ApJ...613..986P}
{Peeters}, E., {Spoon}, H.~W.~W., \& {Tielens}, A.~G.~G.~M. 2004, \apj, 613,
  986

\bibitem[{{Peterson} {et~al.}(2003){Peterson}, {Kahn}, {Paerels}, {Kaastra},
  {Tamura}, {Bleeker}, {Ferrigno}, \& {Jernigan}}]{2003ApJ...590..207P}
{Peterson}, J.~R., {Kahn}, S.~M., {Paerels}, F.~B.~S., {Kaastra}, J.~S.,
  {Tamura}, T., {Bleeker}, J.~A.~M., {Ferrigno}, C., \& {Jernigan}, J.~G. 2003,
  \apj, 590, 207

\bibitem[{{Quillen} {et~al.}(2008)}]{2008ApJS..176...39Q}
{Quillen}, A.~C., {et~al.} 2008, \apjs, 176, 39

\bibitem[{{Rafferty} {et~al.}(2008){Rafferty}, {McNamara}, \&
  {Nulsen}}]{2008ApJ...687..899R}
{Rafferty}, D.~A., {McNamara}, B.~R., \& {Nulsen}, P.~E.~J. 2008, \apj, 687,
  899

\bibitem[{{Rafferty} {et~al.}(2006){Rafferty}, {McNamara}, {Nulsen}, \&
  {Wise}}]{2006ApJ...652..216R}
{Rafferty}, D.~A., {McNamara}, B.~R., {Nulsen}, P.~E.~J., \& {Wise}, M.~W.
  2006, \apj, 652, 216

\bibitem[{{Rieke} {et~al.}(2004)}]{2004ApJS..154...25R}
{Rieke}, G.~H., {et~al.} 2004, \apjs, 154, 25

\bibitem[{{Rigby} \& {Rieke}(2004)}]{2004ApJ...606..237R}
{Rigby}, J.~R., \& {Rieke}, G.~H. 2004, \apj, 606, 237

\bibitem[{{Rosenberg} {et~al.}(2008){Rosenberg}, {Wu}, {Le Floc'h},
  {Charmandaris}, {Ashby}, {Houck}, {Salzer}, \&
  {Willner}}]{2008ApJ...674..814R}
{Rosenberg}, J.~L., {Wu}, Y., {Le Floc'h}, E., {Charmandaris}, V., {Ashby},
  M.~L.~N., {Houck}, J.~R., {Salzer}, J.~J., \& {Willner}, S.~P. 2008, \apj,
  674, 814

\bibitem[{{Roussel} {et~al.}(2001){Roussel}, {Sauvage}, {Vigroux}, \&
  {Bosma}}]{2001A&A...372..427R}
{Roussel}, H., {Sauvage}, M., {Vigroux}, L., \& {Bosma}, A. 2001, \aap, 372,
  427

\bibitem[{{Roussel} {et~al.}(2007)}]{2007ApJ...669..959R}
{Roussel}, H., {et~al.} 2007, \apj, 669, 959

\bibitem[{{Sakon} {et~al.}(2004){Sakon}, {Onaka}, {Ishihara}, {Ootsubo},
  {Yamamura}, {Tanab{\'e}}, \& {Roellig}}]{2004ApJ...609..203S}
{Sakon}, I., {Onaka}, T., {Ishihara}, D., {Ootsubo}, T., {Yamamura}, I.,
  {Tanab{\'e}}, T., \& {Roellig}, T.~L. 2004, \apj, 609, 203

\bibitem[{{Schutte} {et~al.}(1993){Schutte}, {Tielens}, \&
  {Allamandola}}]{1993ApJ...415..397S}
{Schutte}, W.~A., {Tielens}, A.~G.~G.~M., \& {Allamandola}, L.~J. 1993, \apj,
  415, 397

\bibitem[{{Sloan} {et~al.}(1999){Sloan}, {Hayward}, {Allamandola}, {Bregman},
  {Devito}, \& {Hudgins}}]{1999ApJ...513L..65S}
{Sloan}, G.~C., {Hayward}, T.~L., {Allamandola}, L.~J., {Bregman}, J.~D.,
  {Devito}, B., \& {Hudgins}, D.~M. 1999, \apjl, 513, L65

\bibitem[{{Smith} {et~al.}(2007{\natexlab{a}}){Smith}, {Armus}, {Dale},
  {Roussel}, {Sheth}, {Buckalew}, {Jarrett}, {Helou}, \&
  {Kennicutt}}]{2007PASP..119.1133S}
{Smith}, J.~D.~T., {Armus}, L., {Dale}, D.~A., {Roussel}, H., {Sheth}, K.,
  {Buckalew}, B.~A., {Jarrett}, T.~H., {Helou}, G., \& {Kennicutt}, Jr., R.~C.
  2007{\natexlab{a}}, \pasp, 119, 1133

\bibitem[{{Smith} {et~al.}(2007{\natexlab{b}})}]{2007ApJ...656..770S}
{Smith}, J.~D.~T., {et~al.} 2007{\natexlab{b}}, \apj, 656, 770

\bibitem[{{Smith} {et~al.}(2004)}]{2004AJ....128.1558S}
{Smith}, R.~J., {et~al.} 2004, \aj, 128, 1558

\bibitem[{{Snijders} {et~al.}(2007){Snijders}, {Kewley}, \& {van der
  Werf}}]{2007ApJ...669..269S}
{Snijders}, L., {Kewley}, L.~J., \& {van der Werf}, P.~P. 2007, \apj, 669, 269

\bibitem[{{Sparks} {et~al.}(1989){Sparks}, {Macchetto}, \&
  {Golombek}}]{1989ApJ...345..153S}
{Sparks}, W.~B., {Macchetto}, F., \& {Golombek}, D. 1989, \apj, 345, 153

\bibitem[{{Springel} {et~al.}(2005){Springel}, {Di Matteo}, \&
  {Hernquist}}]{2005ApJ...620L..79S}
{Springel}, V., {Di Matteo}, T., \& {Hernquist}, L. 2005, \apjl, 620, L79

\bibitem[{{Stocke} {et~al.}(1991){Stocke}, {Morris}, {Gioia}, {Maccacaro},
  {Schild}, {Wolter}, {Fleming}, \& {Henry}}]{1991ApJS...76..813S}
{Stocke}, J.~T., {Morris}, S.~L., {Gioia}, I.~M., {Maccacaro}, T., {Schild},
  R., {Wolter}, A., {Fleming}, T.~A., \& {Henry}, J.~P. 1991, \apjs, 76, 813

\bibitem[{{Thornley} {et~al.}(2000){Thornley}, {Schreiber}, {Lutz}, {Genzel},
  {Spoon}, {Kunze}, \& {Sternberg}}]{2000ApJ...539..641T}
{Thornley}, M.~D., {Schreiber}, N.~M.~F., {Lutz}, D., {Genzel}, R., {Spoon},
  H.~W.~W., {Kunze}, D., \& {Sternberg}, A. 2000, \apj, 539, 641

\bibitem[{{Treyer} {et~al.}(2010)}]{2010ApJ...719.1191T}
{Treyer}, M., {et~al.} 2010, \apj, 719, 1191

\bibitem[{{Van Kerckhoven} {et~al.}(2000)}]{2000A&A...357.1013V}
{Van Kerckhoven}, C., {et~al.} 2000, \aap, 357, 1013

\bibitem[{{Voit}(1992{\natexlab{a}})}]{1992MNRAS.258..841V}
{Voit}, G.~M. 1992{\natexlab{a}}, \mnras, 258, 841

\bibitem[{{Voit}(1992{\natexlab{b}})}]{1992ApJ...399..495V}
---. 1992{\natexlab{b}}, \apj, 399, 495

\bibitem[{{Voit} \& {Donahue}(1997)}]{1997ApJ...486..242V}
{Voit}, G.~M., \& {Donahue}, M. 1997, \apj, 486, 242

\bibitem[{{Wu} {et~al.}(2010)}]{2010ApJ...723..895W}
{Wu}, Y., {et~al.} 2010, \apj, 723, 895

\bibitem[{{Zabludoff} {et~al.}(1990){Zabludoff}, {Huchra}, \&
  {Geller}}]{1990ApJS...74....1Z}
{Zabludoff}, A.~I., {Huchra}, J.~P., \& {Geller}, M.~J. 1990, \apjs, 74, 1

\bibitem[{{Zakamska}(2010)}]{2010Natur.465...60Z}
{Zakamska}, N.~L. 2010, \nat, 465, 60

\end{thebibliography}

\clearpage

\end{document}